\documentclass[11pt]{article}

 \usepackage{mathrsfs}
 \usepackage[utf8]{inputenc}
 \usepackage{epsfig,graphicx,color,mathtools}
 \usepackage{subfig}
 \usepackage{latexsym}                 
 \usepackage{amsmath,amsfonts,verbatim,xkeyval,bm, amsthm, upgreek, amscd, wasysym, amssymb}
 \allowdisplaybreaks
 \usepackage{cases}

 \usepackage[english]{babel}
 \usepackage{listings}
 \usepackage{multicol}
 \usepackage[colorlinks]{hyperref}  
 \hypersetup{hidelinks}  
 \usepackage{tikz-cd}
 \usepackage{booktabs}
\usepackage[numbers, sort&compress]{natbib}
 \usepackage{sidecap}
 \usepackage[font=small,labelfont=bf]{caption}
 \usepackage{float}
\usepackage{titlesec}
\usepackage{soul}
\usepackage[normalem]{ulem}
\usepackage[title]{appendix}
\newcommand{\stkout}[1]{\ifmmode\text{\sout{\ensuremath{#1}}}\else\sout{#1}\fi}

\titleformat{\paragraph}
{\normalfont\normalsize\bfseries}{\theparagraph}{1em}{}
\titlespacing*{\paragraph}
{0pt}{3.25ex plus 1ex minus .2ex}{1.5ex plus .2ex}
 
\newtheorem{Theorem}{Theorem}[section]

\newtheorem{Definition}[Theorem]{Definition}

\newtheorem{Proposition}[Theorem]{Proposition}

\newtheorem{Observation}{Observation}[section]

\let\Grad=\nabla

\def\vet#1{{\bm #1}}
\def\build#1_#2^#3{\mathrel{
\mathop{\kern 0pt#1}\limits_{#2}^{#3}}}
\def\reali{\mathbb{R}}

\def\e{\mathrm{e}}
\def\i{\mathrm{i}}
\def\AMD{\mathrm{AMD}}

\def\AA{\mathcal{A}}

\def\CC{\mathcal{C}}
\def\DD{\mathcal{D}}
\def\DELTA{\Delta}

\def\Cscr{\mathcal{C}}

\def\Escr{\mathcal{E}}

\def\Gscr{\mathcal{G}}
\def\Hscr{\mathcal{H}}
\def\Kscr{\mathcal{K}}

\def\Oscr{\mathcal{O}}

\def\Sscr{\mathcal{S}}

\def\Zscr{\mathcal{Z}}
\def\Kscr{\mathcal{K}}

\def\rho{\varrho}

\def\poisson#1#2{\lbrace #1,#2 \rbrace}

\def\t#1{\tilde{#1}}

\definecolor{greenCUD}{RGB}{0, 158, 115}

\title{Bifurcation sequences in the secular 3D planetary 3-Body problem:\\ a geometric approach}
\author{
  {\bf Rita Mastroianni}\\
  {\small Advanced Concepts Team, European Space Agency, European Space Technology and Research Centre}\\
  {\small Keplerlaan 1, 2201 AZ Noordwijk}\\
  {\bf Antonella Marchesiello}\\
  {\small Department of Applied Mathematics, Faculty of Information Technology, Czech Technical University in Prague,}\\
  {\small Th\' akurova 9, 160 00 Prague}\\
  {\bf Christos Efthymiopoulos}\\
  {\small Dipartimento di Matematica ``Tullio Levi-Civita'', Universit\`a degli Studi di Padova,}\\
  {\small via Trieste 63, 35121 Padova,}\\
  {\bf Giuseppe Pucacco}\\
  {\small Dipartimento di Fisica, Universit\`a
  degli Studi di Roma ``Tor Vergata'',}\\
  {\small via della Ricerca Scientifica 1, 00133 Roma,}\\
  [1.2ex]
  {\small e-mails:
  {\tt rita.mastroianni@esa.int, marchant@fit.cvut.cz}}\\
  {\small
  {\tt cefthym@math.unipd.it, pucacco@roma2.infn.it}}
}

\date{}
\begin{document}

\pagenumbering{arabic}

\maketitle
\begin{abstract}
We implement the geometric method proposed in (\cite{kum1976},~\cite{cusbat1997},~\cite{marpuc2016}) to analytically predict the sequence of bifurcations leading to a change of stability and/or the appearance of new periodic orbits in the secular 3D planetary three body problem. Stemming from the analysis in~\cite{maseft2023}, we examine various normal form models as regards the extent to which they lead to a phase space dynamics qualitatively similar as that in the complete system. For fixed total angular momentum, the phase space in Hopf variables is the 3D sphere, and the complete sequence of bifurcations of new periodic orbits can be recovered through formulas yielding the tangencies or degenerate intersections between the sphere and the surfaces of a constant second integral of motion of the normal form flow. In particular, we find the critical values of the second integral giving rise to pitchfork and saddle-node bifurcations of new periodic orbits in the system. This analysis renders possible to predict the most important structural changes in the phase space, as well as the emergence of new possible stable periodic planetary orbital configurations which can take place as the mutual inclination between the two planets is allowed to increase. 

\end{abstract}

\section{Introduction}
\label{Intro}
A classical problem of Celestial Mechanics is the \textit{planetary three-body problem}, which deals with the motion of two planets of non-negligible mass under the gravitational attraction of a star,  and including the gravitational perturbation of each planet on the other. In Poincar\'{e} heliocentric canonical variables, the problem is defined by the Hamiltonian
\begin{equation}
\label{ham.3BP}
\Hscr(\vet{r_2},\vet{r_3},\vet{p_2},\vet{p_3})=\frac{\vet{p_2}^2}{2\, m_2}-\frac{\mathcal{G}\, m_0\, m_2}{r_2}+\frac{\vet{p_3}^2}{2 \,m_3}-\frac{\mathcal{G}\, m_0\, m_3}{r_3}+\frac{(\vet{p_2}+\vet{p_3})^2}{2\, m_0}-\frac{\mathcal{G}\, m_2\, m_3}{|\vet{r_2}-\vet{r_3}|}\, ,
\end{equation}
where $m_0 \gg m_2,\, m_3\, $. The parameter $m_0$ represents the mass of the star, and $m_{i}\,$, $\vet{p_i}\,$, $\vet{r_{i}}\,$, $i=2,3$ are the masses, barycentric momenta and heliocentric position vectors of the two planets, with $r_i=|\vet{r_i}|\,$. We use the numbering (2,3) instead of (1,2) for the two planets, to have continuity with the notation of the work \cite{maseft2023} which is the starting point whereby departs our analysis in the present paper.  

The historical importance of the planetary three body problem in the conceptual growth of dynamical systems' theory in general cannot be overemphasized. Stemming from Laplace and Lagrange's investigations on the `secular' (long-term) evolution of the planetary orbits in our own solar system (\cite{lag1772},~\cite{lap1878}), the model was shown by Poincar\'{e} to serve as an archetype for the development of key ideas in dynamical systems, including the importance of periodic orbits in sculpting the phase-space structure, as well as the concepts of (near-)integrability and chaos (\cite{poi1893}). 

In recent years, new interest in the problem has emerged owing to its close relevance to understanding planetary motions in \textit{extrasolar} planetary systems.  We refer the reader to the introduction in  \cite{maseft2023} for an extended literature review. Most recent studies stem from the planar case, i.e., where the orbits of all three bodies are co-planar. A basic known configuration of planetary orbits in such systems is the one of \textit{apsidal corotation resonance} (ACR). In this configuration, the two planets move in osculating Keplerian ellipses whose pericenters undergo precession both with the same `secular' (i.e., slow) frequency, remaining always either anti-aligned (hereafter `state A') or always aligned (`state B'). Mathematically, the ACR states are periodic orbits corresponding to fixed points of the \textit{secular Hamiltonian} of the system (see, for example,~\cite{lauetal2002},~\cite{beaetal2003},~\cite{leepea2003}). This is an integrable Hamiltonian produced from the Hamiltonian (\ref{ham.3BP}) taking the restriction to co-planar orbits, and averaging with respect to the planetary mean anomalies (see section \ref{sec:Ham}).  

While the planar secular planetary three body problem is well understood, several of its phase space features are already non-trivial and give rise to substantial complication in the analysis when passing from the 2D to the full 3D problem. In the latter, the planetary orbits are allowed to have non-zero mutual inclination, whose value becomes an additional parameter of the problem. Several planetary systems have been observed in a state of mutual inclination of several degrees (see \cite{nao2016} for a review). The main change from 2D to the 3D case is that the 3D secular Hamiltonian is \textit{non-integrable}. After performing Jacobi reduction of the nodes, the 3D Hamiltonian is reduced to one with three degrees of freedom (DOF). The conservation of the total angular momentum leads to the existence of one global invariant besides the energy, namely the \textit{Angular Momentum Deficit}  
\begin{equation}
\label{eq.AMD}
\AMD=L_2+L_3-L_z~~.
\end{equation}
In Eq.~\eqref{eq.AMD} $L_2$, $L_3$ are the angular momenta of the circular orbits of semi-major axes $a_2,a_3$ equal to those of the two planets, and $L_z$ the modulus of the total angular momentum normal to the system's Laplace plane. {Ignoring the precession of the common nodal line of the planetary orbits in the Laplace plane, the conservation of the angular momentum essentially means that the system can be formally treated as with 2 DOF. The lack of a third global integral implies that the system can admit also chaotic solutions.} As shown in the recent work \cite{maseft2023}, chaotic motions actually occupy a substantial volume in phase space as the mutual inclination of the two planets approaches the critical limit of birth of the so-called `Lidov-Kozai' regime, where the configuration with one of the two planets in circular orbit changes stability character, turning from stable to unstable. 

On the other hand, numerical investigations (see \cite{maseft2023} and references therein, as well as section \ref{sec:Ham} below) show that the phase space of 3D secular motions acquires a rich structure at levels of the mutual inclination well before the Lidov-Kozai regime. Our focus in the present paper is in the \textit{near-integrable regime} arising at moderate values of the mutual inclination. Numerical experiments show that, while regular, the structure of the phase space in this regime exhibits a substantial departure from the structure of the phase space in the corresponding planar case. The main difference regards the kind of periodic orbits which represent stable possible orbital configurations for the two planets in the two cases. Most notably, in the 3D case we witness the birth of a rich variety of new possible periodic states with features quite distinct from those of the ACR states of the planar problem. Our purpose is to show that such states can be efficiently predicted and/or classified through an analysis of the \textit{sequences of bifurcations} of periodic orbits, which stem from the ACR states (A or B) as we gradually increase the level of mutual orbital inclination. More specifically, fixing  a choice of values of the star's and planets' masses $(m_0,m_2,m_3)$, as well as the planets' heliocentric semi-major axes $(a_2,a_3)$, and gradually increasing the level of mutual inclination of the orbits, we numerically observe the emergence of a certain sequence of bifurcations. This can be visualised through numerical phase portraits (surfaces of section) plotted in suitably defined Poincar\'e variables. Altering then the initial parameters $m_i$ and $a_i$, and repeating the study, we observe that we are led to the possibility of various distinct sequences of bifurcations, some examples of which are seen in the sequence of phase portraits of figures 17 and 18 of~\cite{maseft2023}. As shown below, all such sequences belong to a class consistent with the general topological features of the phase space of the dynamical system here discussed. Our goal then becomes to give an analytical method to predict which particular sequence of bifurcations will emerge as a function of the model's parameters $(m_i,a_i)$ and the value of the system's $\AMD$. On physical ground, this means to predict and classify the most important possible orbital configurations (besides the ACRs) to which a planetary system with substantial level of mutual inclination can settle after its formation.

The main steps of the geometric method here implemented to achieve these goals can be summarized as follows: as a preliminary step, in \textit{Section \ref{sec:Ham}} we follow the formalism introduced in \cite{maseft2023} to arrive at a starting 3D secular Hamiltonian model of a given system with fixed AMD. This is written as 
\begin{equation}\label{h3ddecompo}
\Hscr_{sec}=\Hscr_{planar}(\mathbf{X},\mathbf{Y})+\Hscr_{space}(\mathbf{X},\mathbf{Y};\AMD)~~.
\end{equation}
In Eq.~\eqref{h3ddecompo}, $(\mathbf{X},\mathbf{Y})$ are Poincar\'{e} canonical variables for the two planets $(X_2,Y_2)$, $(X_3,Y_3)$, which scale nearly linearly with the planets' eccentricities. The Hamiltonian $\Hscr_{planar}$ admits as a second integral the quantity
\begin{equation}\label{sig0}
\sigma_0 = (X_2^2+Y_2^2+X_3^2+Y_3^2)/2\, .
\end{equation}
Passing, then, to the 3D case, several integrable models formally similar to $\Hscr_{planar}$ can be constructed, stemming from different (multipolar or in the order of the orbital eccentricities) truncations of the full 3D Hamiltonian $\Hscr_{sec}$. To illustrate all aspects of the geometric method here implemented, in the following sections we focus on three different integrable models of such type. 

i) A basic first model is given by the Hamiltonian
\begin{equation}\label{def.h0}
\Hscr_{int}=\Hscr_{planar}+\Hscr_{0,space} 
\end{equation}
where $\Hscr_{0,space}$ stems from the decomposition $\Hscr_{space}=\Hscr_{0,space}+\Hscr_{1,space}$, such that  $\Hscr_{0,space}$ also admits $\sigma_0$ as a second integral.

ii) Rewriting the Hamiltonian as
\begin{equation}\label{h3ddecompo2}
\Hscr_{sec}=\Hscr_{int}(\mathbf{X},\mathbf{Y})+\Hscr_{1,space}(\mathbf{X},\mathbf{Y};\AMD)~~,
\end{equation}
and using canonical perturbation theory with Lie series (see subsection~\ref{subsec:Norm_Hint1}), we are led to our second integrable model, i.e., the \textit{secular normal form} 
\begin{equation}\label{zint}
\Hscr_{int}^{(1)}=\Hscr_{int}+{1\over 2}<\{\Hscr_{1,space},\chi\}>~~.
\end{equation}
In~\eqref{zint}, $\chi$ is a Lie generating function acting to normalize the Hamiltonian (\ref{h3ddecompo2}) with respect to all terms of $\Hscr_{1,space}$ not `in normal form' (see subsection~\ref{subsec:Norm_Hint1}), and $\{\cdot,\cdot\}$ denotes the Poisson bracket operation. The averaging $<\cdot>$ in the last term of Eq.~\eqref{zint} means to eliminate from $\{\Hscr_{1,space},\chi\}$ all the terms not in normal form. 

iii) Our third model considered, $\tilde{\Hscr}_{int}^{(N_P=3,N_{bk}=4)}$, is similar to the secular normal form $\Hscr_{int}^{(1)}$, but obtained within the framework of the octupole approximation of the initial Hamiltonian truncated up to fourth order in the orbital eccentricities. Our interest in such a model, which becomes precise in cases of so-called `hierarchical' systems (planets in nearly circular orbits with $a_2/a_3<<1$), is motivated by the simplicity in the form of the Hamiltonian, which allows for analytical formulas  {concerning} the dependence of the sequence of bifurcations on the system's parameters. 

Focusing, now, on the above three examples of integrable models approximating the 3D secular dynamics, we implement the  geometric method exposed in detail in \textit{Section \ref{sec:Bif}} to obtain theoretical predictions for the sequences of bifurcations of periodic orbits arising in each model. New periodic orbits appear stemming from either of the ACR modes A or B, through both saddle-node and pitchfork bifurcations. The sequences are parameterized altering the values of each of the integrals of the energy $\Escr$ and angular momentum $\sigma_0$. The method relies on the use of suitable Hopf variables $\sigma_1,\sigma_2,\sigma_3$ defined so that, for fixed $\sigma_0$, the corresponding reduced manifold in phase space is the 3-sphere $\sigma_1^2+\sigma_2^2+\sigma_3^2=\sigma_0^2$. All quasi-periodic motions of the system are represented by the curves of transverse intersections of the 2D surfaces of constant energy (e.g. $\Hscr_{int}(\sigma_0,\sigma_1,\sigma_3)=\mathcal{E}$) with a sphere of fixed $\sigma_0$. All periodic orbits of the system can then be classified in terms of two different kinds of geometric connections between constant energy surfaces and the spheres of different radii $\sigma_0$: i) tangencies, and ii) transverse intersections, where the constant energy condition leads to a surface of degenerate form. Section \ref{sec:Bif} makes a systematic classification of these cases, the corresponding type of stability of the resulting periodic orbits, as well as the various possibilities for the resulting sequences of bifurcations, depending on the chosen integrable model. Whenever possible, we also provide the corresponding analytical formulas pertinent to the above analysis.  {Note that our analysis as above 
is strictly valid in the regular set of initial conditions which avoid the singularities mentioned in~\cite{paletal2013},~\cite{paletal2015} (see also ~\cite{ferosa1994}). In ~\cite{paletal2013}, special motions of the inner body leading to singularities are considered. They could be: i) circular trajectories, ii) coplanar motions and iii) rectilinear motions. In our work, we do not include iii), since it would imply $\e_2=1$, while we are interested in motions such that $\e_2\in$ $[0, 1)$. The other two cases will be discussed in Section~\ref{subsec:Hopf}.}

To verify the validity of the formulas deduced in section \ref{sec:Bif}, \textit{Section \ref{sec:Apply}} implements the method to each of the models $\Hscr_{int}$, $\Hscr_{int}^{(1)}$, or $\tilde{\Hscr}_{int}^{(N_P=3,N_{bk}=4)}$. Besides showing the ability of the method to predict the correct sequence of bifurcations observed in each model, a key result is that even small differences in the chosen model can lead to drastic differences as regards the observed sequence of bifurcations. We explain the geometric origin of such differences. Comparing to the initial non-integrable Hamiltonian (Eq.~\eqref{h3ddecompo}), we find that the model $\Hscr_{int}^{(1)}$, which represents a higher order normal form for the secular Hamiltonian, yields also the best predictions as regards both the correct thresholds in parameter values and the form of the sequence of bifurcations. Of course, the question of how precisely a secular normal form model can approximate the true dynamics can only be answered through case by case study, i.e, studying various cases of exoplanetary systems. While this is beyond our present scope, we stress that the method exposed in Section \ref{sec:Bif} is equally applicable to all such cases. 

\textit{Section~\ref{sec:Conclusions}} gives a summary of the conclusions from the present study.

\section{Hamiltonian model and numerical phase portraits}
\label{sec:Ham}
\subsection{Secular Hamiltonian}
\label{subsec:Hamsec}
As in section $2.2$ of~\cite{maseft2023}, we derive a secular model $\Hscr_{sec}$ for the Hamiltonian~\eqref{ham.3BP} by performing averaging `by scissors' with respect to the fast angles. Let $(a,\e,\i,M,\omega,\Omega)$ be the Keplerian elements of a body (semi-major axis, eccentricity, inclination, mean anomaly, argument of the periastron, argument of the nodes). Denoting $\lambda=M+\omega+\Omega$, $\varpi=\omega+\Omega$ the mean longitude and longitude of the periastron respectively, the secular model is obtained averaging $\Hscr$ with respect to the fast angles $M_2\,$, $M_3\,$:
\begin{equation}\label{hamscis}
\begin{aligned}
\Hscr_{sec}&={1\over 4\pi^2}
\int_{0}^{2\pi}\!\!\!\int_{0}^{2\pi} \Hscr(\vet{r_2},\vet{r_3},\vet{p_2},\vet{p_3}) \,dM_{2}\,dM_{3} \\
&= -{\mathcal{G} m_0 m_2\over 2a_2}-{\mathcal{G} m_0 m_3\over 2a_3}+{\mathcal{G}  m_2^2\over 2a_2}+{\mathcal{G} m_3^2\over 2a_3}-{\mathcal{G} m_2 m_3\over a_3}
+\mathcal{R}_{sec}(a_2,a_3,\e_2,\e_3,\i_2,\i_3,\omega_2,\omega_3,\Omega_2-\Omega_3),
\end{aligned}
\end{equation}
where
\begin{equation}\label{remainder}
\mathcal{R}_{sec}= {1\over 4\pi^2}
\int_{0}^{2\pi}\int_{0}^{2\pi}-\frac{\mathcal{G}\, m_2\, m_3}{r_3}\left(-\frac{1}{2}\frac{r_2^2}{r_3^2}+\frac{3}{2}\frac{(\vet{r}_{2}\cdot\vet{r}_{3})^2}{r_{3}^4} +\ldots\right) \,dM_{2}\,dM_{3}\, .
\end{equation}
Keeping only the lowest order term in the integrand of~\eqref{remainder}, proportional to $(r_2/r_3)^2\,$, leads to the so-called \textit{quadrupole} approximation; the next truncation (up to terms proportional to $(r_2/r_3)^3\,$) is the \textit{octupole} approximation, etc. The integrals of any multipole approximation  can be computed in so-called \textit{closed form} (i.e. without expansions in the eccentricities), by avoiding completely the series reversion of Kepler's equation, using, instead, the change of variables $M_2\rightarrow u_2$ (eccentric anomaly), $M_3\rightarrow f_3\,$ (true anomaly).

Following, now, section $2.3$ of~\cite{maseft2023}, we first Jacobi-reduce the Hamiltonian (\ref{hamscis}), introducing two `book-keeping symbols'  {($\varepsilon$, $\eta$)} which keeps track of the order of smallness in eccentricity and mutual inclination of all the terms in the Hamiltonian expansion. 
 {We briefly recall in the following the main passages of this procedure. We first introduce the following canonical transformation (see also~\cite{libhen2007})
\begin{equation}\label{Delaujac}
\begin{aligned}
&\Lambda_j=L_j\, , & \qquad\qquad &\lambda_j=M_j+\omega_j+\Omega_j\, ,\\
&W_j=L_j-G_j\,   , & \qquad\qquad &w_j=-\omega_j\, ,\\
&R_2=L_2-H_2\, , & \qquad\qquad & \theta_{r2}=\Omega_3-\Omega_2\, , \\
&R_3=L_2+L_3-H_2-H_3=\AMD\, , & \qquad\qquad &\theta_{r3}=-\Omega_3\, ,
\end{aligned}
\end{equation}
where $j=2,3$ and 
\begin{align}\label{Delau}
&L_j=m_j\sqrt{\mathcal{G}\,m_0\,a_j}\, , & &l_j=M_j\, ,\notag\\
&G_j=L_j\sqrt{1-\e^2_j}\, , & &g_j=\omega_j\, , \\
&H_j=G_j\cos(\i_j)\, , & &h_j=\Omega_j\, ,\notag
\end{align}
are the Delaunay variables, and then we express the Hamiltonian given by Eq.~\eqref{hamscis} as:
$$
\Hscr_{sec}\equiv \Hscr_{sec}(\Lambda_2, \Lambda_3, W_2, W_3, R_2, R_3, w_2, w_3, \theta_{r2}).
$$
From Hamilton's equation of the latter, it is easy to recover the constancy of the semi-major axes $a_2$, $a_3$ and of the $\AMD$. The crucial invariance property of the Hamiltonian is given by the equation $\dot{\theta}_{r2}=(\partial \Hscr_{sec}/\partial R_2)|_{\theta_{r2}=\pi}=0$, corresponding to the invariance in time of the relation $\theta_{r2}=\Omega_3-\Omega_2 = \pi$ for all trajectories. This means that this substitution can be made directly in the above Hamiltonian. However, after this substitution, while the evolution of the eccentricity vector can be obtained by Hamilton's equation, the evolution of inclination should be obtained directly from the conservation of the angular momentum (see Eq. (17) of~\cite{maseft2023}).}

 {
In the practical implementation, we can leave the dependence of the Hamiltonian $\Hscr_{sec}$ on the action variables $(\Lambda_j, W_j, R_j)$ through the elements $(a_j, \e_j, \i_j) $ $j=2,3$ and we observe that, after the substitution $\theta_{r2}=\pi$, the Hamiltonian depends on the inclinations \textit{only} through the trigonometric combination $\cos(\i_2+\i_3)$. This geometrical fact is crucial for the following reduction. In fact, we can introduce the \textit{book-keeping control identities}: 
\begin{align}\label{riduzione.cos.sen}
&\cos(\i_2)\cos(\i_3)\!=\!\varepsilon^2\eta\left(\cos(\i_2)\cos(\i_3)-1\right)+1\, , &\!\!\!\! 
&\sin(\i_2)\sin(\i_3)\!=\!\varepsilon^2\eta\sin(\i_2)\sin(\i_3)\, ,
\end{align}
where $\Oscr(\varepsilon^2)$ stands for `second order in the eccentricities and inclinations' while $\eta$ is the book-keeping symbol separating the Hamiltonian terms depending on powers of the quantity $\cos(\i_2 + \i_3 )$ from
those not depending on the mutual inclination $\i_{mut}$. Moreover, we use the substitution rule $\sin(\i_2)\sin(\i_3)=\cos(\i_2)\cos(\i_3)+F(a_2, a_3, \e_2, \e_3; L_z)$
(where the function $F$ is explicitly reported in Eq. (19) of~\cite{maseft2023}), determined by the constancy of the angular momentum. Substituting the above expressions into the Hamiltonian, and truncating the resulting expression up to a preselected \textit{maximum order in book-keeping} $N_{bk}$, by symmetry the terms in equal powers of the products $\sin(\i_2)\sin(\i_3)$ and $\cos(\i_2)\cos(\i_3)$ are opposite, and thus they are canceled. Hence, the Hamiltonian resumes the form:
\begin{equation}\label{hambketa}
\mathcal{H}_{sec}=\sum_{s_1=0}^{N_{bk}/2}\eta^{s_1}\varepsilon^{2s_1}\mathcal{H}_{sec,s_1}(\e_2,\e_3,w_2,w_3;a_2,a_3,L_z)~~. 
\end{equation} We finally set $\varepsilon=1$ and $\eta=1$ (as implied by the identities \eqref{riduzione.cos.sen}) of the book-keeping parameters, and write the truncated (up to book-keeping order $N_{bk}$) Hamiltonian as:
\begin{equation}\label{hamplsp}
\mathcal{H}_{sec}=\mathcal{H}_{planar}+\mathcal{H}_{space} 
\end{equation}
where
\begin{equation}\label{hamplspanal}
\begin{aligned}
&\mathcal{H}_{planar}= \mathcal{H}_{sec,0}(\e_2,\e_3,w_2-w_3;a_2,a_3)\, , 
\qquad &\mathcal{H}_{space}= \sum_{s_1=1}^{N_{bk}/2}\mathcal{H}_{sec,s_1}(\e_2,\e_3,w_2,w_3;a_2,a_3,L_z)~.
\end{aligned}
\end{equation}
Finally, we expand the Hamiltonian in the orbital eccentricities, as explained in Step $3$ of~\cite{maseft2023}. The angle $\theta_{r3}$ is ignorable, hence, the action variable $R_3$ is constant of motion. Considering $R_3$ as a parameter, we can substitute it in terms of the (also constants) $(L_2,L_3,L_z)$ via the relation $R_3=L_2+L_3-L_z$. Then, we obtain the following decomposition of the Hamiltonian: 
}
\begin{equation}
\begin{aligned}
&\Hscr_{sec}( w_2, w_3,W_2,\,W_3; L_z)=\Hscr_{planar}(w_2-w_3,W_2,W_3)
+\Hscr_{space}(w_2,w_3,W_2,W_3; L_z)\, .
\end{aligned}
\end{equation}
 {The Hamiltonian can now be formally considered as with 2 DOF. Furthermore,} the spatial part $\Hscr_{space}$ can be split as $\Hscr_{space}=\Hscr_{0,{space}}+\Hscr_{1,{space}}$, where $\Hscr_{0,{space}}$ contains terms independent of the angles or depending trigonometrically on the difference $w_2-w_3=\omega_3-\omega_2$, and $\Hscr_{1,{space}}$ contains terms depending trigonometrically on combinations of the angles $w_2,w_3$ other than their difference. Then we write
\begin{equation}\label{hamdecompo}
\begin{aligned}
\Hscr_{sec}&=\underbrace{\Hscr_{planar}( w_2-w_3,W_2,W_3) + \Hscr_{0,space}(w_2-w_3,W_2,W_3; L_z)}_{\textbf{\textit{ Integrable part}}:=\Hscr_{int}} + \Hscr_{1,space}(w_2,w_3,W_2,W_3; L_z)\, .
\end{aligned}
\end{equation}
The first two terms in the above expression give rise to  {the Hamiltonian}
\begin{equation}\label{hamint}
\begin{aligned}
&\Hscr_{int}( w_2-w_3,W_2,W_3;\AMD)= \Hscr_{planar}( w_2-w_3,W_2,W_3)+ \Hscr_{0,space}( w_2-w_3,W_2,W_3;\AMD)\, ,
\end{aligned}
\end{equation}
where the $\AMD$ is defined as in Eq.~\eqref{eq.AMD}.  {The Hamiltonian $\Hscr_{int}$ admits a second integral of motion $J=(W_2+W_3)/2$. After a linear symplectic change of variables (see Eq.~\eqref{trasf.can.int.plane} in subsection \ref{subsec:Phaseequiv} below), the Hamiltonian can be formally regarded as 1 DOF. The fixed points of the 1 DOF Hamiltonian actually correspond to periodic orbits of the original symplectic flow of $\Hscr_{int}$ in all four variables $(w_2,w_3,W_2,W_3)$. The bifurcation chains and stability of such periodic orbits are our main subject of interest in the sequel.}

\subsection{Numerical phase portraits}
\label{subsec:Poinc}
The only exact integral of motion for the Hamiltonian $\Hscr_{sec}$ is the energy integral $\Hscr_{sec}(X_2,Y_2,X_3, Y_3=0;\AMD)=\Escr$. Exploiting the constancy of the energy, we can define a suitable Poincar\'e surface of section $\mathcal{P}(\Escr;\AMD)$ capturing all the important structural changes in the phase space under the Hamiltonian $\Hscr_{sec}$ as the energy $\Escr$ increases. As demonstrated in ~\cite{maseft2023}, increasing the energy $\Escr$ in the Jacobi-reduced Hamiltonian means orbits of higher mutual inclination at a given level of eccentricities, or higher eccentricity at a given level of mutual inclination. We define the surface of section through the relations 
\begin{equation}\label{pcsec}
 \begin{aligned}
\mathcal{PS}_{\Hscr_{sec}}(\Escr;\AMD)=
\bigg\{ &(X_2,Y_2,X_3,Y_3)\in\mathbb{R}^4:~ \Hscr_{sec}(X_2,Y_2,X_3, Y_3=0;\AMD)=\Escr\, ,\,Y_3=0\, ,\\
&\dot{Y}_3=-{\partial\Hscr_{sec}\over\partial X_3}\big|_{Y_3=0}\geq 0\, ,
\cos(\i_{max})\leq\cos(\i_{mut})(X_2,Y_2,X_3,Y_3=0;\AMD)\leq 1\bigg\}\, ,
\end{aligned}
\end{equation}
where the variables $(X_j, Y_j)$ are Poincaré canonical coordinates and momenta related to the variables $(w_j,W_j)$ through the canonical transformation
\begin{align}\label{Poincare1}
& X_j= -\sqrt{2 \,W_j}\cos(w_j)\, , & & Y_j=\sqrt{2\,W_j}\sin(w_j)\, , & &j=2,3~~.
\end{align} 
The mutual inclination $\i_{mut}=\i_2+\i_3$ for fixed $\AMD$ is given by 
\begin{equation}\label{cosimutua}
\begin{aligned}
\cos(\i_{mut})&=
\frac{L_z^2-\Lambda_2^2-\Lambda_3^2+\Lambda_2^2\,\e_2^2+\Lambda_3^2\,\e_3^2}{2\Lambda_2\Lambda_3\sqrt{1-\e_2^2}\sqrt{1-\e_3^2}}=\frac{L_z^2-G_2^2-G_3^2}{2 G_2 G_3}~~
\end{aligned}
\end{equation}
where $L_z=\Lambda_2+\Lambda_3-\AMD$. The maximum possible mutual inclination consistent with the given $\AMD$ is
\begin{equation}\label{imutuamax}
\i_{max}=
\cos^{-1}\left(\frac{L_z^2-\Lambda_2^2-\Lambda_3^2}{2\Lambda_2\Lambda_3}\right)~~.
\end{equation}

\begin{figure}[h]
\begin{center}
\frame{\includegraphics[scale=1.]{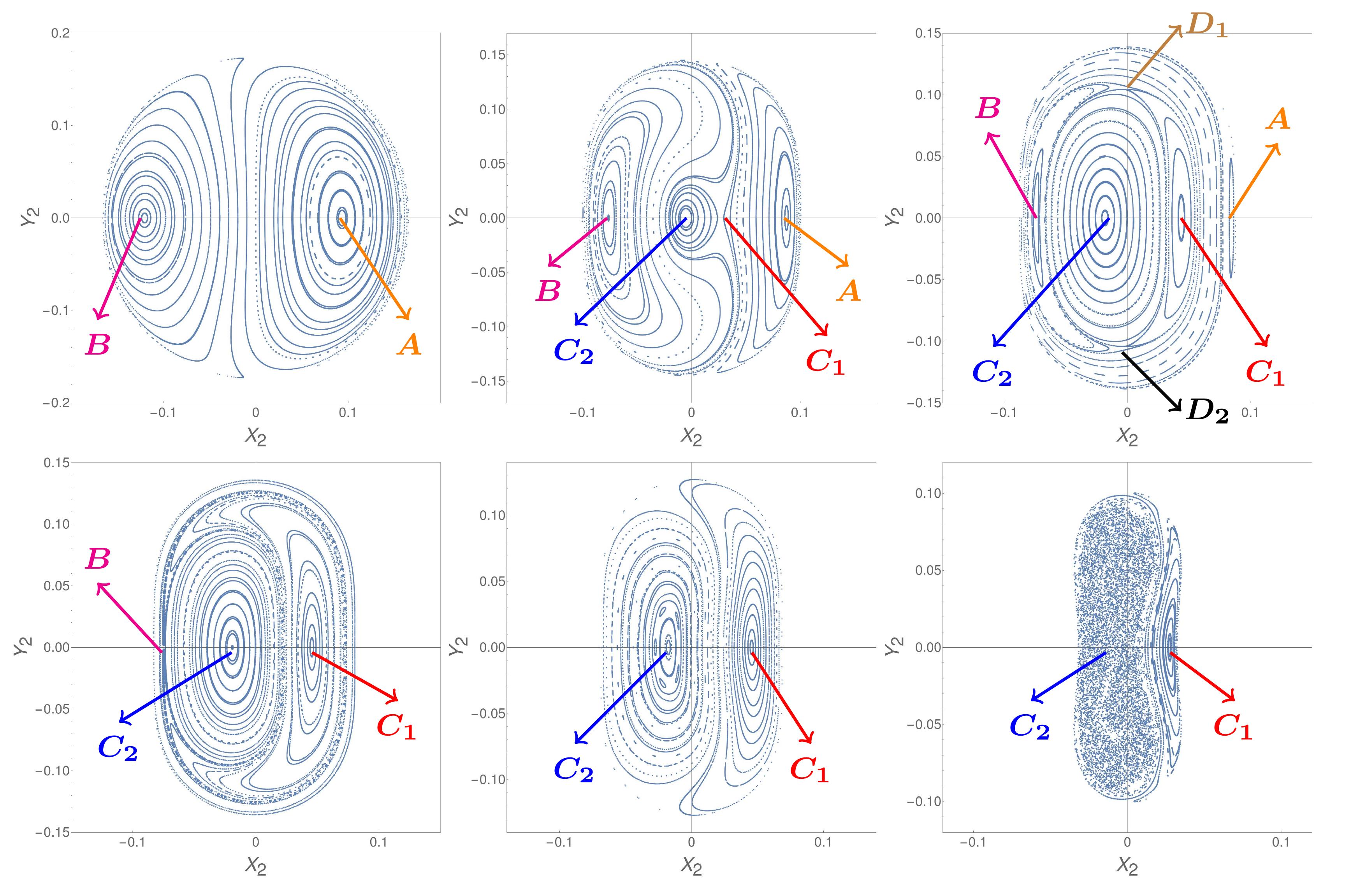}}
\caption{Poincar\'{e} surfaces of section $\mathcal{PS}_{\Hscr_{sec}}(\Escr;\AMD)$ in the plane $(X_2, Y_2)$ with $\AMD$ fixed and different values of energy. The surfaces of section are computed by a numerical integration of trajectories in the Hamiltonian truncated at multipolar degree $N_{P}=5$, order $N_{bk}=12$ in the eccentricities, and energies (from left to right) $\Escr= -6.77\cdot 10^{-5},-2.53\cdot 10^{-5}, -1.92\cdot 10^{-5}, {-1.73\cdot 10^{-5},} -1.17\cdot 10^{-5}$,$-2.61\cdot 10^{-6}$. }
\label{Fig.npol5eps12}
\end{center}
\end{figure}

The phase portrait corresponding to the Poincar\'e surface of section at a fixed level of energy $\Escr$ is obtained numerically, by choosing several initial conditions $(X_2,Y_2)\in\mathcal{D}(\Escr)\subset\mathbb{R}^2$, where $\mathcal{D}(\Escr)$ is the domain of permissible initial conditions consistent with the definition of the surface of section as defined in Eq.~\eqref{pcsec}. Figure~\ref{Fig.npol5eps12} shows an example of computation of phase portraits, referring to the mass, periods and AMD parameters as estimated for the $\upsilon$-Andromed{\ae} system reported in section $2.4$ of~\cite{maseft2023}. We use a model truncated at multipolar order $N_P = 5$ and of book-keeping order $N _{bk} = 12$. This essentially reproduces Figs. $5$ and $9$ of ~\cite{maseft2023}, but illustrating the surface of section in the canonical variables $(X_2, Y_2)$ instead of their `physical' analogues $(\e_2 \cos(\omega_2 ), \e_2 \sin(\omega_2 ))$. 

Figure \ref{Fig.npol5eps12} allows to recognize one out of all in-principle possible chains of bifurcations that can emerge in 3D secular models of the type discussed above. Such models may differ in the parameters (masses and semi-major axes of the planets, value of the AMD) as well as on the truncation orders (in the multipoles and/or the orbital eccentricities and inclinations). Our aim in the next section is to provide a systematic method to predict which out of all possibilities actually takes place in a system with given parameters. In the case of Fig.~\ref{Fig.npol5eps12}, a quick description of the numerically obtained sequence of bifurcations is as follows: for low $\Escr$, i.e.,  {for} nearly planar orbits (left panel  {in the first row of Fig.}~\ref{Fig.npol5eps12}), the phase portrait exhibits two stable fixed points which correspond to the ACR modes A (right fixed point) and B (left fixed point, reversed with respect to the left-right position of the same fixed points in figure 9 of \cite{maseft2023}). In between the centers A and B we have a continuous transition of quasi-periodic orbits represented in the surface of section by closed curves surrounding either the fixed point A or B. No unstable points or separatrices exist between the stable points, while the apparent separation of the two domains is only an artifact of the projection of the phase space, whose real topology is the one of the 2-sphere (see below), to the plane $(X_2,Y_2)$. The second panel  {in the first row of} Fig.~\ref{Fig.npol5eps12} shows, now, the birth of the families called `C1' (unstable) and `C2' (stable) in \cite{maseft2023} through a saddle-node bifurcation.  {We denote by $\Escr_{C}$ the critical value of the energy in which the saddle-node bifurcation takes place. Increasing the value of the energy $\Escr$, we see (third panel of first row) that} the unstable orbit C1 becomes stable through a pitchfork bifurcation giving rise to two `off-axis' ($Y_2\neq 0$) unstable orbits {, namely $D_1$ and $D_2$,} which (as the energy $\mathcal{E}$  {further} increases) join each other through an inverse pitchfork bifurcation turning the left fixed point of the middle panel of Fig.~\ref{Fig.npol5eps12} from stable to unstable.  {More precisely, at a critical value of the energy, the fixed points $D_1$ and $D_2$ collide with the B-mode, who becomes unstable (see first panel, second row).} For still higher energy $\mathcal{E}$, the C2 orbit finally turns also to unstable via the `Lidov-Kozai' mechanism.  {We denote by $\Escr_{C,2}$ the critical value of the energy leading to the Lidov-Kozai instability}.  {In the Lidov-Kozai regime,} a large volume of the phase space is dominated by chaotic motions, excluded by any of the integrable models used in the sequel to approximate the flow under the complete Hamiltonian.  {We call \textit{transition regime} the one holding at energies in the interval $\Escr_{C}\leq \Escr \leq \Escr_{C,2}$, representing the transition between the planar-like regime and the Lidov-Kozai one (see section 3.3 of~\cite{maseft2023}).}

\section{Sequences of bifurcations: geometric method of determination}
\label{sec:Bif}

\subsection{Preliminary definitions. Equivalence of phase portraits}
\label{subsec:Phaseequiv}
The Poincar\'e surfaces of section $\mathcal{PS}_{\Hscr_{sec}}(\Escr;\AMD)$ in Fig.~\ref{Fig.npol5eps12} were obtained by fixing the value of the energy $\Escr$ and taking a set of initial conditions for each energy. 
 {
\begin{Definition}
Let $P$ be a periodic orbit of the flow under $\Hscr_{sec}$ at the energy value $\Hscr_{sec}=\Escr$. To the orbit $P$ corresponds a fixed point in the Poincar\'e surfaces of section $\mathcal{PS}_{\Hscr_{sec}}(\Escr;\AMD)$ with coordinates $X_{2,P}(\Escr),Y_{2,P}(\Escr)$. The orbit $P$ exists, in general, within two energy limits $\Escr_{min,P}\leq\Escr\leq\Escr_{max,P}$. We call $\Escr_{min,P}$ and $\Escr_{max,P}$ the \textit{bifurcation limits} of the periodic orbit $P$. The $\mbox{graphs of~}$ $X_{2,P}(\Escr)$, $Y_{2,P}(\Escr)$ as a function of $\Escr$ are called \textit{characteristic curves} $C_P$ of the periodic orbit $P$. 
\end{Definition}
}
 {
The procedure to compute the bifurcation limits of particular periodic orbits is detailed in Section $2.4$ of~\cite{maseft2023}. In particular, a fixed value of $\AMD$ determines the maximal allowed value of the mutual inclination for any planetary orbits of the system. Consider the domain in mutual inclination $[0, \i_{mut_{max}}]$. A certain value of the energy $\Escr$ establishes, within that domain, a range of allowed mutual inclinations (see figure 8 of ~\cite{maseft2023}). To find it, it is sufficient to look for tangencies between the manifold of constant energy to the section $Y_3=0$, i.e. 
\begin{equation*}
\mathcal{M}(\Escr)=\bigg\{(X_2,Y_2,X_3)\in\mathbb{R}^3: \Hscr_{sec}(X_2,X_3,Y_2,Y_3=0;\AMD)=\Escr\bigg\} 
\end{equation*} 
and the ellipsoidal surface 
\begin{equation*}
\mathcal{I}_{\i_{mut}}=\bigg\{(X_2,Y_2,X_3)\in\mathbb{R}^3: \frac{L_z^2-\Lambda_2^2\left(1-{X_2^2+Y_2^2\over 2\Lambda_2}\right)^2
-\Lambda_3^2\left(1-{X_3^2\over 2\Lambda_3}\right)^2}
{2\Lambda_2\Lambda_3\left(1-{X_2^2+Y_2^2\over 2\Lambda_2}\right)\left(1-{X_3^2\over 2\Lambda_3}\right)}=\cos(\i_{mut}) \bigg\}~~,
\end{equation*}
with angular momentum $L_z=\Lambda_2+\Lambda_3-\AMD$ (see Eq. (29) of~\cite{maseft2023}). Once established, the correspondence between energies $\Escr$ and mutual inclinations $\i_{mut}$ allows to find the bifurcation limits $\Escr_{min, P}$, $\Escr_{max, P}$ for a periodic orbit $P$.
}

\begin{figure}[!h]
\begin{center}
\fbox{\includegraphics[scale=.5]{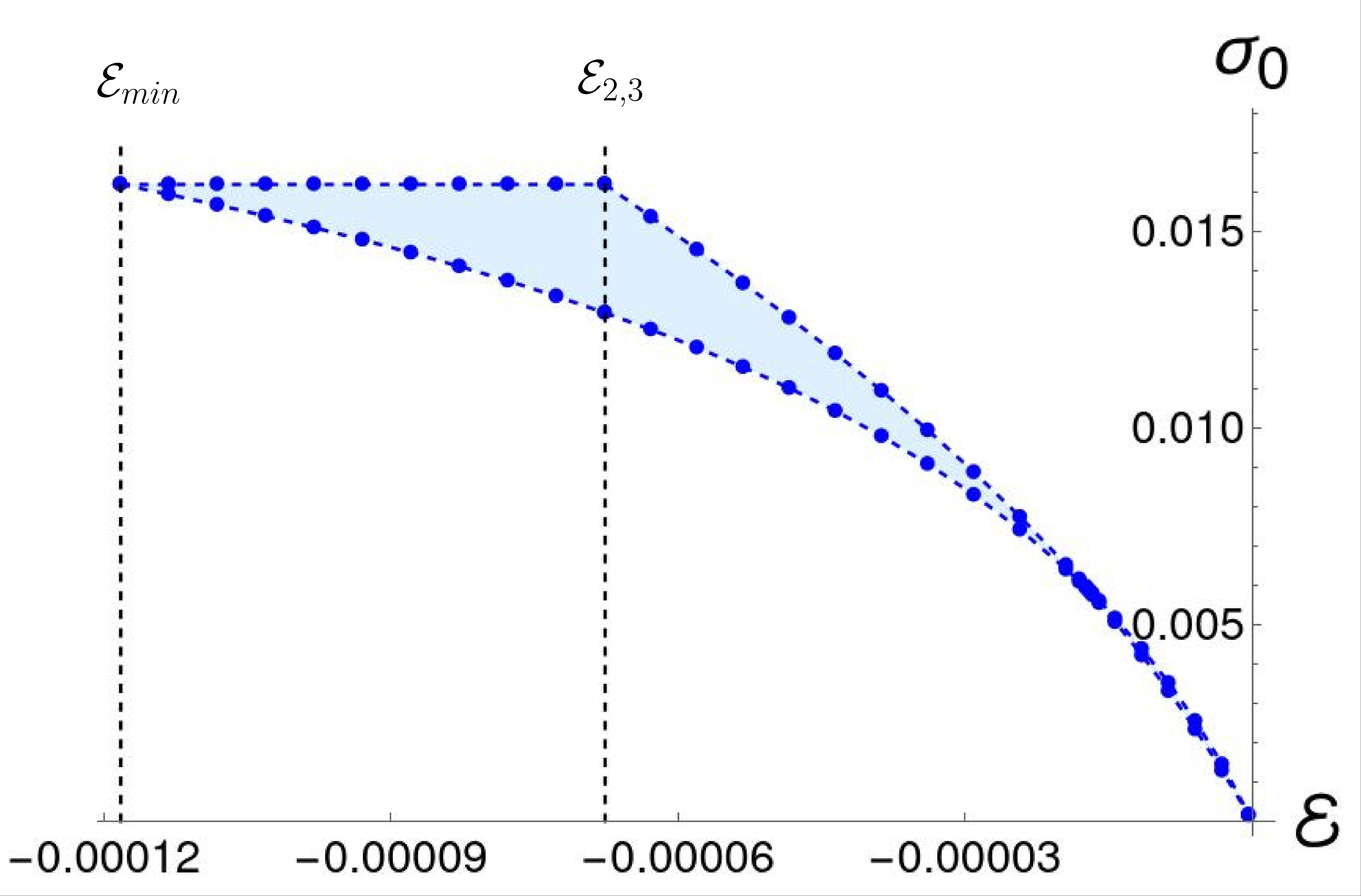}}
\caption{The permissible domain in $\Escr$ and $\sigma_0$ consistent with a fixed value of the system's $\AMD$. The numerical curves shown above refer to the example of the integrable model $\mathcal{Z}=\Hscr_{int}$. Fixing a value of $\sigma_0$, the two limiting curves yield the corresponding limiting energies $\Escr_L(\sigma_0)$, $\Escr_R(\sigma_0)$. Alternatively, fixing a value of the energy we obtain the limits $\sigma_{0,min}(\Escr),\sigma_{0,max}(\Escr)$. The above limits can be computed analytically as explained in \cite{maseft2023}  {(computing the tangencies between the sphere of radius $\sigma_0$ described in~\eqref{sphere} and the constant energy surface reported in~\eqref{energy.curve})}. In summary, there exist two critical values of the energy $\Escr_{min}$, $\Escr_{2,3}$ such that, for any value of the energy in the interval $\Escr_{min}<\Escr<\Escr_{2,3}$, $\sigma_0$ is limited from below by a minimum value $\sigma_{0,min}(\Escr)$, while the limit from above $\sigma_{0,max}$ is posed only by the globally maximum value $\sigma_{0,max}=\sigma_{0,\AMD}=\AMD$. The limiting values $\Escr=\Escr_{min}$ and $\Escr=\Escr_{2,3}$ correspond to the co-planar ACR orbits A (anti-aligned) and B (aligned) respectively. On the other hand, for energies larger than $\Escr_{2,3}$ there are no possible co-planar orbits, and $\sigma_{0,max}$ as well becomes a decreasing function of $\Escr$. Overall, as we move from left to right or top to bottom, we obtain orbits of higher mutual inclination. For the Hamiltonian $\Hscr_{int}$ (see text for parameter values) we compute the values $\Escr_{min}=-1.18\cdot 10^{-4} $ and $\Escr_{2,3}=-6.77\cdot 10^{-5}$ and $\sigma_{0_{max}}=\AMD=0.0162044$.  
}
\label{Fig.Evssigma0}
\end{center}
\end{figure}

\vspace{-15pt}  
 {
\begin{Definition}
Consider all possible periodic orbits $P_n$, $n=1,2,\ldots$, yielding fixed points on the section $\mathcal{PS}_{\Hscr_{sec}}(\Escr;\AMD)$ of a given character of Floquet $S_n$ (with values `stable' or 'unstable'), and sorted in ascending values of $\Escr_{min,P_n}$.  The set $\mathcal{B}\equiv\{(\Escr_{min,P_n},\Escr_{max,P_n},C_{P_n},S_n), n=1,2,\ldots\}$ is hereafter called a \textit{bifurcation sequence}. 
\end{Definition}
}
 {The periodic orbit $P_n$ is called Floquet unstable if the Floquet matrix of the linearized flow around the periodic orbit has at least one Floquet exponent with positive real part.  Otherwise, the periodic orbit is called Floquet stable (see \cite{tes2012})}.

The definition of a bifurcation sequence $\mathcal{B}$ as above requires parameterizing all bifurcation limits in terms of the corresponding energies 
$(\Escr_{min,P_n},\Escr_{max,P_n})$. To predict the structural changes in the corresponding phase portraits after the n-th bifurcation in the sequence, we employ integrable approximations to the dynamics, as the models  
$\Hscr_{int}$, $\Hscr_{int}^{(1)}$, or $\tilde{\Hscr}_{int}^{(N_P=3,N_{bk}=4)}$ mentioned in the introduction. An integrable approximation $\mathcal{Z}$ to the Hamiltonian $\Hscr_{sec}$ is defined as a Hamiltonian which i) admits $\sigma_0$ (Eq.(\ref{sig0})) as a second integral, and ii) has flow close to the flow of $\Hscr_{sec}$. The closeness of the two flows can be established by appropriate norms in the functional space containing the derivatives of $\Hscr_{sec}$ and $\mathcal{Z}$. 
\vspace{-3pt}  
 {
\begin{Definition}
Let $\mathcal{Z}$ be an integrable approximation to $\Hscr_{sec}$. Define the family of curves 
\begin{align}
\label{poincsig0}
\mathcal{C_Z}(\sigma_0;\Escr;\AMD)=
\bigg\{ 
&(X_2,Y_2)\in\mathbb{R}^2:
~\mathcal{Z}(X_2,Y_2,X_3=X_{3,+},Y_3=0;\AMD)=\Escr, \nonumber\\
&X_2^2+Y_2^2+X_{3,+}^2=2\sigma_0,
~\dot{Y}_3=-{\partial\mathcal{Z}\over\partial X_3}\big|_{Y_3=0,X_3=X_{3,+}}\geq 0, 
\\
&\cos(\i_{max})\leq\cos(\i_{mut})(X_2,Y_2,X_3=X_{3,+},Y_3=0;\AMD)\leq 1\bigg\}~~,
\nonumber
\end{align}
 {where $X_{3,+}=\sqrt{2\sigma_0-X_2^2-Y_2^2}\,$.}
A \textit{$\sigma_0-$fixed phase portrait of $\mathcal{Z}$} is defined by the family of curves 
\begin{equation}\label{PoincZsig}
\mathcal{P_Z}(\sigma_0) = 
\bigg\{\mathcal{C_Z}(\sigma_0;\Escr;\AMD),~~
\Escr_{L}(\sigma_0,\AMD)\leq\Escr\leq
\Escr_{R}(\sigma_0,\AMD)
\bigg\}\, ,
\end{equation}
where the functions $\Escr_{L}(\sigma_0,\AMD), \Escr_{R}(\sigma_0,\AMD)$ are defined by the requirement that the angular momentum deficit of the system is equal to the given value $\AMD$ (see \cite{maseft2023} as well as the caption in Fig.~\ref{Fig.Evssigma0}). A \textit{$\Escr-$fixed phase portrait of $\mathcal{Z}$} is defined by the family of curves 
\begin{equation}\label{PoincZene}
\mathcal{P_Z}(\Escr) = 
\left\{\mathcal{C_Z}(\sigma_0;\Escr;\AMD),~~
\sigma_{0,min}(\Escr,\AMD)\leq\sigma_0\leq
\sigma_{0,max}(\Escr,\AMD)
\right\}
\end{equation}
where the functions $\sigma_{0,min}(\Escr,\AMD), \sigma_{0,max}(\Escr,\AMD)$ are defined through the inverse of the functions $\Escr_{L}(\sigma_0,\AMD), \Escr_{R}(\sigma_0,\AMD)$. 
\end{Definition}
}

\begin{figure}[!h]
\begin{center}
\frame{\includegraphics[width=0.8\textwidth]{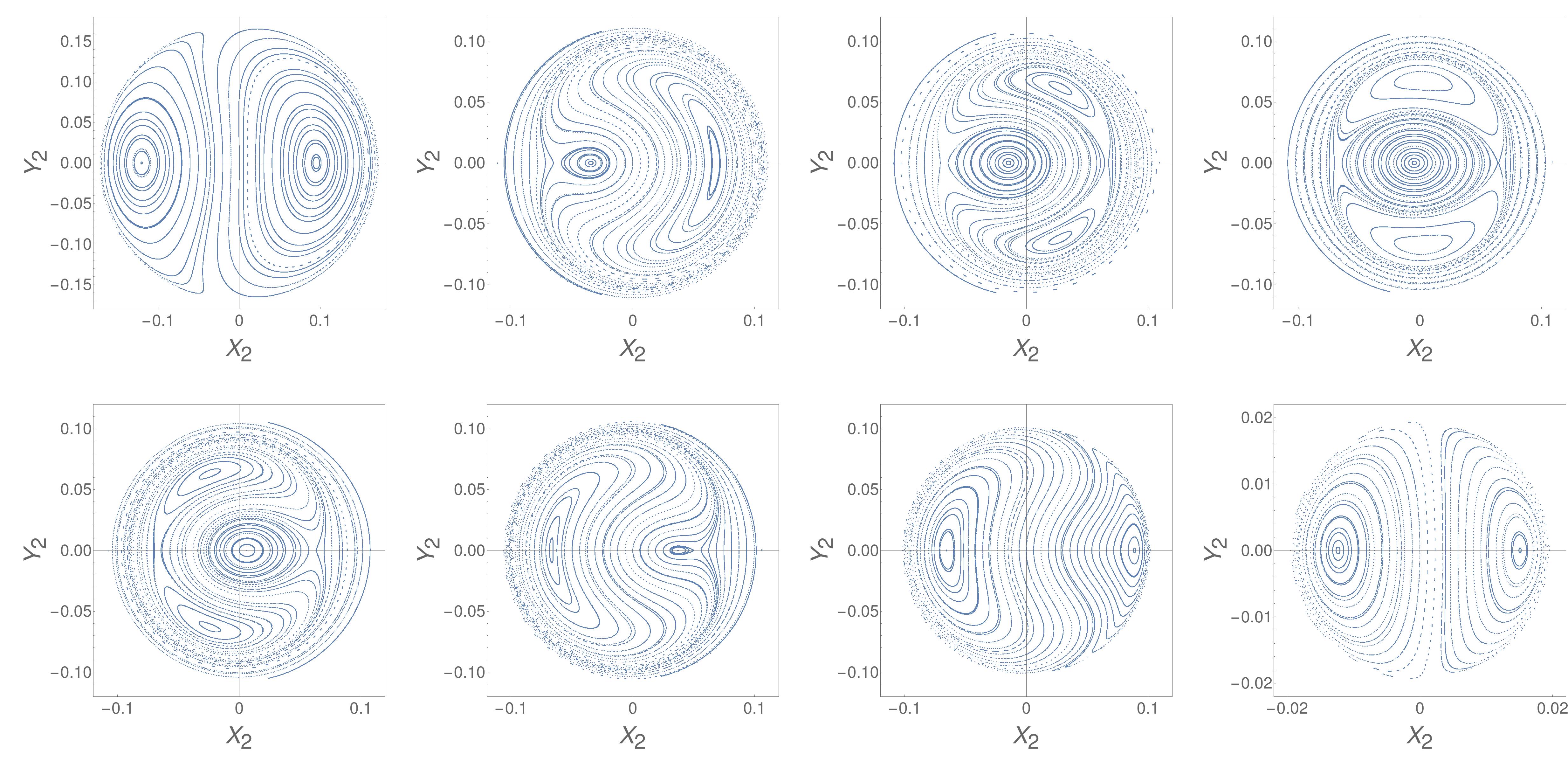}}
\frame{\includegraphics[width=0.8\textwidth]{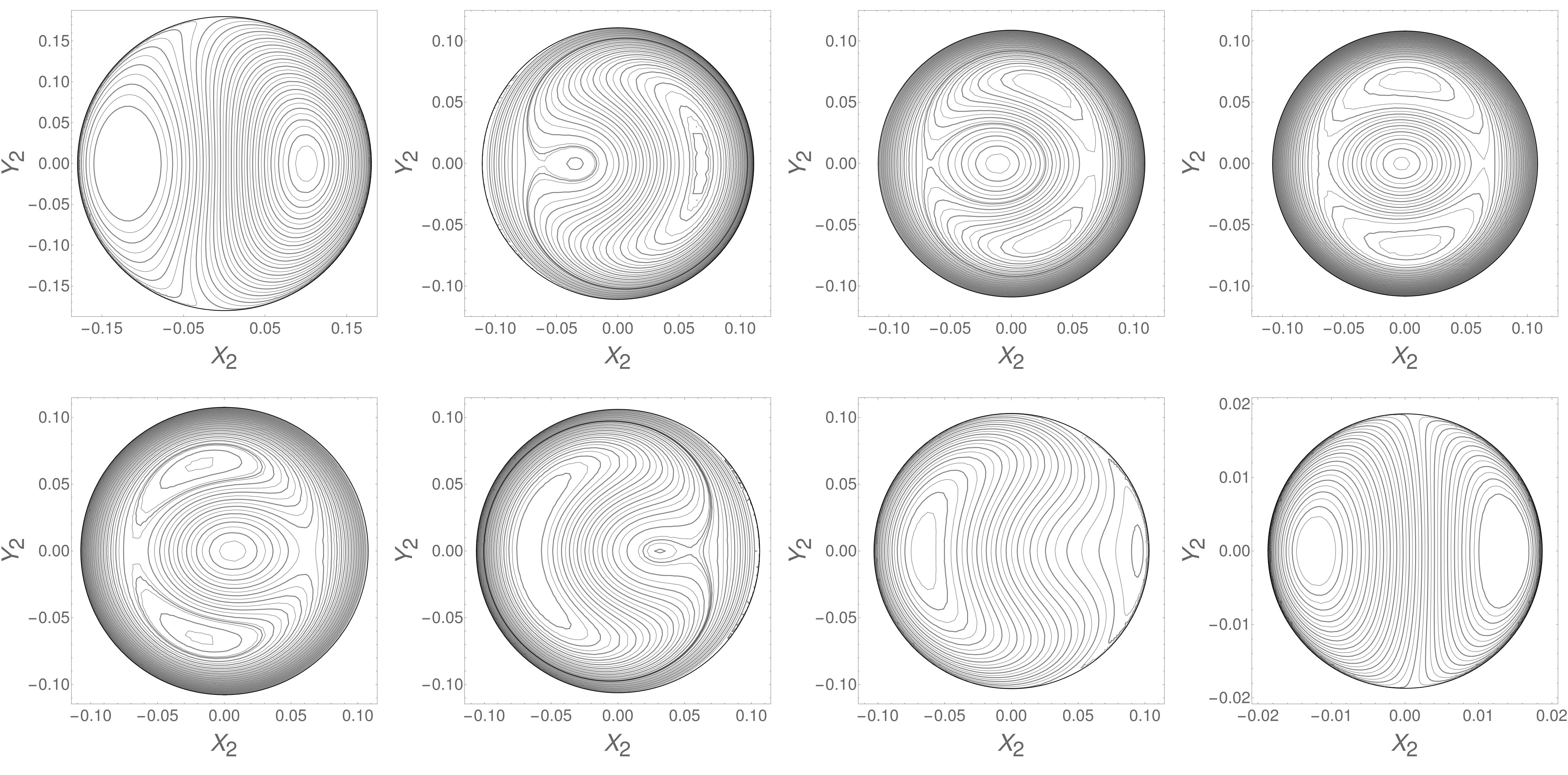}}
\caption{\textit{Top frame:} Poincar\'{e} surfaces of section $\mathcal{PS}_{\Hscr_{int}}(\Escr;\AMD)$ for the integrable Hamiltonian model $\Hscr_{int}$ at the values of the energies (from top to bottom, left to right) $\Escr= -6.77\cdot 10^{-5},-1.8\cdot 10^{-5},-1.74\cdot 10^{-5},-1.7\cdot 10^{-5}, -1.67\cdot 10^{-5}, -1.6\cdot 10^{-5}$,$-1.43\cdot 10^{-5}, -4.05\cdot 10^{-7}$. \textit{Bottom frame:} Phase portraits by contour plots $\mathcal{P}_{\Hscr_{int}}(\sigma_0;\AMD)$ with decreasing values of $\sigma_0$ (from top to bottom, left to right) $\sigma_0= 1.62\cdot 10^{-2}, 6.15\cdot 10^{-3}, 5.93\cdot 10^{-3}, 5.85\cdot 10^{-3}, 5.78\cdot 10^{-3}, 5.62\cdot 10^{-3}, 5.304\cdot 10^{-3}, 1.74\cdot 10^{-4}$. The two alternative representations of the phase portraits, corresponding to $\mathcal{P_Z}(\Escr)$ (top) or $\mathcal{P_Z}(\sigma_0)$ (bottom), for $\mathcal{Z}=\Hscr_{int}$, yield an equivalent bifurcation sequence.} 
\label{Fig.npol5eps12INTEGsigma0}
\end{center}
\end{figure}

By the above definitions the following hold:
 {
\begin{Proposition}
\label{i)}
$\Escr-$fixed phase portraits of $\mathcal{Z}$, i.e. $\mathcal{P_Z}(\Escr)$, are equivalent to the Poincar\'{e} surfaces of section $\mathcal{PS}_{\mathcal{Z}}(\Escr;\AMD)$.
\end{Proposition}
\begin{proof}
To demonstrate the equivalence of the phase portraits $\mathcal{P_Z}(\Escr)$ and $\mathcal{PS}_{\mathcal{Z}}(\Escr;\AMD)$, define the canonical transformation
\begin{equation}
\label{trasf.can.int.plane}
\begin{aligned}
& \psi=w_2-w_3\, , & &\Gamma=\frac{W_2-W_3}{2}\, , \\
& \varphi=w_2+w_3\, , & &J=\frac{W_2+W_3}{2}~~.
\end{aligned}
\end{equation}
The integrable Hamiltonian $\mathcal{Z}$ in the new variables reads (apart from a constant)
\begin{equation}
\label{Ham.new.var}
\mathcal{Z}(\psi, \varphi,\Gamma, J)=\mathcal{Z}(\psi, \Gamma; J)~~,
\end{equation}
i.e., the angle $\varphi$ is ignorable. In view of Eq.~\eqref{Ham.new.var}, $\mathcal{Z}$ can be regarded as defining either the complete flow in four variables $(\psi,\varphi,\Gamma,J)$, or, alternatively, the flow of a one degree of freedom Hamiltonian in the variables $( \psi, \Gamma)$, with $J$
serving as parameter. Hence, the level curves of constant energy $\mathcal{Z}(\psi,\Gamma;J)=\Escr$ computed in the plane $( \psi, \Gamma)$ for various values of the energy and keeping fixed the value of $J$ are geometrically equivalent to the invariant curves in the Poincar\'{e} section under the complete 2-DOF flow. In particular, any closed curve $\gamma$ (1-torus) in the top frame of Fig.~\ref{Fig.npol5eps12INTEGsigma0} represents a 2-torus of the full flow, obtained by the product $\gamma\times$ the 1-torus defined by the solution for $\varphi$ of the differential equation $\dot{\varphi}=(\partial\mathcal{Z}/\partial J)_{\psi_\gamma(t),\Gamma_\gamma(t)}$, where $\psi_\gamma(t),\Gamma_\gamma(t)$ is the solution of the Hamiltonian flow for the variables $(\psi,\Gamma)$ along the curve $\gamma$. In the same way, any fixed point $(\psi_0,\Gamma_0)$ in the same plots represents a periodic orbit $\varphi(t)=\varphi(0)+(\partial\mathcal{Z}/\partial J)_{\psi=\psi_0,\Gamma=\Gamma_0}$ of the complete flow. Hence, the problem of the number and bifurcations of new periodic orbits in the complete 2-DOF flow is reduced to the problem of the number and bifurcations of new fixed points in the reduced 1-DOF flow. 
\end{proof}
 {Note that, in the above proposition, `equivalence’ means that each member of the family of the 2D tori of the Hamiltonian flow of $\Zscr$, regarded as a function of four variables, can be mapped one-to-one to a member of the family of  contours $\gamma(\Escr, J)$.}
\begin{Observation}
\label{i)}
Since the flow of $\mathcal{Z}$ is close to the flow of $\Hscr_{sec}$ in a suitable domain of initial conditions the phase portraits $\mathcal{P}_{\mathcal{Z}}(\Escr)$ are expected to approximate as well the phase portraits of the full non-integrable Hamiltonian $\Hscr_{sec}$, obtained through the Poincar\'{e} surfaces of section $\mathcal{PS}_{\Hscr_{sec}}(\Escr;\AMD)$. 
\end{Observation}
\begin{Observation}
\label{ii)}
For fixed integrable model $\mathcal{Z}$, the phase portraits $\mathcal{P_Z}(\Escr)$ and $\mathcal{P_Z}(\sigma_0)$ yield an identical sequence of bifurcations. 
\end{Observation}
This is an obvious consequence of the definition of bifurcation sequence together with the fact that the sets $\mathcal{P_Z}(\Escr)$ and $\mathcal{P_Z}(\sigma_0 {)}$, spanning the whole permissible domain in $\Escr$ and $\sigma_0$ (Fig.~\ref{Fig.Evssigma0}), contain the same curves $\mathcal{C_Z}(\sigma_0,\Escr;\AMD)$.
An example is provided in Fig.~\ref{Fig.npol5eps12INTEGsigma0}.
}

Note that, due to  {Observation~\ref{i)}}, only the representation $\mathcal{P}_{\mathcal{Z}}(\Escr)$ can be compared directly with the numerical phase-portraits under the full (non-integrable) Hamiltonian $\Hscr_{sec}$. On the other hand, due to  {Observation~\ref{ii)}}, the theoretical analysis of the bifurcation sequences can be performed exploring the geometric properties of the representation $\mathcal{P_Z}(\sigma_0)$ in suitable Hopf variables (see next subsection).  {Observation~\ref{ii)} then implies that the two different representations of the phase portraits show an identical sequence of bifurcations, namely $\mathcal{B}$.}

\subsection{Hopf representation of the phase space}
\label{subsec:Hopf}
Having established the equivalence of all kinds of phase portraits defined in the previous subsection, we now focus on the theoretical interpretation of the structural changes in the phase portraits $\mathcal{P_Z}(\sigma_0)$ (e.g. those of Fig.~\ref{Fig.npol5eps12INTEGsigma0}, bottom frame), as $\sigma_0$ is altered within the established permissible domain of values. To this end, it turns convenient to express the Hamiltonian flow in terms of Hopf variables $(\sigma_1,\, \sigma_2,\, \sigma_3\,)$ defined by: 
\begin{equation}\label{def.Hopf}
\sigma_1=X_2 X_3+Y_2 Y_3\, ,\quad
\sigma_2=Y_2 X_3-Y_3 X_2\, ,\quad
\sigma_3=\frac{1}{2}\left(X_2^2+Y_2^2-X_3^2-Y_3^2\right)~ ,
\end{equation}
satisfying the Poisson algebra $\poisson {\sigma_i} {\sigma_j} =-2\,\epsilon_{i j k}\sigma_k\, $, where $\epsilon_{i j k}$ is the Levi-Civita symbol and $i,j,k=1,2,3$. Furthermore, we introduce the variable 
\begin{equation}\label{def.sigma0}
\sigma_0=\frac{1}{2}\left(X_2^2+Y_2^2+X_3^2+Y_3^2\right)
\end{equation}
which is a Casimir invariant of the previous algebra (all Poisson brackets $\{\sigma_i,\sigma_0\}$, $i=1,2,3$, vanish). From the definition~\eqref{def.Hopf} it follows that 
\begin{equation}\label{def.Hopf.Delau}
\begin{aligned}
&\sigma_1=2\sqrt{J+\Gamma}\sqrt{J-\Gamma}\cos(\psi)\, ,\, &
&\sigma_2=-2\sqrt{J+\Gamma}\sqrt{J-\Gamma}\sin(\psi)\, ,\,\\
&\sigma_3=W_2-W_3=2\Gamma\, , & &
\end{aligned}
\end{equation}
 {where the variables $(\psi, \varphi, \Gamma, J)$ are introduced in Eq.~\eqref{trasf.can.int.plane}.}
We also have the relation $\sigma_0=W_2+W_3=2J$, as well as
\begin{equation}\label{cond.sphere}
\sigma_1^2+\sigma_2^2+\sigma_3^2=\sigma_0^2=4J^2\, .
\end{equation}
The latter expression implies that, for any fixed value of the constant $J$, the flow under any of the integrable models $\Zscr$ approximating the Hamiltonian, represented in a Euclidean space with axes $(\sigma_1,\sigma_2,\sigma_3)$, is restricted to the 2-sphere of radius $\sigma_0$. 

Given the values of $(\sigma_1,\sigma_2,\sigma_3)$, the values of $\Gamma$, $J$ and $\psi$ can be computed unequivocally using the relations (\ref{def.Hopf.Delau}) and (\ref{cond.sphere}). Furthermore, the Hamiltonian $\Zscr$ is even in the difference $w_2-w_3$, hence it does not depend on $\sigma_2$, i.e., $\Zscr=\Zscr(\sigma_1,\sigma_3;\sigma_0)$. Fixing a value of $\sigma_0$ (i.e. of the integral $J=\sigma_0/2$), the phase flow under $\Zscr$ {, continuous in time,} can be expressed in the Hopf variables via the equations
\begin{equation}\label{sigmaflow}
\begin{aligned}
&\dot{\sigma}_1=\{\sigma_1,\sigma_3\}{\partial\Zscr\over\partial\sigma_3},~~~~~ & &
\dot{\sigma}_2=
\{\sigma_2,\sigma_1\}{\partial\Zscr\over\partial\sigma_1}+
\{\sigma_2,\sigma_3\}{\partial\Zscr\over\partial\sigma_3},~~~~~
\dot{\sigma}_3=-\{\sigma_1,\sigma_3\}{\partial\Zscr\over\partial\sigma_1}~~. 
\end{aligned}
\end{equation}
In particular, the level curves of Fig.~\ref{Fig.npol5eps12INTEGsigma0} can be represented by the intersection of the constant energy surface $\Zscr(\sigma_1,\sigma_3;\sigma_0)=\Escr$ with the sphere (\ref{cond.sphere}). Altering the energy $\Escr$ in the interval $\Escr_L(\sigma_0)\leq\Escr\leq\Escr_R(\sigma_0)$, while keeping $\sigma_0=2J$ constant yields a family of closed curves on the sphere which can be mapped to the invariant curves of the Poincar\'e surface of section of Fig.~\ref{Fig.npol5eps12INTEGsigma0} through the relations $X_2=-\sqrt{2(\Gamma+J)}\cos(\psi-\pi)$, $Y_2=\sqrt{2(\Gamma+J)}\sin(\psi-\pi)$, with $\Gamma=\sigma_3/2$, $\cos\psi=\sigma_1/\sqrt{\sigma_0^2-\sigma_3^2}$, $\sin\psi=\sigma_2/\sqrt{\sigma_0^2-\sigma_3^2}$. Equivalently, keeping the energy $\Escr$ fixed and altering $\sigma_0$ in the interval $\sigma_{0,min}(\Escr)\leq\sigma_0\leq\sigma_{0,max}(\Escr)$ allows to recover the family of all invariant curves in Fig.~\ref{Fig.npol5eps12INTEGsigma0}. 

 {The special motions leading to singularities in ref.~\cite{paletal2013} can be seen in the Hopf variables in the following way. Concerning i) circular trajectories: the inner circular trajectory $\e_2=0$ is described by a point in the sphere, more precisely by the south pole $(\sigma^{(S)}_0, \sigma^{(S)}_1=0, \sigma^{(S)}_2=0, \sigma^{(S)}_3=-\sigma^{(S)}_0)$. In the Poincaré surface of section $Y_3=0$ $\dot{Y}_3\geq 0$ this would lead to a curve passing through the origin. Instead, the outher circular trajectory $\e_3=0$ is described by a north pole in the Hopf variables, i.e. $(\sigma^{(N)}_0, \sigma^{(N)}_1=0, \sigma_2^{(N)}=0, \sigma^{(N)}_3=\sigma^{(N)}_0)$ and by an open curve in the Poincaré surface of section. This has been already analyzed in Fig. 11 of~\cite{maseft2023}. Moreover, recalling the expression of the cosine of the mutual inclination as function of the eccentricity (see Eq. (29) of~\cite{maseft2023}) and considering that it is in $[-1,1]$, it is possible to find the limits for such a kind of orbits. On the other hand, the ii) coplanar motions can be limited by the fixed value of the $\AMD$ (see Fig. 8 of~\cite{maseft2023}, where the existence of coplanar orbits in a given interval of energies is provided).}

\subsection{Geometrical representation of the sequence of bifurcations}
\label{subsec:Hopf_analysis_integ}
Consider a fixed  {integrable} approximation $\mathcal{Z}$ to the Hamiltonian $\Hscr_{sec}$. Let
\begin{equation}
\label{sphere}
\mathcal{S}_{\sigma_0}=\lbrace(\sigma_1, \sigma_2, \sigma_3)\in\reali^3\, : \, \sigma_1^2+\sigma_2^2+\sigma_3^2=\sigma_0^2 \rbrace
\end{equation}
be the sphere of fixed radius $\sigma_0$ and 
\begin{equation}
\label{energy.curve}
\Cscr_{\sigma_0,\, \Escr} =\lbrace(\sigma_1, \sigma_2, \sigma_3)\in\reali^3\, : \, \Zscr(\sigma_0, \sigma_1, \sigma_3)=\Escr\rbrace\, ,
\end{equation} 
the constant energy surface in the space $(\sigma_1,\sigma_2,\sigma_3)\in\mathbb{R}^3$. By altering the parameters $(\sigma_0,\Escr)$ we have three possibilities: 

i) $\mathcal{S}_{\sigma_0}$ and $\mathcal{C}_{\sigma_0,\, \Escr}$ have no common points. This corresponds to values $(\sigma_0,\Escr)$ outside the permissible area by the curves of Fig.~\ref{Fig.Evssigma0}. 

ii) $\mathcal{S}_{\sigma_0}$ and $\mathcal{C}_{\sigma_0,\, \Escr}$ intersect transversally. This yields, in general, closed curves in the sphere $\Sscr_{\sigma_0}$ representing quasi-periodic orbits of the Hamiltonian $\Zscr$. In degenerate cases, the closed curves reduce to points representing periodic orbits of the flow under $\Zscr$.

iii) $\mathcal{C}_{\sigma_0,\, \Escr}$ arrives tangently to one or more points of $\mathcal{S}_{\sigma_0}$. The point of tangency is a periodic orbit of the flow under $\Zscr$. 

The whole sequence of bifurcations of new fixed points as in Fig.~\ref{Fig.npol5eps12INTEGsigma0} can be computed exploiting the above properties as follows: since the Hamiltonian does not depend on $\sigma_2$, for any fixed value of $\sigma_0$ all possible kinds of intersection or tangency of the surface $\mathcal{C}_{\sigma_0,\, \Escr}$ with $\mathcal{S}_{\sigma_0}$ can be classified through their projection to the plane $\sigma_2=0$. In particular, all points of tangency are along the meridian of the sphere $\mathcal{S}_{\sigma_0}^{(\sigma_2=0)}$ contained in the plane $(\sigma_1,\sigma_3)$ with $\sigma_2=0$. Altering the energy in the interval $\Escr_{L}(\sigma_0)\leq\Escr\leq\Escr_{R}(\sigma_0)$ we then compute the curve $\mathcal{C}_{\sigma_0,\Escr}^{(\sigma_2=0)}$ at which the surface $\mathcal{C}_{\sigma_0,\Escr}$ intersects the  {plane $\sigma_2=0$}, for values of $\Escr$ within the permissible limits discussed in the previous subsection. We then distinguish two cases, both satisfying the equation
\begin{equation}
\label{CPkummer}
\Grad\mathcal \Zscr= {2}\mu\,
\begin{pmatrix}
\sigma_1\\ \sigma_2 \\ \sigma_3
\end{pmatrix}
\end{equation}
for some $\mu\in\mathbb R$.

\subsubsection{Case 1: Non-degenerate tangency of $\mathcal{S}_{\sigma_0}$ and $\mathcal{C}_{\sigma_0,\, \Escr}$}
\begin{figure}
\begin{center}
\includegraphics[width=0.8\textwidth]{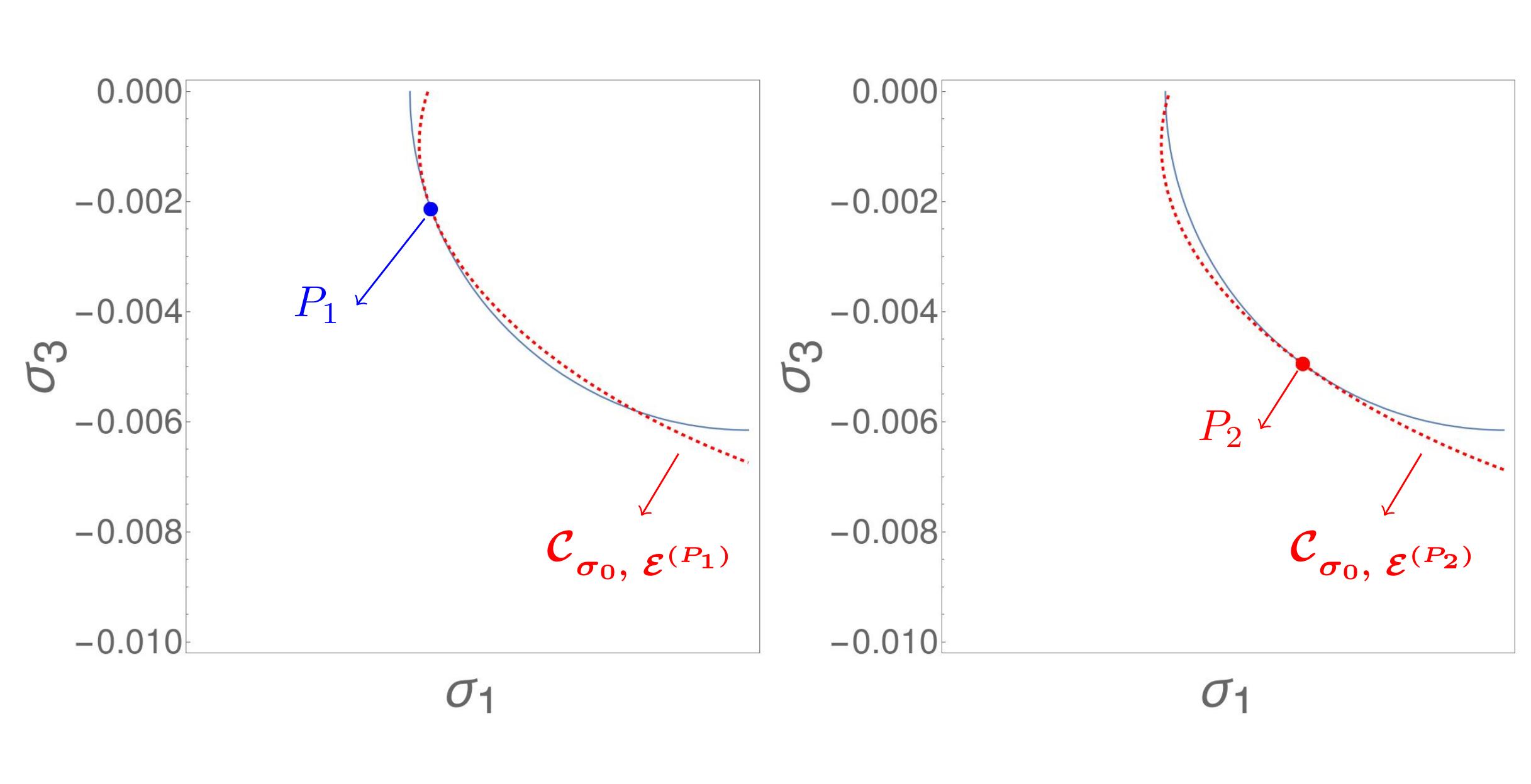}
\caption{An example of inner tangency between $\mathcal{S}_{\sigma_0}$ and $\mathcal{C}_{\sigma_0,\, \Escr^{(P_1)}}$ at the point $P_1$ (left), and outer tangency at the point $P_2$ (right). The example is taken from the analysis of the integrable model $\Zscr=\Hscr_{int}$ (see second panel of the second row of Fig.~\ref{Fig.Poin_Hopf_INTEG}, corresponding to the phase portrait $\mathcal{P_Z}(\sigma_0)$ with $ {\sigma_0}= 6.15\cdot 10^{-3}$). }
\label{Fig.zoom_Hopf_INTEG}
\end{center}
\end{figure}

Non-degenerate tangency points of the surfaces $\mathcal{S}_{\sigma_0}$ and $\mathcal{C}_{\sigma_0,\, \Escr}$ are contained in the curves $\mathcal{S}_{\sigma_0}^{(\sigma_2=0)}$ and $\mathcal{C}_{\sigma_0,\Escr}^{(\sigma_2=0)}$ and satisfy the condition
$$
\sigma_2=0,~~\Grad_{\sigma_1,\sigma_3}\left(\Zscr\right)_{\sigma_2=0}\neq 0,~~~
\mathrm{rank}\begin{pmatrix}
\displaystyle{2\sigma_1} & \displaystyle{2\sigma_3} \\
\displaystyle{\frac{\partial \Zscr}{\partial \sigma_1}} & \displaystyle{\frac{\partial \Zscr}{\partial \sigma_3}}
\end{pmatrix}=1~~,
$$
(i.e. Eq.~\eqref{CPkummer} with $\mu\neq 0$) implying 
\begin{align}
\label{CP1}
\sigma_2=0,~~~\sigma_3\frac{\partial \Zscr}{\partial \sigma_1}=\sigma_1\frac{\partial \Zscr}{\partial \sigma_3}~.
\end{align}
By Eq.~\eqref{CP1} we have that the point of tangency is a fixed point of the flow (\ref{sigmaflow}) since
\begin{align*}
&\frac{\dot{\sigma}_1}{2}=\sigma_2 \frac{\partial\Zscr}{\partial \sigma_3} =0 \, ,& &\frac{\dot{\sigma}_2}{2}=\sigma_3\frac{\partial \Zscr}{\partial\sigma_1}-\sigma_1\frac{\partial\Zscr}{\partial\sigma_3}=0\, , &\frac{\dot{\sigma}_3}{2}=-\sigma_2 \frac{\partial\Zscr}{\partial\sigma_1}=0 \, .
\end{align*}
The critical points (CP) of tangency are called of the `first kind' ( {CPI}), according to the terminology introduced in~\cite{kum1976}, and they represent periodic orbits under the complete 2-DOF flow of the Hamiltonian (\ref{Ham.new.var}) (see subsection \ref{subsec:Phaseequiv}). We distinguish two subcases (see Fig.~\ref{Fig.zoom_Hopf_INTEG}): \\
\noindent
\textit{Inner tangency:}  {in the space of variables $(\sigma_1, \sigma_2, \sigma_3)$, there exists} a neighborhood  enclosing the point of tangency,  {where} the curve $\mathcal{C}_{\sigma_0,\Escr^{(P_1)}}^{(\sigma_2=0)}$ is contained within the disc limited by the circle $\mathcal{S}_{\sigma_0}^{(\sigma_2=0)}$ (point $P_1$ in Fig.~\ref{Fig.zoom_Hopf_INTEG}). Then, in the same neighborhood the constant energy surface $\mathcal{C}_{\sigma_0,\,\Escr^{(P_1)}}$ intersects the sphere $\mathcal{S}_{\sigma_0}$ at two closed loops joining each other at the singular point $P_1$. The fixed point $P_1$ corresponds to a Floquet-unstable periodic orbit, and the two loops to the separatrices asymptotically connected to the periodic orbit.\\

\noindent
\textit{Outer tangency:}  {there exists} a neighborhood enclosing the point of tangency  {where} the curve $\mathcal{C}_{\sigma_0,\Escr^{(P_2)}}^{(\sigma_2=0)}$ is not contained within the disc limited by the circle $\mathcal{S}_{\sigma_0}^{(\sigma_2=0)}$ (point $P_2$ in Fig.~\ref{Fig.zoom_Hopf_INTEG}). Then, the constant energy surface $\mathcal{C}_{\sigma_0,\,\Escr^{(P_2)}}$ has no common points with the sphere $\mathcal{S}_{\sigma_0}$ in the same neighborhood other than $P_2$. The fixed point corresponds to a Floquet-stable periodic orbit. For values of the energy $\Escr\approx\Escr^{(P_2)}$ we obtain surfaces $\mathcal{C}_{\sigma_0,\, \Escr}$ which intersect the sphere at closed curves near to, and surrounding $P_2$. These are invariant curves of the reduced flow representing 2-tori of the complete flow around the stable periodic orbit.

\subsubsection{Case II: degenerate transverse intersection of $\mathcal{S}_{\sigma_0}$ with $\mathcal{C}_{\sigma_0,\, \Escr}$}
\label{subsec:CPII}

\begin{figure}[!h]
\begin{center}
\includegraphics[scale=.82]{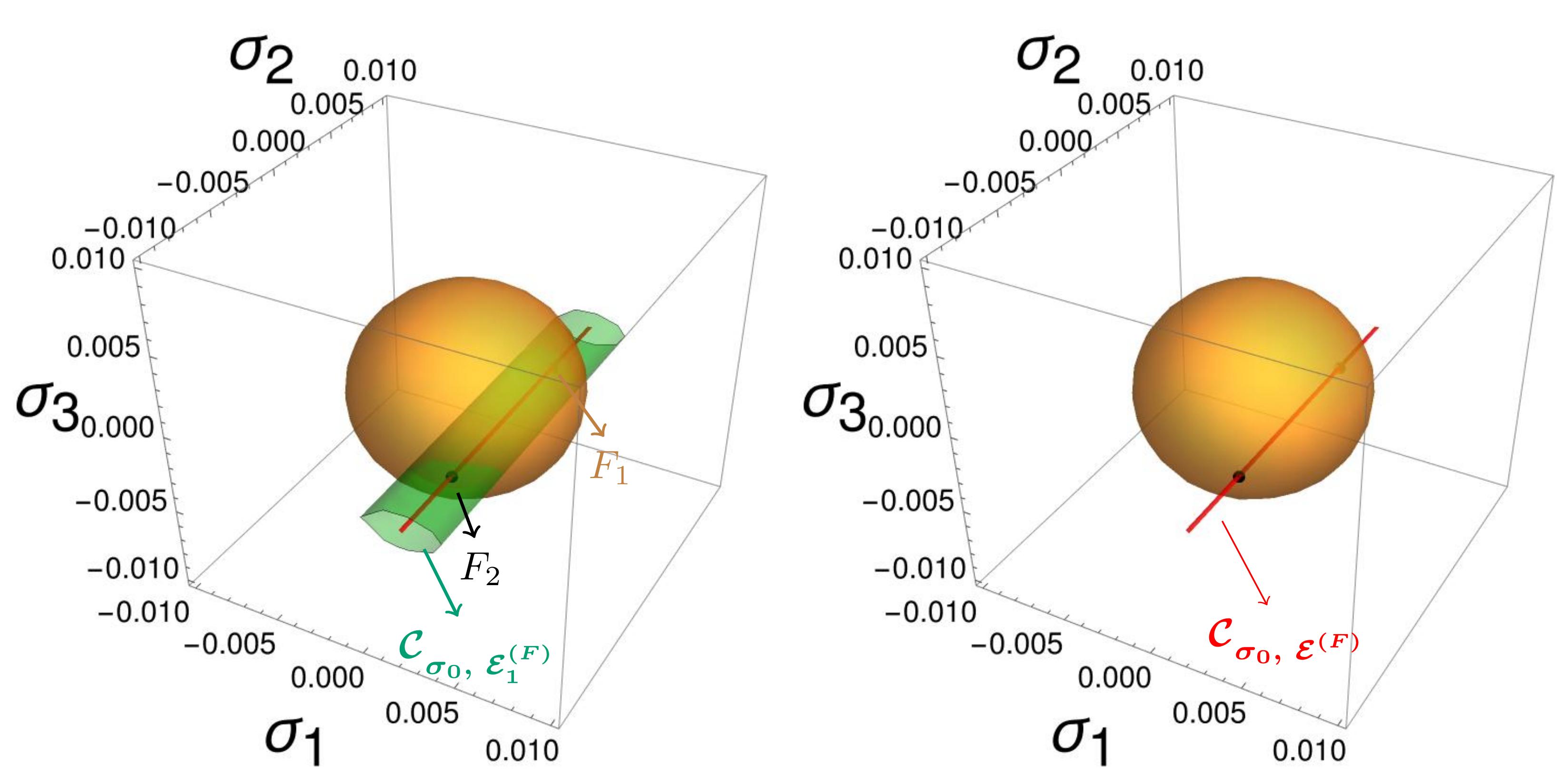}\\
\includegraphics[scale=.78]{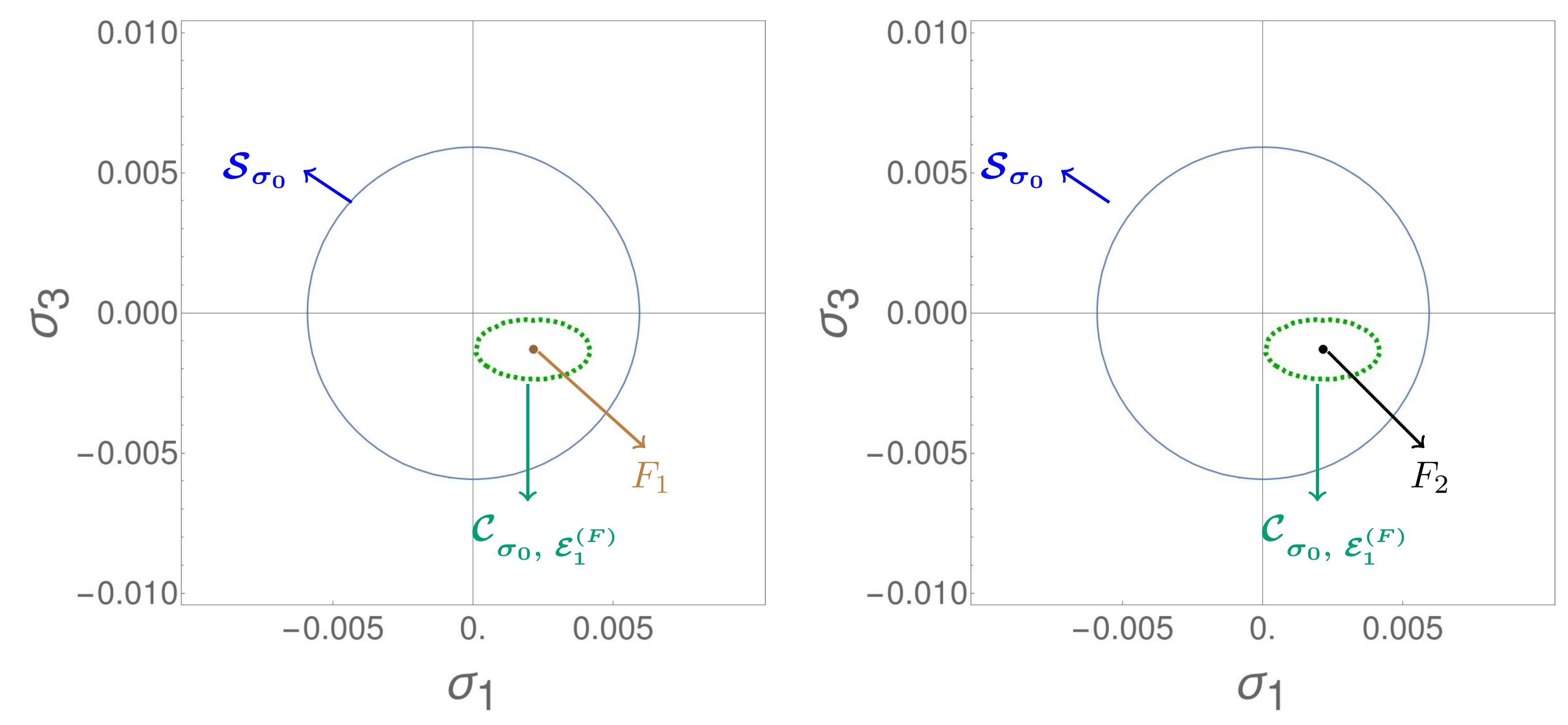}
\end{center}
\caption{Graphical representation of the critical points of second kind $F_1$ and $F_2$  {in the} elliptic case (see text). The example is taken from the choice of integrable model $\Zscr=\Hscr_{int}$, and it corresponds to the phase portrait $\mathcal{P_Z}(\sigma_0)$ with $\sigma_0=5.93\cdot 10^{-3}$ (the third panel in the top row of Fig.~\ref{Fig.Poin_Hopf_INTEG}).  {The top} left panel shows the intersection of the sphere $\mathcal{S}_{\sigma_0}$ with an energy surface $\mathcal{C}_{\sigma_0,\, \Escr}$ for $\Escr\approx\Escr^{(F)}$ and the  {right top} panel for $\Escr=\Escr^{(F)}$.  
 {The bottom} panels  {show} the corresponding projections to the plane $\sigma_2=0$ (intersection of the curves $\mathcal{C}_{\sigma_0,\, \Escr}^{(\sigma_2=0)}$ with the circles $\mathcal{S}_{\sigma_0}^{(\sigma_2=0)}$). We have  $\Escr=\Escr_1^{(F)}=-1.7362\cdot 10^{-5}$ and $\Escr^{(F)}=-1.7366\cdot 10^{-5}$, in which the corresponding surfaces are a cylinder ( {top} left panel) reducing to a straight line ( {top right} panel).  {The curve $\mathcal{C}_{\mathcal{Z}}(\sigma_0, \Escr_{1}^{(F)})$ in the corresponding phase portraits $\mathcal{P_Z}(\sigma_0)$ (see third panel of first row of Fig.~\ref{Fig.Poin_Hopf_INTEG}) is plotted in green.} } 
\label{Fig.ellissi_INTEG_3D2D_1}
\end{figure}


\begin{figure}[!h]
\begin{center}
\includegraphics[scale=.82]{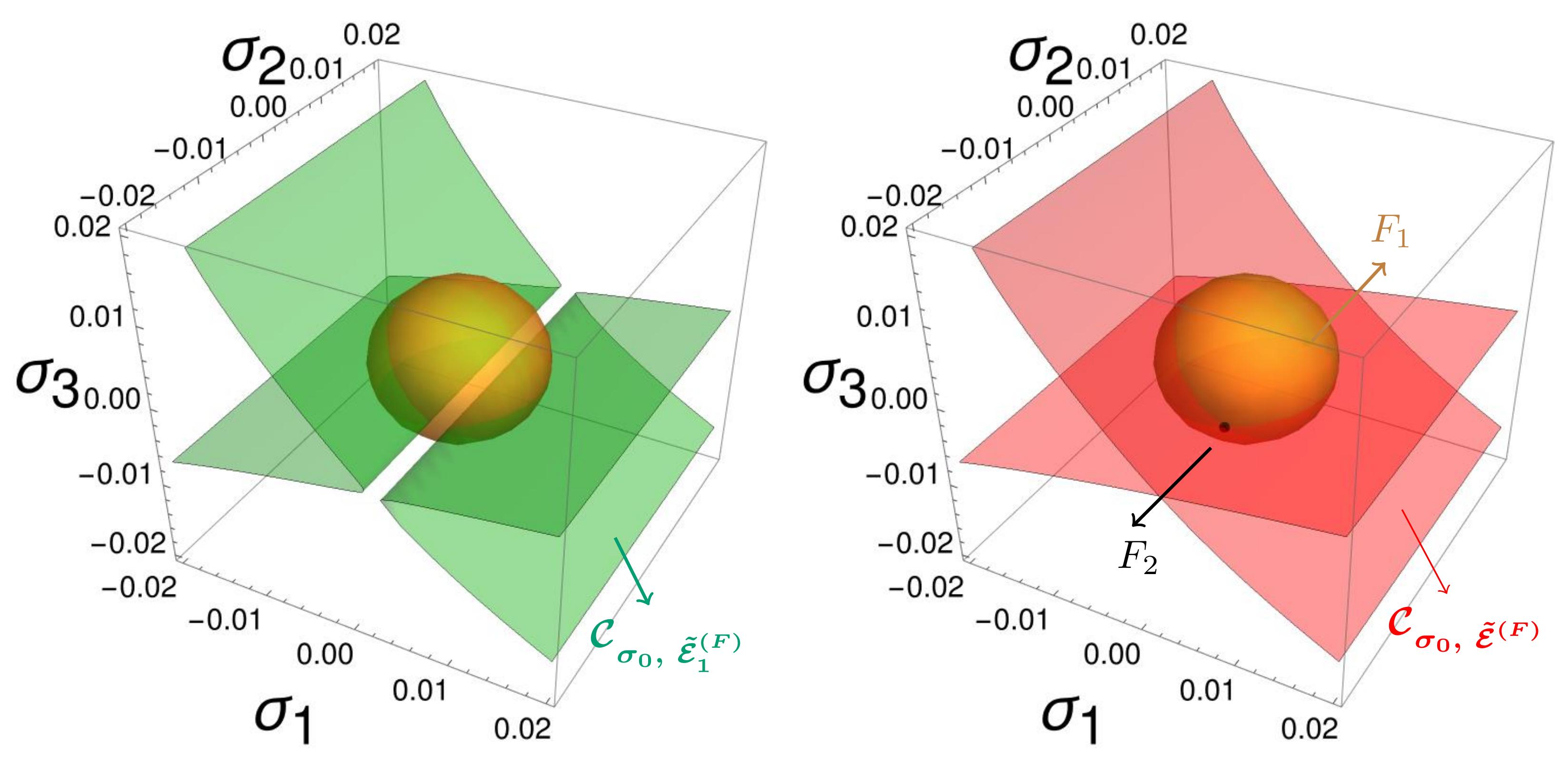}
\\
\includegraphics[scale=.78]{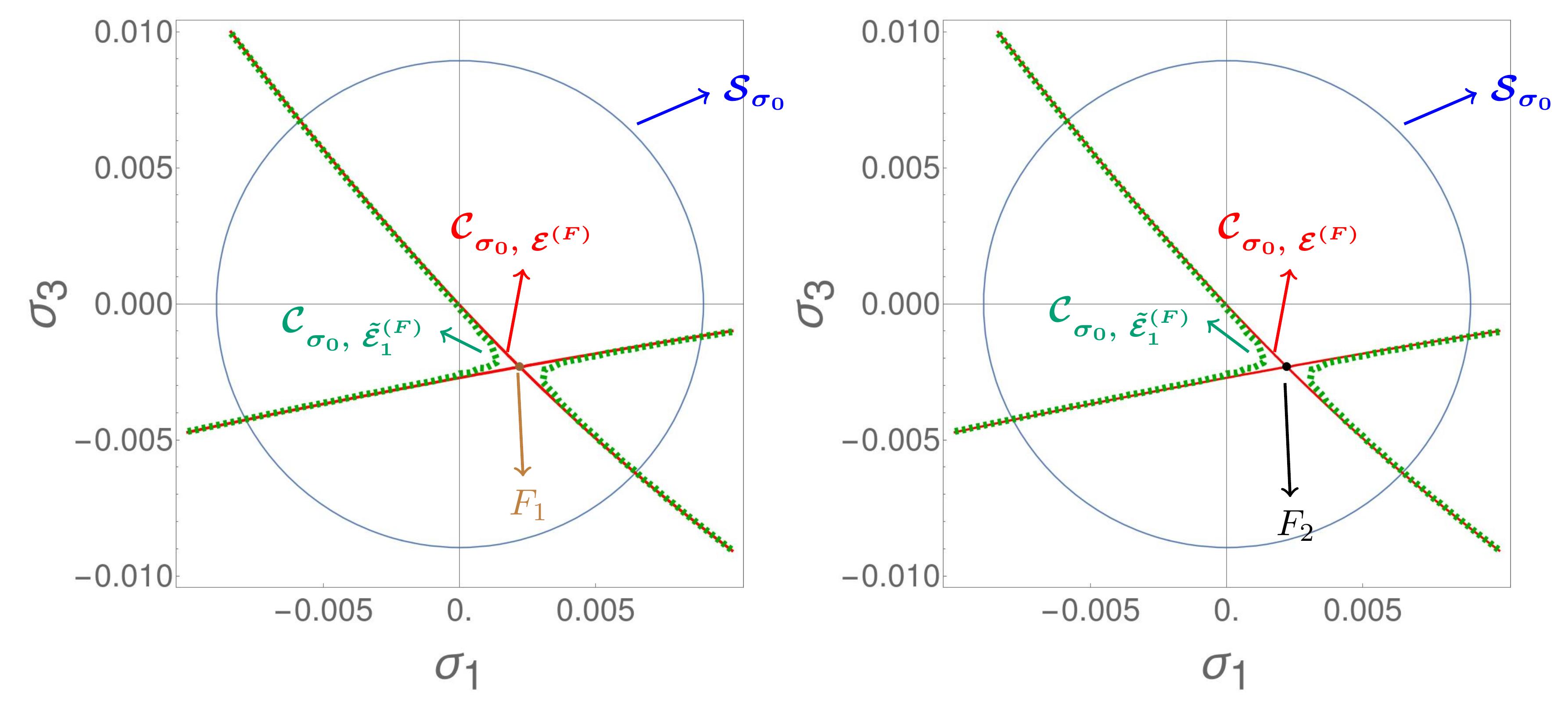}
\end{center}
\caption{Graphical representation of the critical points of second kind $F_1$ and $F_2$  {in the} hyperbolic case (see text). The example is taken from the choice of integrable model $\Zscr=\Hscr_{int}^{(1)}$, and it corresponds to the phase portrait $\mathcal{P_Z}(\sigma_0)$ with $\sigma_0=8.94\cdot 10^{-3}$ (see Fig.~\ref{Fig.Poin_Hopf_INTEG_NORMAL}).  {The top} left panel shows the intersection of the sphere $\mathcal{S}_{\sigma_0}$ with an energy surface $\mathcal{C}_{\sigma_0,\, \Escr}$ for $\Escr\approx\tilde{\Escr}^{(F)}$and the  {right top} panel for $\Escr= {\tilde{\Escr}}^{(F)}$.
 {The bottom} panels  {show} the corresponding projections to the plane $\sigma_2=0$ (intersection of the curves $\mathcal{C}_{\sigma_0,\, \Escr}^{(\sigma_2=0)}$ with the circles $\mathcal{S}_{\sigma_0}^{(\sigma_2=0)}$). We have  $\Escr=\tilde{\Escr}_1^{(F)}=-2.874\cdot 10^{-5}$ and $\tilde{\Escr}^{(F)}=-2.87453\cdot 10^{-5}$, where the corresponding surfaces are hyperbolic sheets ( {top} left panel) reducing to two sheets intersecting at the line $F_1F_2$ in the  {top right} panel.  {The curve $\mathcal{C}_{\mathcal{Z}}(\sigma_0, \t{\Escr}_{1}^{(F)})$ in the corresponding phase portraits $\mathcal{P_Z}(\sigma_0)$ (see second panel of second row of Fig.~\ref{Fig.Poin_Hopf_INTEG_NORMAL}) is plotted in green.}}
\label{Fig.ellissi_INTEG_3D2D_2}
\end{figure}
At those points $(\sigma_{1,F},\sigma_{3,F})$ where
\begin{align}
\label{CP2}
\left(\frac{\partial \Zscr}{\partial \sigma_1}\right)_{\sigma_1=\sigma_{1,F},\sigma_3=\sigma_{3,F};\sigma_0}=\left(\frac{\partial \Zscr}{\partial \sigma_3}\right)_{\sigma_1=\sigma_{1,F},\sigma_3=\sigma_{3,F};\sigma_0}=0\, 
\end{align}
(that is, Eq.~\eqref{CPkummer} with $\mu= 0$)
the surface $\mathcal{C}_{\sigma_0,\, \Escr^{(F)}}$, with $\Escr^{(F)}=\Zscr(\sigma_{1,F},\sigma_{3,F};\sigma_0)$ reduces either to a single straight line parallel to the $\sigma_2-$axis, or to two sheets intersecting along a line parallel to the $\sigma_2-$axis. If the line intersects  {transversely} the sphere $\mathcal{S}_{\sigma_0}$ (see Fig.~\ref{Fig.ellissi_INTEG_3D2D_1} and~\ref{Fig.ellissi_INTEG_3D2D_2}) we obtain two fixed points of the flow $F_1,F_2$ on the sphere, i.e. two periodic orbits of the complete flow. Such critical points (or orbits) are called of the `second kind' ( {CPII}). Their stability depends on the sign-definiteness of the quadratic form $(\partial^2\Zscr/\partial\sigma_1^2)\delta\sigma_1^2+2(\partial^2\Zscr/\partial\sigma_1\partial\sigma_3)\delta\sigma_1\delta\sigma_3+(\partial^2\Zscr/\partial\sigma_3^2)\delta\sigma_3^2$ where the partial derivatives are evaluated at $\sigma_1=\sigma_{1,F}$, $\sigma_3=\sigma_{3,F}$. If the quadratic form is sign-definite, the points $F_1,F_2$ are linearly stable. For values of the energy $\Escr$ near $\Escr^{(F)}$ the surface $\mathcal{C}_{\sigma_0,\, \Escr}$ has the form of a cylinder with elliptic cross-section which surrounds the axis $\mathcal{C}_{\sigma_0,\, \Escr^{(F)}}$, and its intersections with the sphere are elliptic-like curves which surround the fixed points $F_1$ or $F_2$ and represent 2-tori of the complete flow around the periodic orbits corresponding to the points $F_1$ or $F_2$ (see Fig.~\ref{Fig.ellissi_INTEG_3D2D_1}). In the case, however, where the sign of the quadratic form is indefinite, the points $F_1,F_2$ represent unstable periodic orbits, and the surface $\mathcal{C}_{\sigma_0,\, \Escr^{(F)}}$, together with nearby surfaces for energies close to $\Escr^{(F)}$ intersect the sphere forming saddles around each of the points $F_1$ or $F_2$ (see Fig.~\ref{Fig.ellissi_INTEG_3D2D_2}). 

\subsection{Analytical formulas}
\label{subsec:formule_Anto_Gius}
In the present section we implement our geometric method to analyze the sequence of bifurcations produced in the case in which the planetary system has a small AMD value, i.e. the planetary orbits (2D or 3D) never become very inclined and very eccentric.  {In this case, we can use a low-order truncation of the Hamiltonian model, leading to simplified expressions which can be treated by explicit analytical formulas. Note that the issue of the proper truncation of the Hamiltonian has been studied as regards i) the maximum multipole order (\cite{foretal2000}), or ii) the truncation order in the orbital eccentricities and inclinations (\cite{henlib2004}). Furthermore the accuracy of the integrable approximation with respect to the full secular Hamiltonian has been studied in subsection 3.2 of~\cite{maseft2023}.}  

 {More specifically, we consider below the low-order truncation model of the secular Hamiltonian written generically in Hopf variables as} 
\begin{equation}\label{Ham.K_prel}
\Kscr_{I}(\sigma_1, \sigma_3; \sigma_0)= A\sigma_1^2+C\sigma_3^2+B\sigma_1\sigma_3+(D_1
\sigma_0 + \tilde{\DELTA}_1)\sigma_1+(D_3 \sigma_0 + \tilde{\DELTA}_3)\sigma_3~.
\end{equation}
Without loss of generality, through a rotation we can set $B=0\,$, hence getting the following generic Hamiltonian
\begin{equation}\label{Ham.K}
\Kscr_{I}( {\tilde{\sigma}_1, \tilde{\sigma}_3; \tilde{\sigma}_0})= \AA {\tilde{\sigma}_1}^2+\CC {\tilde{\sigma}_3}^2+(\DD_1
 {\tilde{\sigma}_0} + \DELTA_1) {\tilde{\sigma}_1}+(\DD_3  {\tilde{\sigma}_0} + \DELTA_3) {\tilde{\sigma}_3} 
\end{equation}
with $\AA$, $\CC$, $\DD_1$, $\DD_3$, $\DELTA_1$, $\DELTA_3$ control parameters. 
The rotation leading to $B=0$ is a linear change of coordinates
$(\sigma_0,\sigma_1,\sigma_2,\sigma_3)\to(\tilde{\sigma}_0,\tilde{\sigma}_1,\tilde{\sigma}_2,\tilde{\sigma}_3)
$ such that $\sigma_0=\tilde{\sigma}_0\,$, $\sigma_2=\tilde{\sigma}_2$ and
\begin{align}\label{rotation}
&
\sigma_1=\alpha\tilde{\sigma}_1+\beta\tilde{\sigma}_3, && 
\sigma_3=-\beta\tilde{\sigma}_1+\alpha\tilde{\sigma}_3
\end{align}
for some sine and cosine coefficients $\alpha$ and $\beta$ computed so that
$$
\alpha^2+\beta^2=1\qquad  \mathrm{and} \qquad B \alpha^2 + 2 A \alpha\beta - 2 C \alpha\beta - B\beta^2 = 0~~.
$$
Resulting from a rotation, the coordinates $(\tilde{\sigma}_0,\tilde{\sigma}_1,\tilde{\sigma}_2,\tilde{\sigma}_3)$ still satisfy the condition  {given by Eq.}~\eqref{cond.sphere},  {i.e., $(\tilde{\sigma}_1,\tilde{\sigma}_2,\tilde{\sigma}_3)$ are on a sphere of radius $\tilde{\sigma}_0$,} as well as the Poisson algebra $\poisson{\tilde{\sigma}_i}{\tilde{\sigma}_j}=-2\epsilon_{ijk}\tilde{\sigma}_k$, $i,j,k=1,2,3\,$. Moreover $\sigma_0=\tilde{\sigma}_0\,$ is also a Casimir invariant of the algebra in the new variables. Hence, computing the values of $\tilde{\sigma}_0$ for which the CP occurs (i.e, solving Eqs.~\eqref{eq.CP1} and~\eqref{sol.CP2}) is equivalent to finding the radius $\sigma_0$ for which the sphere and the surface $ {\Cscr_{\sigma_0,\Escr}}$ are tangent. For the sake of simplicity, in the following examples we perform the rotation (\ref{rotation}) whenever needed and then simply rename the variables $(\tilde{\sigma}_0,\tilde{\sigma}_1,\tilde{\sigma}_2,\tilde{\sigma}_3)$ in Eq.~\eqref{Ham.K} as $({\sigma}_0,{\sigma}_1,{\sigma}_2,{\sigma}_3)$.

The advantage on considering a low-order system is that implementing the geometric method of the previous sections leads to analytical formulas allowing to predict the form and sequence of the bifurcations of periodic orbits stemming from the basic ACR states. 

To find the non-degenerate critical points of the type  {CPI} of the Hamiltonian \eqref{Ham.K} we solve Eq.~\eqref{CPkummer}, with the assumption that $(\DD_1
\sigma_0 + \DELTA_1)$ and $(\DD_3 \sigma_0 + \DELTA_3)$ are not vanishing. This gives
\begin{eqnarray}
\label{CP1_coord}
\sigma_1^{( {CPI})} (\mu,\sigma_0)&=&- \frac{ \DD_1\sigma_0+\Delta_1}
                     {2(\AA-\mu)} \nonumber \, ,\\
     \sigma_2^{( {CPI})} (\mu,\sigma_0)&=&      0 \, ,\\         
\sigma_3^{( {CPI})}(\mu,\sigma_0)&=& -\frac{ \DD_3 \sigma_0+ \Delta_3}
	             {2(\CC-\mu)} \nonumber \,.
\end{eqnarray}
\noindent {If $(\DD_1
\sigma_0 + \DELTA_1)=0$, Hamiltionian  \eqref{Ham.K} becomes symmetric with respect to the reflection $\sigma_1\rightarrow -\sigma_1$, corresponding to the $2:2$ resonance (see~\cite{marpuc2016}). Then equation \eqref{CPkummer} is solved by 
\begin{equation}
\mu=\mathcal{A}\;\;,\,\, \sigma_3^{( {CPI})}(\mu,\sigma_0)= -\frac{ \DD_3 \sigma_0+ \Delta_3} {2(\CC-\mu)}.
\end{equation}
Similarly, if $(\DD_3\sigma_0 + \DELTA_3)=0$ we find the solution to be
\begin{equation}
\mu=\CC\;\;, \,\,\sigma_1^{( {CPI})} (\mu,\sigma_0)=- \frac{ \DD_1\sigma_0+\Delta_1}{2(\AA-\mu)}.
\end{equation}
}\noindent
 {Exchanging $\sigma_3\rightarrow\sigma_1$, the case $(\DD_3\sigma_0 + \DELTA_3)=0$ corresponds to  $(\DD_1\sigma_0 + \DELTA_1)=0$. The bifurcation sequences of the Hamiltonian \eqref{Ham.K} are well understood in this last case (see~\cite{marpuc2016}). In the present work, instead, we focus on the study of the generic case in which $(\DD_j\sigma_0 + \DELTA_j)\neq0$, $j=1,3$.}

\subsubsection{ {Computation of the bifurcation value of $\sigma_0$}}
\label{subsub:anal}

 {The value of $\mu$ corresponding to the solutions of Eq.~\eqref{CPkummer} is given by} the constraint
$$ {\mathcal S} (\mu,\sigma_0) = (\sigma_1^{( {CPI})}(\mu,\sigma_0))^2 + (\sigma_3^{( {CPI})}(\mu,\sigma_0))^2 - \sigma_0^2 = 0\, , $$
 {where the expressions for $\sigma_1^{( {CPI})}$ and  $\sigma_3^{( {CPI})}$ are provided in Eq.~\eqref{CP1_coord}.}
 {There exist at least two solutions to the above equations for $\mu$ (see~\cite{kum1976}). Thus}, we  look for the  {critical values of $\sigma_0$ determining a change in the number of real solutions in $\mu$ from $2$ to $4$, or viceversa.}

The numerator of ${\cal S} (\mu,\sigma_0) = 0$ corresponds to an equation of 4th degree in $\mu$, namely
\begin{equation}\label{eqEquilambda}
4 (\AA-\mu )^2 (\CC-\mu )^2-(\AA-\mu )^2\, \mathcal T_3(\sigma_0)- (\CC-\mu )^2\, \mathcal T_1(\sigma_0)=0\, ,
\end{equation}
where
\begin{equation}
\label{def_T1T3}
\mathcal T_1(\sigma_0)= \left(\DD_1+\frac{\Delta_1}{\sigma_0}\right)^2, \;\; 
\mathcal T_3(\sigma_0)=\left(\DD_3+\frac{\Delta_3}{\sigma_0}\right)^2\, .
\end{equation}
When the number of real solutions of the above equation changes, we have the corresponding {\it bifurcation value} of $\sigma_0$. 
The discriminant  of \eqref{eqEquilambda} w.r.t. $\mu$ is a function in $\sigma_0$  whose simple zeros provide the required change in the number of
 solutions. It is given by
\begin{equation}
\begin{aligned}
\mathcal Q(\sigma_0)&=64 (\AA - \CC)^2\, \mathcal{T}_1(\sigma_0) \,\mathcal {T}_3(\sigma_0)\Big[\left(4 (\AA-\CC)^2-\mathcal{T}_1(\sigma_0)\right)^3  -3\mathcal{T}_3(\sigma_0) \Big(16 (\AA-\CC)^4\\
&\phantom{=} + 28 \mathcal{T}_1(\sigma_0) (\AA-\CC)^2+ \mathcal{T}_1(\sigma_0)^2\Big)+3 \mathcal{T}_3(\sigma_0)^2 \left(4 (\AA-\CC)^2 - \mathcal{T}_1(\sigma_0)\right) -\mathcal{T}_3(\sigma_0)^3\Big] \,  {.}\label{discriminant}
\end{aligned}
\end{equation}
Notice that  {the values of $\sigma_0$ such that} $\mathcal{T}_1(\sigma_0) =0$ or $\mathcal{T}_3(\sigma_0) =0$ do not correspond to a change in the sign of the discriminant {, since $\mathcal{T}_1(\sigma_0)$ and $\mathcal{T}_3(\sigma_0)$ are quadratic expressions.} At  $\AA=\CC$ the discriminant is identically zero for every value of $\sigma_0$; this is the situation at which the ellipses degenerate to circles. Therefore, tangencies with the phase space can occur either at an infinite number of points or at no point. Higher order normal forms would be needed to describe the bifurcations.  We do not consider this case here,  {which, anyway, has no particular physical significance and would appear only through a coincidence of the values of some Laplace coefficients, for finely tuned values of the planetary masses and/or semi-major axes.}

 {Dividing Eq.~\eqref{discriminant} by $\mathcal {T}_3(\sigma_0)$, the discriminant with respect to $\mathcal T_3(\sigma_0)$ turns to be always negative. This means that, apart from the solutions $\mathcal{T}_3(\sigma_0) =0$, the equation}
$\mathcal{Q}(\sigma_0)=0$  {has only one real solution for} $\mathcal T_3(\sigma_0)$,  {namely}
 {$$\mathcal T_3 (\sigma_0)=4(\AA-\CC)^2-\mathcal T_1(\sigma_0)+ 2^{2/3}\, 3\Big((\AA-\CC)^2\,\mathcal T_1(\sigma_0)^2\Big)^{1/3} - 2^{1/3}\,6 \Big((\AA-\CC)^{4}\,\mathcal T_1(\sigma_0)\Big)^{1/3}\,.$$}
 {Recalling the definitions of $\mathcal{T}_1(\sigma_0)$ and $\mathcal{T}_3(\sigma_0)$ (Eq.~\eqref{def_T1T3}), it is easy to see that the above expression gives an equation for the bifurcation values of $\sigma_0$, namely}
\begin{equation}\label{eq.CP1}
f_{1}(\sigma_0)=0\, ,
\end{equation}
where
\begin{equation}
f_1(\sigma_0)=-4(\AA-\CC)^2+\mathcal T_1(\sigma_0)+ \mathcal T_3 (\sigma_0)- 2^{2/3}\, 3\Big((\AA-\CC)^2\,\mathcal T_1(\sigma_0)^2\Big)^{1/3} + 2^{1/3}\,6 \Big((\AA-\CC)^{4}\,\mathcal T_1(\sigma_0)\Big)^{1/3}\,.
\end{equation} 
 {For real not negative roots $\sigma_0$ of Eq.~\eqref{eq.CP1}, the discriminant with respect to $\mu$ of Eq.~\eqref{eqEquilambda} (i.e. Eq.~\eqref{discriminant}) vanishes. This implies that a pair of real solutions in $\mu$ of Eq.~\eqref{eqEquilambda} appears/disappears. Correspondingly, a pair of equilibria appears/disappears for the  Hamiltonian~\eqref{Ham.K}, i.e., we obtain a bifurcation.  In general, we cannot predict the number of non-negative real solutions of Eq.~\eqref{eq.CP1} for $\sigma_0$, as this depends on the specific values of $\AA,\CC, \mathcal D_j, \Delta_j$. However, since Eq.~\eqref{eqEquilambda} is of fourth degree in $\mu$ and we know that at least two real solutions for $\mu$ always exist (\cite{kum1976}), we can conclude that for the values of $\sigma_0$ that solve Eq.~\eqref{eq.CP1}, the number of fixed points of the type CPI of system~\eqref{Ham.K}  changes from $2$ to $4$, or vice versa.}

Instead,  {if $\AA\CC\neq0$,} the bifurcation values of $\sigma_0$ for fixed points of the type  {CPII} can be found by solving
\begin{equation}\label{EqCP2}
f_{2}(\sigma_0)=-4+\frac{\mathcal T_1(\sigma_0)}{\AA^2}+\frac{\mathcal T_3(\sigma_0)}{\CC^2}=0\,.
\end{equation}
 {In the case $\AA=0$ or $\CC=0$, there are no isolated equilibria of the second kind (i.e. CPII). This is a degenerate case that we do not consider in this work.}
 {If $\CC^2\DD_1^2+\AA^2(-4\CC^2+\DD_3^2)\neq0$, we find two solutions for Eq.~\eqref{EqCP2}, given by}
\begin{equation}
\label{sol.CP2}
\begin{split}
&\sigma_0^{( {CPII},1)}=-\frac{\CC^2\DD_1\DELTA_1+\AA^2\DD_3\DELTA_3+\AA\CC\sqrt{4\CC^2\DELTA_1^2-(\DD_3\DELTA_1+2\AA\DELTA_3-\DD_1\DELTA_3)(\DD_3\DELTA_1-(2\AA+\DD_1)\DELTA_3)}}{\CC^2\DD_1^2+\AA^2(-4\CC^2+\DD_3^2)}\, ,\\
&\sigma_0^{( {CPII},2)}=-\frac{\CC^2\DD_1\DELTA_1+\AA^2\DD_3\DELTA_3-\AA\CC\sqrt{4\CC^2\DELTA_1^2-(\DD_3\DELTA_1+2\AA\DELTA_3-\DD_1\DELTA_3)(\DD_3\DELTA_1-(2\AA+\DD_1)\DELTA_3)}}{\CC^2\DD_1^2+\AA^2(-4\CC^2+\DD_3^2)}\, ,
\end{split}
\end{equation}
 {provided that $$4\CC^2\DELTA_1^2-(\DD_3\DELTA_1+2\AA\DELTA_3-\DD_1\DELTA_3)(\DD_3\DELTA_1-(2\AA+\DD_1)\DELTA_3)\geq0.$$}
 {Instead, we find
\begin{equation} \label{sol.CP2-1}
\sigma_0^{( {CPII})}=-\frac{(\CC^2 \Delta_1^2 +
 \AA^2 \Delta_3^2)|\CC|}{2 (\AA^2\DD_3\Delta_3 |\CC| \pm \CC^2 \Delta_1 |\AA| \sqrt{4 \CC^2 - \DD_3^2)}} \,
\end{equation}
if  $\DD_1=\pm\frac{|\AA|}{|\CC|}\sqrt{(4\CC^2-\DD_3^2)}$, respectively. In this case, necessarily  $\AA^2\DD_3\Delta_3 |\CC| \pm \CC^2 \Delta_1 |\AA| \sqrt{4 \CC^2 - \DD_3^2}\neq0$ and $4 \CC^2 - \DD_3^2\geq 0$.}
 {Otherwise if
$\AA^2\DD_3\Delta_3 |\CC| \pm \CC^2 \Delta_1 |\AA| \sqrt{4 \CC^2 - \DD_3^2}=0$
there is no solution of Eq.~\eqref{EqCP2} for $\sigma_0$.}

Geometrically,  {the fixed points of type CPII correspond to solutions of Eq.~\eqref{CPkummer} at which the surface $\mathcal C_{\sigma_0,\mathcal E} $ of constant energy of the Hamiltonian \eqref{Ham.K} degenerates to a straight line parallel to the $\sigma_2-$axis or to two sheets intersecting along a line parallel to the $\sigma_2-$axis. At the values of $\sigma_0^{(CPII)}$ given by \eqref{sol.CP2} or \eqref{sol.CP2-1}, this line is tangent to the sphere $\mathcal S_{\sigma_0^{(CPII)}}$. In particular, on the plane $\sigma_2=0$, the center of the family of ellipses or hyperbolas corresponding to $\mathcal C_{\sigma_0^{(CPII)},\,\mathcal E}$  lies on the circle
$\mathcal{S}_{\sigma_0^{(CPII)}}^{(\sigma_2=0)}$, i.e.  $\sigma_1^2+\sigma_3^2=\left(\sigma_0^{(CPII)}\right)^2.$
The appearance/disappearance of the CPII occurs when the center of the ellipses or hyperbolas enters or leaves this circle (see Section~\ref{subsec:CPII}). }

\section{Application to different integrable models approximating the secular Hamiltonian}
\label{sec:Apply}
 {In this section, we apply the method of the previous section in order to compute the sequence of bifurcations in three different Hamiltonian models approximating the extrasolar planetary system, i.e., the Hamiltonians $\Hscr_{int}$, $\Hscr_{int}^{(1)}$ and $\tilde{\Hscr}_{int}^{(N_P=3,\, N_{bk}=4)}$. As already mentioned, $\Hscr_{int}$ represents the integrable approximation of the full secular Hamiltonian analyzed in Section~\ref{sec:Ham} (see Eq.~\eqref{h3ddecompo2}). Since we use a secular Hamiltonian model truncated at multipolar order $N_P = 5$ and of book-keeping order (in eccentricity and inclination) $N_{bk} = 12$ (see Section~\ref{subsec:Poinc}), the integrable Hamiltonian $\Hscr_{int}$, expressed in the Hopf variables contains powers in $\sigma_i$, $i=0,\dots,3$ up to $6-$th degree. For this reason, in the analyses of Section~\ref{subsec:ApplyHint}, as well as of Section~\ref{subsec:Norm_Hint1}, we can apply the general discussion of Sections~\ref{subsec:Phaseequiv}-\ref{subsec:Hopf_analysis_integ}, but we cannot apply the analytical formul{\ae} described in Section~\ref{subsec:formule_Anto_Gius}. In fact, the exact equality given in Eq.~\eqref{Ham.K_prel}, that is the basis for the development of the analytical formul{\ae}, would be valid just for a Hamiltonian model developed up to fourth order in eccentricity. In the case of $\Hscr_{int}$ and $\Hscr_{int}^{(1)}$, Eq.~\eqref{Ham.K_prel} is just an approximation and not an exact equality. The approximation is sufficient to understand the global behavior of the system and the shape of the curves through which the bifurcation sequences are computed, but not to provide quantitative estimates on the bifurcation values. Instead, the analytical formul{\ae} of subsection \ref{subsec:formule_Anto_Gius} can be used in Section~\ref{subsec:NormOct}, where the equality is valid. 
}

\subsection{Integrable Hamiltonian $\Hscr_{int}$}
\label{subsec:ApplyHint}
As a first example of application of the geometric method exposed to Section \ref{sec:Bif}, we consider the integrable Hamiltonian $\Hscr_{int}$ (Eq.~\eqref{def.h0}). 
 {Apart from constants, the integrable Hamiltonian can be written as}
\begin{equation}
\label{Ham.integ.}
\Hscr_{int} = A\sigma_1^2+C\sigma_3^2+B\sigma_1\sigma_3+D(\sigma_0)\sigma_1+E(\sigma_0)\sigma_3+F(\sigma_0) + \Oscr\left((\sigma_i)^2\right)\, , 
\end{equation}
with $i=0,\,\dots,3$,
where the symbol $\Oscr((\sigma_i)^2)$ correspond to fourth order in the eccentricities.
 {The numerical value of the coefficients are given in Appendix~\ref{appendix:num_val_H_int}. For those values of the coefficients $A,B,C$, the condition $B^2<4A C\,$ is satisfied}, hence the quadratic form $A\sigma_1^2+C\sigma_3^2+B\sigma_1\sigma_3$ yields ellipses.  Thus, for any permissible value $\Escr$, the curve $\mathcal{C}_{\sigma_0,\, \Escr}^{(\sigma_2=0)}$ is ellipse-like.

\begin{figure}[htp]
\begin{center}
\includegraphics[width=1\textwidth]{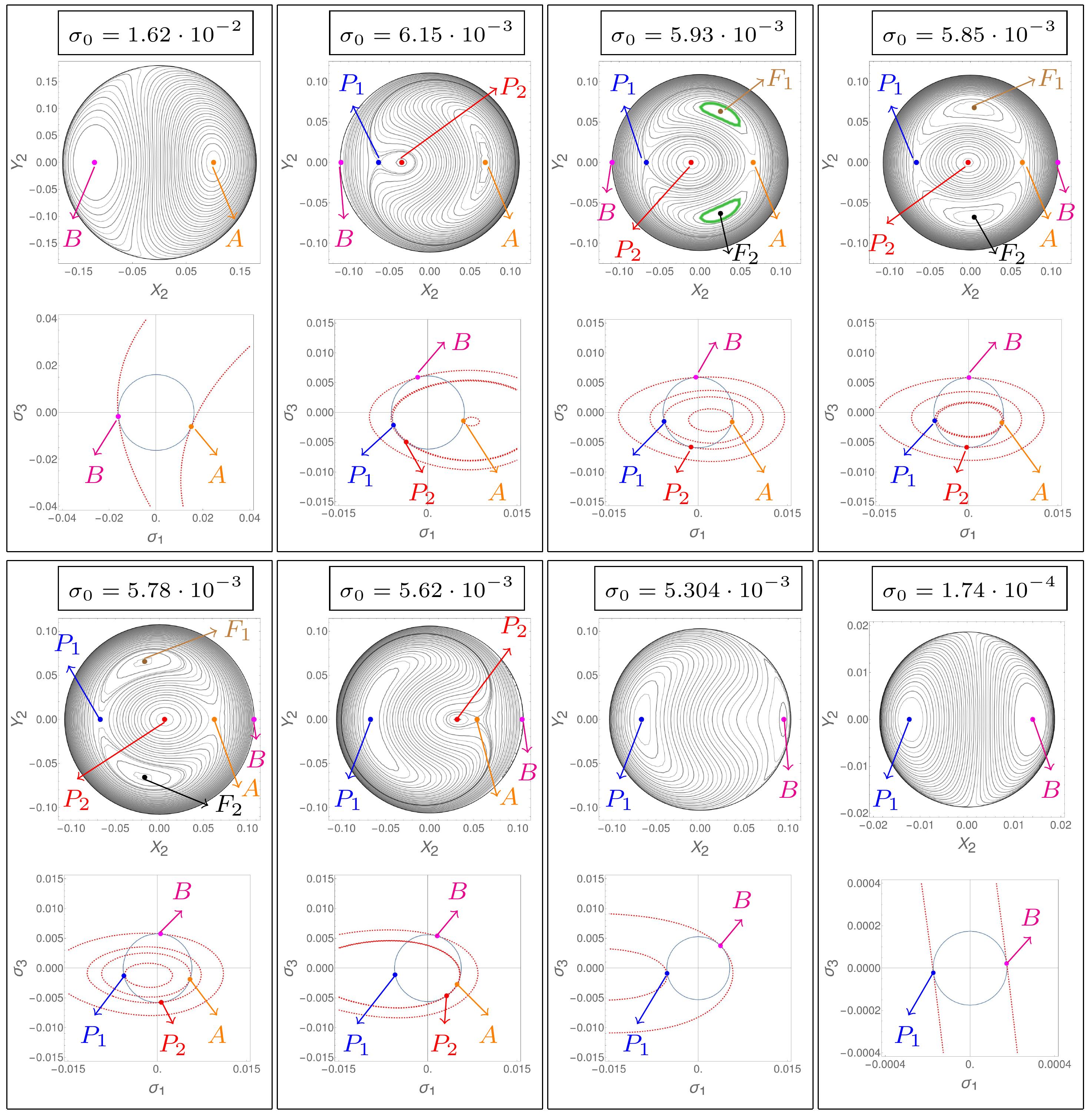}
\caption{The first and third rows show the contour plots $\mathcal{P}_{\Hscr_{int}}(\sigma_0;\AMD)$ (same as in Fig.~\ref{Fig.npol5eps12INTEGsigma0}), where the position of the fixed points (indicated by colored points) has been computed by the tangency method  {explained in Section~\ref{subsec:Hopf_analysis_integ}, i.e. computing, for each panel, the tangencies between the sphere $\mathcal{S}_{\sigma_0}$ of fixed radius $\sigma_0$ and $\mathcal{C}_{\sigma_0, \Escr}.$ Computing these tangencies, we obtain critical points of the first kind (i.e. A, B, $P_1$, $P_2$)  satisfying the condition given by Eq.~\eqref{CP1}, as well as critical points of the second kind (the points $F_1$ and $F_2$) satisfying ~\eqref{CP2}}. The second and fourth rows show the intersections of the spheres $\mathcal{S}_{\sigma_0}$ (blue curve $\mathcal{S}_{\sigma_0}^{(\sigma_2=0)}$) with the energy surfaces $\mathcal{C}_{\sigma_0,\Escr}$(red dashed curves $\mathcal{C}_{\sigma_0,\Escr}^{(\sigma_2=0)}$)  {in} the plane $(\sigma_1,\sigma_3)$ for $\sigma_2=0$, for different values of the energy $\Escr$ and values of $\sigma_0$ as indicated in each panel.}
\label{Fig.Poin_Hopf_INTEG}
\end{center}
\end{figure}
Figure \ref{Fig.Poin_Hopf_INTEG} shows the phase portraits $\mathcal{P}_{\Hscr_{int}}(\sigma_0)$ in the above model. Altering the value of $\sigma_0$ as indicated in each panel, we obtain fixed points of both classes  {CPI} and  {CPII} in the phase portraits (first and third row). These are compared with the fixed points found through the curves $\mathcal{S}_{\sigma_0}^{(\sigma_2=0)}$ and $\mathcal{C}_{\sigma_0,\Escr}^{(\sigma_2=0)}$. As $\sigma_0$ decreases, implying a smaller value of the total angular momentum, the general level of mutual inclination of the orbits increases (from top left to bottom right in Fig.~\ref{Fig.Poin_Hopf_INTEG}). The top left panel shows the starting point of the observed bifurcation sequence, where we have the two fixed points A and B corresponding to the basic ACR states. These are tangency points of $\mathcal{S}_{\sigma_0}$ with $\mathcal{C}_{\sigma_0,\, \Escr}$ at the values $\sigma_0=\sigma_{0_{max}}(=1.62\cdot 10^{-2}$) and the energies $\Escr=\Escr_{min}$ for the mode A, and $\Escr=\Escr_{2,3}$ for the mode B (cf. with caption in Fig.~\ref{Fig.Evssigma0}). Both A and B are points of outer tangency of the curves $\mathcal{S}_{\sigma_0}^{(\sigma_2=0)}$ and $\mathcal{C}_{\sigma_0,\Escr}^{(\sigma_2=0)}$, hence linearly stable. It is easy to verify that $\psi^{(A)}=0$ and $\psi^{(B)}=\pi$, leading to $\varpi_3=\varpi_2+\pi$ (perihelia anti-aligned) and $\varpi_3=\varpi_2$ (perihelia aligned). Hence, the associated orbits yield two coplanar ellipses with, respectively, anti-aligned and aligned pericenters precessing with the same frequency. 

Decreasing the value of $\sigma_0$, at the second frame on the top of Fig.~\ref{Fig.Poin_Hopf_INTEG}, a \textit{saddle-node} bifurcation takes place, giving rise to two new fixed points, i.e., the periodic orbits $P_1$ and $P_2$. $P_1$ corresponds to an inner tangency of the curves $\mathcal{S}_{\sigma_0}^{(\sigma_2=0)}$ with $\mathcal{C}_{\sigma_0,\Escr}^{(\sigma_2=0)}$ (unstable), while $P_2$ corresponds to an outer tangency (stable) (see Fig.~\ref{Fig.zoom_Hopf_INTEG}). Besides these orbits, we still have present in the phase portraits the stable fixed points A and B (outer tangencies) of the ACR states. However, the ellipse yielding the outer tangency of the fixed point A gradually shrinks in size as $\sigma_0$ decreases. Then, at a critical value of $\sigma_0=\sigma_0^{(A)}$ the ellipse shrinks to a point, while, for still lower values, as in the third top panel of Fig.~\ref{Fig.Poin_Hopf_INTEG}, the tangent ellipse re-emerges, being now contained in the disc limited by  $\mathcal{S}_{\sigma_0}^{(\sigma_2=0)}$, yielding now a fixed point A of inner tangency, hence unstable. At the critical value $\sigma_0^{(A)}$ we then have a pitchfork bifurcation, accompanied with the birth of two new stable fixed points ($F_1$ and $F_2$), not shown in the second and fourth rows of Fig.~\ref{Fig.Poin_Hopf_INTEG} since they are of the type  {CPII}, i.e., out of the plane $\sigma_2=0$. For values of $\sigma_0$ smaller than $\sigma_0^{(A)}$, there are surfaces $\mathcal{C}_{\sigma_0,\, \Escr}$ of the form of the elliptic cylinders, as in the  {top left} panel of Fig.~\ref{Fig.ellissi_INTEG_3D2D_1}, which intersect transversely with the sphere $\mathcal{S}_{\sigma_0}$. The intersection yield invariant curves forming the islands of stability around the fixed points $F_1$ and $F_2$. The latter correspond to the value of the energy $\Escr^{(F)}$ at which the elliptic cylinders collide to straight lines ( {top right} panel in Fig.~\ref{Fig.ellissi_INTEG_3D2D_1}). In the third, fourth and fifth panel we find elliptic cylinders for energies $\Escr>\Escr^{(F)}$. However, decreasing further $\sigma_0$, the fixed points $F_1$, $F_2$ eventually collide with $P_1$ (sixth panel in Fig.~\ref{Fig.Poin_Hopf_INTEG}), and this collision terminates the $F$-family of periodic orbits by an inverse pitchfork bifurcation which renders the point $P_1$ stable. Note that this process is equivalent to the one of the birth of the F-family, since, at the point of the collision the tangency corresponding to the point $P_1$ turns from inner to outer, i.e., the point $P_1$ turns from unstable to stable. Finally, decreasing $\sigma_0$ still further, the points $P_2$ and A similarly disappear by an inverse saddle-node bifurcation. Hence,  the only surviving periodic orbits for small $\sigma_0$ are the stable orbits $P_1$ and B.

As regards its comparison with the full Hamiltonian, simple visual comparison between  Figs.~\ref{Fig.Poin_Hopf_INTEG} and \ref{Fig.npol5eps12} shows that the model $\Hscr_{int}$ does not capture the correct sequence of bifurcations of periodic orbits seen in the phase portraits obtained by the full 3D secular Hamiltonian model $\Hscr_{sec}$. This is remedied, however, computing a higher order secular normal form, as discussed in the next subsection.  

\subsection{Secular normal form $\Hscr_{int}^{(1)}$}
\label{subsec:Norm_Hint1}
A second integrable model, giving rise to a better integrable approximation of the Hamiltonian model $\Hscr_{sec}$, can be derived by constructing a \textit{secular normal form} of the Hamiltonian $\Hscr_{sec}$. As shown below, a first order normal form leads to an integrable model able to qualitatively recover the correct sequence of bifurcations found in the full model $\Hscr_{sec}$.

To compute the normal form, we make use of the `book-keeping method' discussed in \cite{eft2012}. Starting from the Hamiltonian of Eq.~\eqref{hamdecompo}
 {
\begin{equation*}
\Hscr_{sec}(w_2, w_3, W_2, W_3)=\Hscr_{int}( w_2-w_3, W_2, W_3)+\Hscr_{1,space}( w_2,w_3,W_2, W_3)\, ,
\end{equation*}
the integrable part of the Hamiltonian $\Hscr_{int}$ is decomposed as
\begin{equation*}
\label{H0_int_norm_1}
\Hscr_{int}=Z_0(  W_2, W_3)+\lambda Z_{0,1}( w_2-w_3, W_2, W_3) ,
\end{equation*}
where 
\begin{equation}
\label{Z0_1}
Z_0( W_2, W_3 )=c + a W_2 + b W_3~~
\end{equation}
is the normal form term (with $c$ a constant value) and $\lambda$ is a `book-keeping' symbol, numerically equal to $\lambda=1$, used to organize all the normal form series terms in groups of similar order of smallness. Using the `book-keeping' notation, we rewrite the Hamiltonian as
\begin{equation}
\label{Ham0_norm_1}
\Hscr^{(0)}=\Hscr_{sec}(w_2, w_3, W_2, W_3)=Z_0(  W_2, W_3)+\lambda Z_{0,1}( w_2-w_3, W_2, W_3)+\lambda \Hscr_{1,space}(  w_2,w_3, W_2, W_3)\, , 
\end{equation}
where $\Hscr_{1,space}(  w_2,w_3, W_2, W_3)$ contains trigonometric terms depending on linear combinations of the angles $k_1w_2+k_2w_3$ with $k_1+k_2\neq 0$. Using angles $(\psi, \varphi)$ (see Eq.~\eqref{trasf.can.int.plane}) this would have implied that all the terms contained in $\Hscr_{1,space}$ depend on the angle $\varphi$. We then define the Lie generating function $\chi$ as the solution of the following homological equation:
\begin{equation}
\label{homo_0}
\poisson{Z_0}{\chi} + \lambda \Hscr_{1, space}= 0\,.
\end{equation}
}
Thus, writing the Taylor-Fourier expansion of $\Hscr_{1,space}$ as
\begin{align*}
\Hscr_{1,space}(  \psi, \varphi, W_2, W_3)=\sum_{l, m, n, k} 
\theta_{l,m,n,k} W_2^l W_3^ m \e^{i n \psi}\e^{i k \varphi}=
\sum_{l, m, n, k} 
\theta_{l,m,n,k} W_2^l W_3^ m \e^{i\left( (n+k) w_2 - (n-k) w_3\right)}
\end{align*}
the solution of the homological equation~\eqref{homo_0} reads
$$
\chi(  \psi, \varphi, W_2, W_3)=\lambda\!\!\!\!\!\sum_{\substack{l, m, n, k\\k\neq 0 \bigwedge \\a(n+k)\neq b(n-k)} }
\!\!\!\!\!\frac{\theta_{l,m,n,k}}{i\left(a(n+k)-b(n-k)\right)} W_2^l W_3^ m \e^{i n \psi}\e^{i k \varphi}\,  .
$$
 {Note that the generating function $\chi $ can be defined for all systems with masses $m_j$ and semi-major axes $a_j$, $j=2,3$ such that we are not close to a \textit{secular resonance}, i.e., a commensurability of the form $a(n+k)=b(n-k)$. For any finite truncation of the Hamiltonian, $\Hscr_{1,space}$ contains harmonics of the form $\cos((n+k)w_2-(n-k)w_3)$ with a rather low order $|n-k|+|n+k|$, which is anyway limited from above, hence the measure of the parameters $(a_j,m_j)$ leading to secular (near-)resonance is small. }

Using $\chi$, the one-step normalized Hamiltonian is given by
\begin{align}\label{ham1lie}
\Hscr^{(1)}=\exp L_{\chi} \Hscr^{(0)}\, ,
\end{align}
where $\exp L_{\chi} \cdot=\sum_{j\geq 0} \left(L_{\chi}^{j}\cdot\right)/j!$ is the Lie series operator, $L_{\chi}\cdot=\poisson{\cdot}{\chi}$ is the Lie derivative with respect to the dynamical function $\chi$ and $\poisson{\cdot}{\cdot}$ denotes the Poisson bracket. Expanding Eq.~\eqref{ham1lie} in powers of the book-keeping symbol $\lambda$, we find 
\begin{equation}\label{ham1exp}
\begin{aligned}
\Hscr^{(1)}&=\underbrace{Z_0+\overbracket{\lambda Z_{0,1}}^{\Oscr(\lambda)}}_{\Hscr_{int}} + \underbrace{\overbracket{\lambda \Hscr_{1,space}}^{\Oscr(\lambda)} + \overbracket{\poisson{Z_0}{\chi }}^{\Oscr(\lambda)}}_{=\,0}+ \overbracket{\lambda\poisson{Z_{0,1}}{\chi}}^{\Oscr(\lambda^2)} + \underbrace{\overbracket{\lambda\poisson{\Hscr_{1,space}}{\chi}}^{\Oscr(\lambda^2)} + \frac{1}{2}\overbracket{\poisson{\poisson{Z_0}{\chi}}{\chi}}^{\Oscr(\lambda^2)}}_{=\, \frac{\lambda}{2}\poisson{\Hscr_{1,space}}{\chi}}+\Oscr(\lambda^3)\\
&= \Hscr_{int}+\left\langle\frac{\lambda}{2}\poisson{\Hscr_{1,space}}{\chi}\right\rangle_{\varphi}+\Hscr^{(1)}_{rest}+\Oscr(\lambda^3)\, ,
\end{aligned}
\end{equation} 
where
\begin{align}
\label{ham1rest}
\Hscr^{(1)}_{rest}= \lambda\poisson{Z_{0,1}}{\chi} +
\frac{\lambda}{2}\poisson{\Hscr_{1,space}}{\chi}
-
\left\langle\frac{\lambda}{2}\poisson{\Hscr_{1,space}}{\chi}\right\rangle_{\varphi}\, .
\end{align}
The average $\langle\cdot\rangle_\varphi$ means all terms independent of the angles or trigonometric terms depending only on the difference $\psi=w_2-w_3$. Finally, we set
\begin{align}
\label{Ham_integ_norm}
\Hscr^{(1)}_{int}=\Hscr_{int}+\Bigg[\left\langle\frac{\lambda}{2}\poisson{\Hscr_{1,space}}{\chi}\right\rangle_{\varphi}\Bigg]_{\leq N_{bk}}~~,
\end{align}
where the operation $[\cdot]_{N_{bk}}$ means to truncate the enclosed expression at the maximum adopted order of expansion in the eccentricities and inclinations ($N_{bk}=12$ in our examples below). The integrable Hamiltonian model $\Hscr^{(1)}_{int}$ is hereafter referred to as the first order \textit{secular normal form} (or simply the `secular normal form'). 

Phase portraits under the secular normal form $\Hscr_{int}^{(1)}$ can be obtained using the change of variables~\eqref{trasf.can.int.plane} and~\eqref{Poincare1} and computing Poincar\'e surfaces of section $\mathcal{PS}_{\Hscr_{int}^{(1)}}(\Escr;\AMD)$ at a fixed level of energy $\Escr$ through
\begin{equation}\label{poinchint1}
\begin{aligned}
\mathcal{PS}_{\Hscr_{int}^{(1)}}(\Escr;\AMD)=
\bigg\{ &(X_2,Y_2,X_3,Y_3)\in\mathbb{R}^4:~ \Hscr^{(1)}_{int}(X_2,Y_2,X_3, Y_3=0;\AMD)=\Escr\, ,\,Y_3=0\, ,\\
&\dot{Y}_3=-{\partial\Hscr^{(1)}_{int}\over\partial X_3}\big|_{Y_3=0}\geq 0\, ,
\cos(\i_{max})\leq\cos(\i_{mut})(X_2,Y_2,X_3,Y_3=0;\AMD)\leq 1\bigg\}\, .
\end{aligned}
\end{equation}

\begin{figure}[!h]
\begin{center}
\frame{\includegraphics[width=1\textwidth]{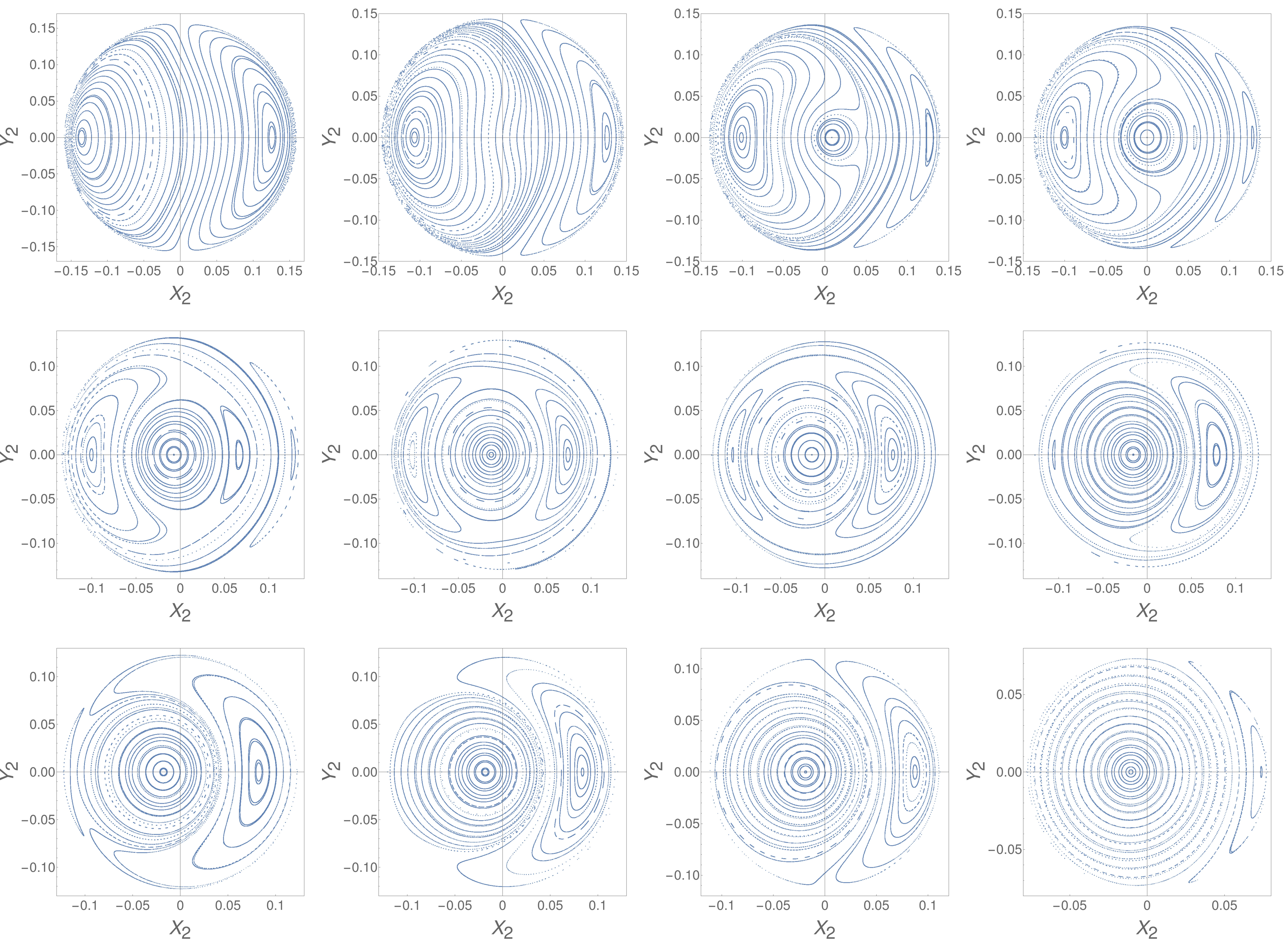}}
\caption{Poincar\'{e} surfaces of section $\mathcal{PS}_{\Hscr_{int}^{(1)}}(\Escr;\AMD)$ in the plane $(X_2, Y_2)$ with $\AMD$ fixed.  The surfaces of section are computed by a numerical integration of trajectories of the normalized integrable Hamiltonian ${\Hscr}^{(1)}_{int}$ (Eq.~\eqref{Ham_integ_norm}) truncated at multipolar degree $N_{P}=5$, order $N_{bk}=12$ in the eccentricities, and energies (from top to bottom, from left to right) $\Escr= -5.82\cdot 10^{-5},-4.3\cdot 10^{-5},-3.59\cdot 10^{-5},-3.34\cdot 10^{-5}, -3.03\cdot 10^{-5}, -2.68\cdot 10^{-5}$,$-2.51\cdot 10^{-5}, -2.42\cdot 10^{-5}, -2.12\cdot 10^{-5}, -1.99\cdot 10^{-5}, -1.44\cdot 10^{-5}, -3.3\cdot 10^{-6}$. In this case the energies for which we can have co-planar orbits range between $\Escr_{min}= -1.71\cdot 10^{-4}$ and $\Escr_{2,3}=-5.82\cdot 10^{-5}\,$. In the plot we consider the phase-portraits for energies $\Escr\geq\Escr_{2,3}$, for which no-coplanar orbits exist.}
\label{Fig.npol5eps12INTEG_NORMAL}
\end{center}
\end{figure}
Figure \ref{Fig.npol5eps12INTEG_NORMAL} shows the surfaces of section for different (increasing, from left to right, from top to bottom) levels of energy. By visual inspection, the sequence of bifurcations produced by the secular normal form $\Hscr_{int}^{(1)}$ differ considerably from those of the integrable Hamiltonian $\Hscr_{int}$ before the normalization (top row of Fig.~\ref{Fig.npol5eps12INTEGsigma0}). The origin of the difference becomes clear by the analysis through Hopf variables as discussed below. On the other hand, comparison with the phase portraits of the full secular Hamiltonian $\Hscr_{sec}$ (Fig.~\ref{Fig.npol5eps12}) shows that the secular normal form reproduces qualitatively the same sequence of bifurcations as in the complete model, except for the last bifurcation (transition from the fourth to the fifth panel in Fig.~\ref{Fig.npol5eps12}), connected to the birth of the Lidov-Kozai periodic orbits. Such orbits are associated with the $\e_2^2\cos(2w_2)$ or $\e_3^2\cos(2w_3)$ terms in the secular Hamiltonian, which are not present in the secular normal form. 

We now apply the geometric method of section \ref{sec:Bif} to the secular normal form $\Hscr_{int}^{(1)}$ {, given by Eq.~\eqref{Ham_integ_norm},} to recover the sequence of bifurcations observed in the regime of energies $\Escr_C\leq \Escr \leq \Escr_{C,2}$, where $\Escr_C$ is the energy of the saddle-node bifurcation of the orbits $P_1$ and $P_2$, and $\Escr_{C,2}$ is the energy of bifurcation of the Lidov-Kozai orbits  {(see Section~\ref{subsec:Poinc})}. Similarly to Fig.~\ref{Fig.Poin_Hopf_INTEG}, in the case of the secular normal form Fig.~\ref{Fig.Poin_Hopf_INTEG_NORMAL} yields the phase portraits of the secular normal form expressed in Hopf variables $\Hscr_{int}^{(1)}(\sigma_1, \sigma_3; \sigma_0)$. Each upper panel in Fig.~\ref{Fig.Poin_Hopf_INTEG_NORMAL} corresponds to a fixed value of $\sigma_0$ decreasing from left to right and from top to bottom, and the invariant curves in the same portrait correspond to different values of the energy $\Escr=\Hscr_{int}^{(1)}$ within the allowed limits computed in the same way as before for the Hamiltonian $\Hscr_{int}$  {(i.e., as already done in Fig.~\ref{Fig.Evssigma0} for $\Hscr_{int}$, it is sufficient to fix a value of $\sigma_0$ and find the limits on the energies trough the computations of the tangencies between Eq.~\eqref{sphere} and Eq.~\eqref{energy.curve}, with $\mathcal{Z}=\Hscr_{int}^{(1)}$)}. The lower panel below each phase portrait shows the intersections or tangencies of various curves $\mathcal{C}_{\sigma_0,\, \Escr}^{(\sigma_2=0)}$ with the circle $\mathcal{S}_{\sigma_0}^{(\sigma_2=0)}$ for the corresponding value of $\sigma_0$. 

Comparing Figs.~\ref{Fig.Poin_Hopf_INTEG} and \ref{Fig.Poin_Hopf_INTEG_NORMAL}, we immediately notice that, for the same model parameters, the curves $\mathcal{S}_{\sigma_0}^{(\sigma_2=0)}$ are hyperbola-like in the case of the secular normal form, instead of ellipse-like, as in the Hamiltonian $\Hscr_{int}$.

 {Apart from constants, the secular normal form reads}
\begin{equation}
\label{Ham.integ.norm}
\Hscr_{int}^{(1)} = A^{(1)}\sigma_1^2+C^{(1)}\sigma_3^2+B^{(1)}\sigma_1\sigma_3+D^{(1)}(\sigma_0)\sigma_1+E^{(1)}(\sigma_0)\sigma_3+F^{(1)}(\sigma_0)+
\Oscr\left((\sigma_i)^2\right)\, , 
\end{equation}
with $i=0,\,\dots,3$.
 {The numerical value of the coefficients are given in Appendix~\ref{appendix:num_val_H_int_norm}. For those values of the coefficients $A^{(1)}, B^{(1)}, C^{(1)}$, the condition $\left(B^{(1)}\right)^2>4A^{(1)} C^{(1)}\,$ is satisfied},
implying that the quadratic form $A^{(1)}\sigma_1^2+C^{(1)}\sigma_3^2+B^{(1)}\sigma_1\sigma_3$ yields now a family of hyperbolas instead of ellipses. This difference brings essential differences in the structure of the phase portraits of the secular normal form $\Hscr_{int}^{(1)}$ compared to the integrable model $\Hscr_{int}$.  
\begin{figure}
\centering
\includegraphics[width=0.95\textwidth]{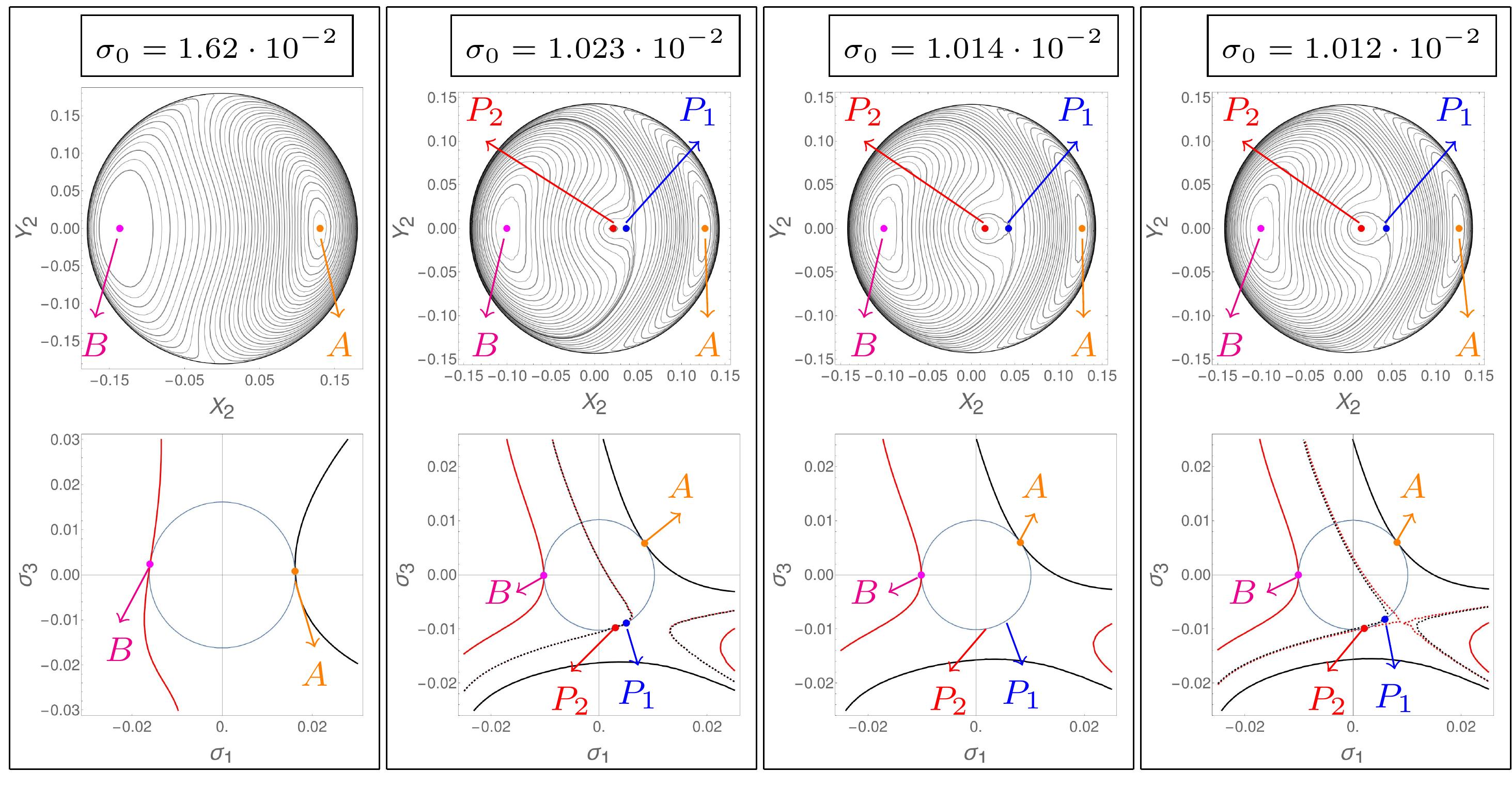}
\includegraphics[width=0.95\textwidth]{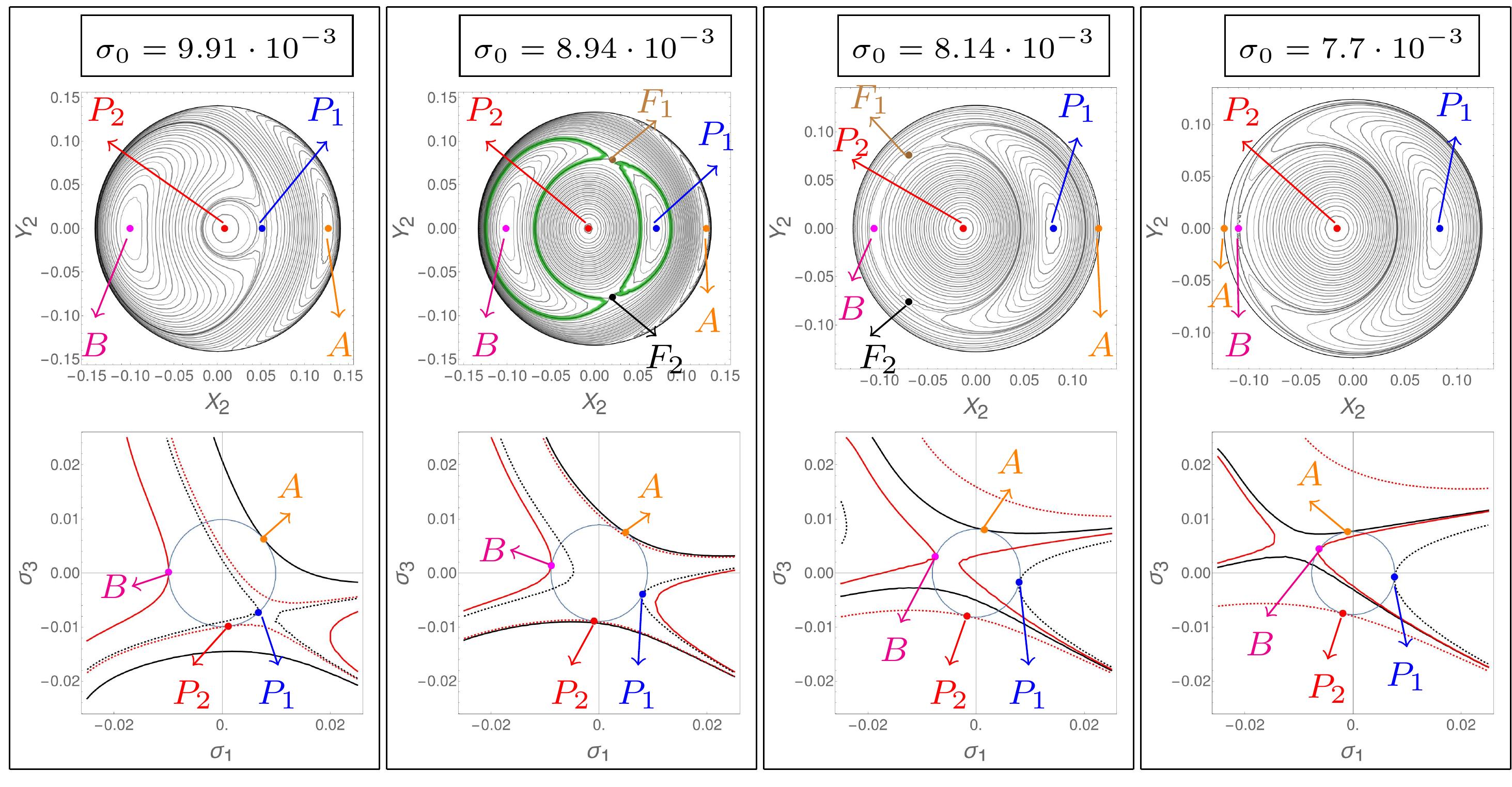}
\caption{Upper panels: the contour plots $\Hscr^{(1)}_{int}(X_2,Y_2;\sigma_0)=\Escr$  {(where $\Hscr^{(1)}_{int}$ written in Eq.~\eqref{Ham_integ_norm})}, with $Y_3=0$, $\dot{Y_3}\geq 0$ for decreasing values of $\sigma_0$ from top to bottom, left to right. Lower panels: intersections of and tangencies between the circles  $\mathcal{S}_{\sigma_0}^{(\sigma_2=0)}$ (blue) and of the curves  $\mathcal{C}_{\sigma_0,\, \Escr}^{(\sigma_2=0)}$ in the plane $(\sigma_1,\sigma_3)$, for different values of the energy $\Escr$ and $\sigma_0$ fixed as in the upper frame corresponding to each lower frame. The points of tangency corresponding to  {CPI}-type fixed points of the secular normal form are calculated by Eq.~\eqref{CP1}. The curves  $\mathcal{C}_{\sigma_0,\, \Escr}^{(\sigma_2=0)}$ passing through the tangency points are hyperbola-like (compare with the corresponding curves in Fig.~\ref{Fig.Poin_Hopf_INTEG}, which are ellipse-like).}
\label{Fig.Poin_Hopf_INTEG_NORMAL}
\end{figure}
\begin{figure}
\ContinuedFloat
\centering
\includegraphics[width=0.95\textwidth]{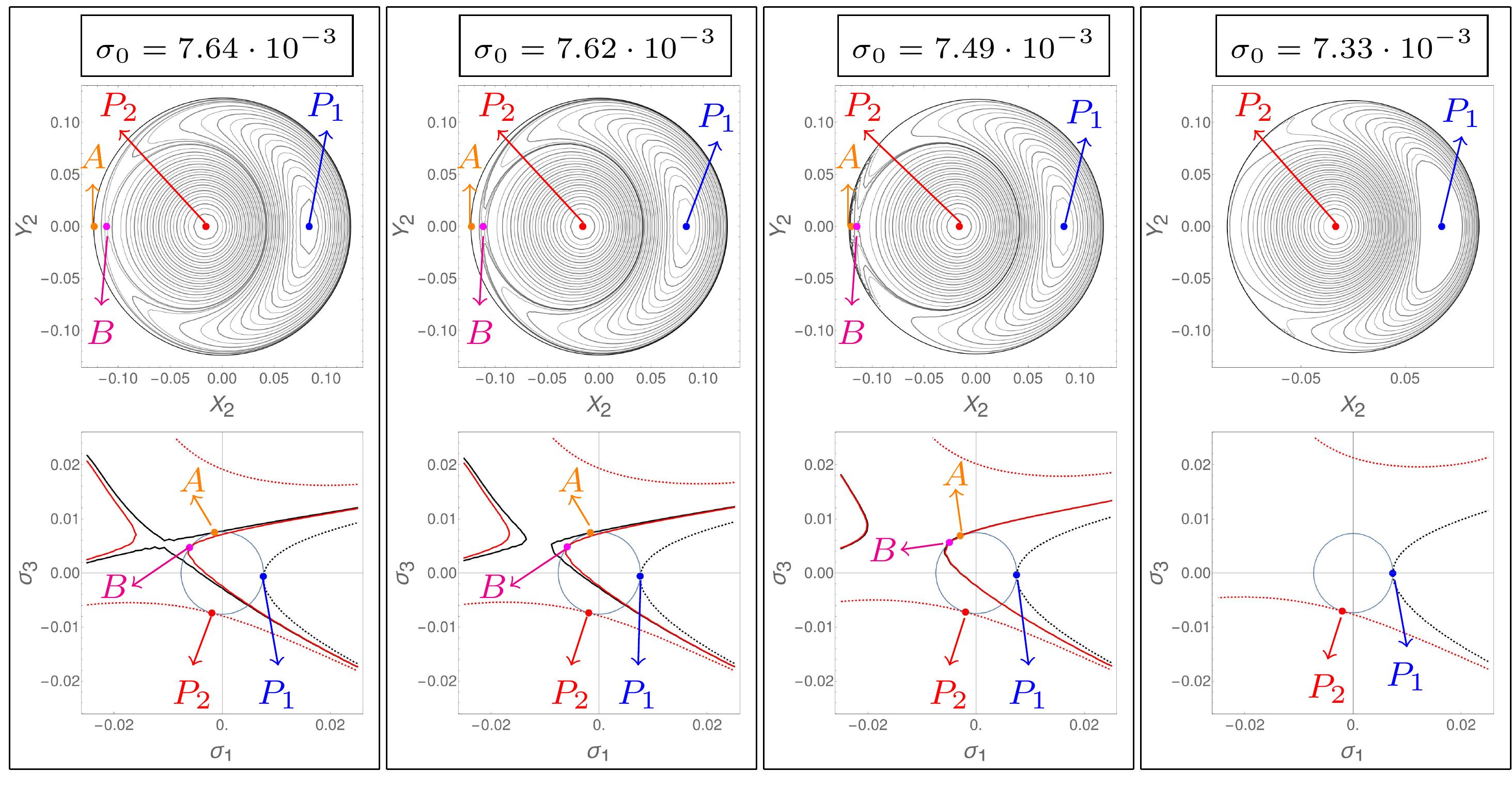}
\caption{Continued.}
\end{figure}
\begin{figure}[!h]
\begin{center}
\includegraphics[width=1.\textwidth]{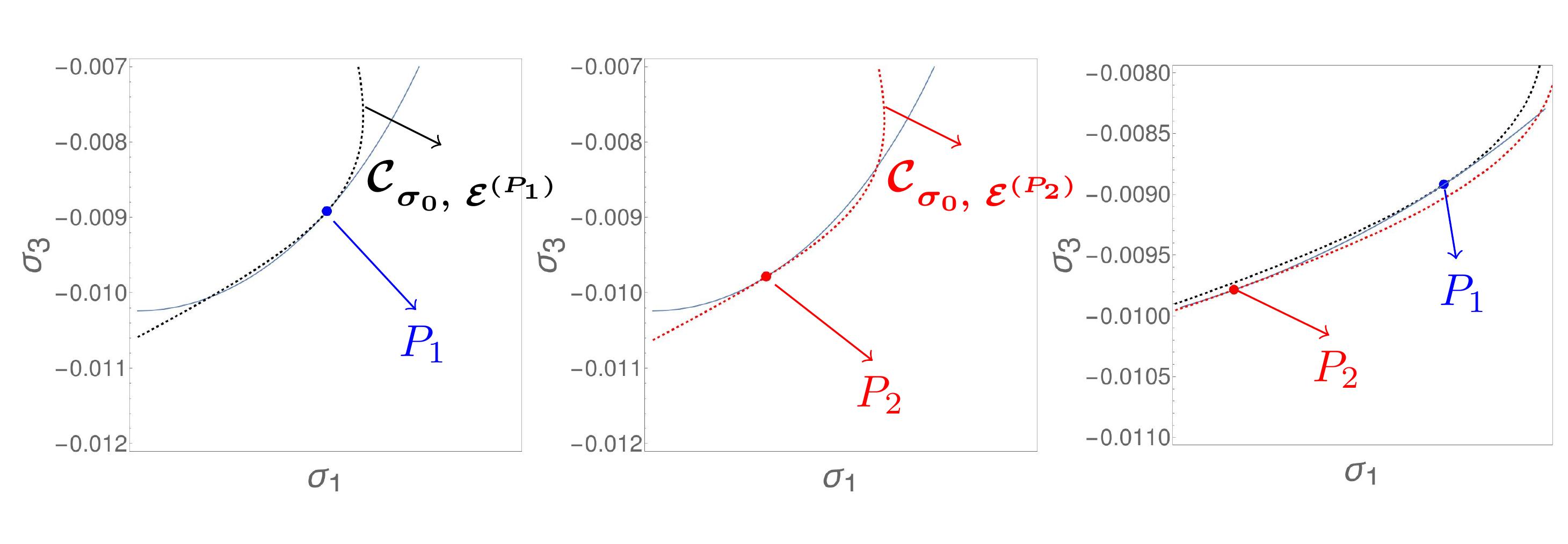}
\caption{Zoom of the second plot of the second row of Fig.~\ref{Fig.Poin_Hopf_INTEG_NORMAL} (i.e. of the intersection between $\mathcal{S}_{\sigma_0}^{(\sigma_2=0)}$ and $\mathcal{C}_{\sigma_0,\, \Escr}^{(\sigma_2=0)}$ in the plane $(\sigma_1,\sigma_3)$ for $\sigma_0= 1.023\cdot 10^{-2}$). The blue and red points are, respectively, unstable and stable since they are given by an inner and outer tangency between $\mathcal{S}_{\sigma_0}^{(\sigma_2=0)}$ and $\mathcal{C}_{\sigma_0,\, \Escr}^{(\sigma_2=0)}$. The left panel shows the inner tangency between $\mathcal{S}_{\sigma_0}^{(\sigma_2=0)}$ and $\mathcal{C}_{\sigma_0,\, \Escr^{(P_1)}}^{(\sigma_2=0)}$ in $P_1$.  {The middle panel} shows the outer tangency between $\mathcal{S}_{\sigma_0}^{(\sigma_2=0)}$ and $\mathcal{C}_{\sigma_0,\, \Escr^{(P_2)}}^{(\sigma_2=0)}$ in $P_2$ (in our example $\Escr^{(P_1)}= -3.78701$ and $\Escr^{(P_2)}= -3.78861$).  {The right panel shows a zoom of the same plots in the domain around the points $P_1$ and $P_2$. Since for this value of $\sigma_0$ the saddle-node bifurcation has already occurred, the two hyperbolas are very close to each other but different, giving rise to two different tangencies at the points $P_1$ and $P_2$.}}
\label{Fig.zoom_Hopf_INTEG_NORMAL}
\end{center}
\end{figure}

\begin{figure}[!h]
\begin{center}
\includegraphics[scale=1.]{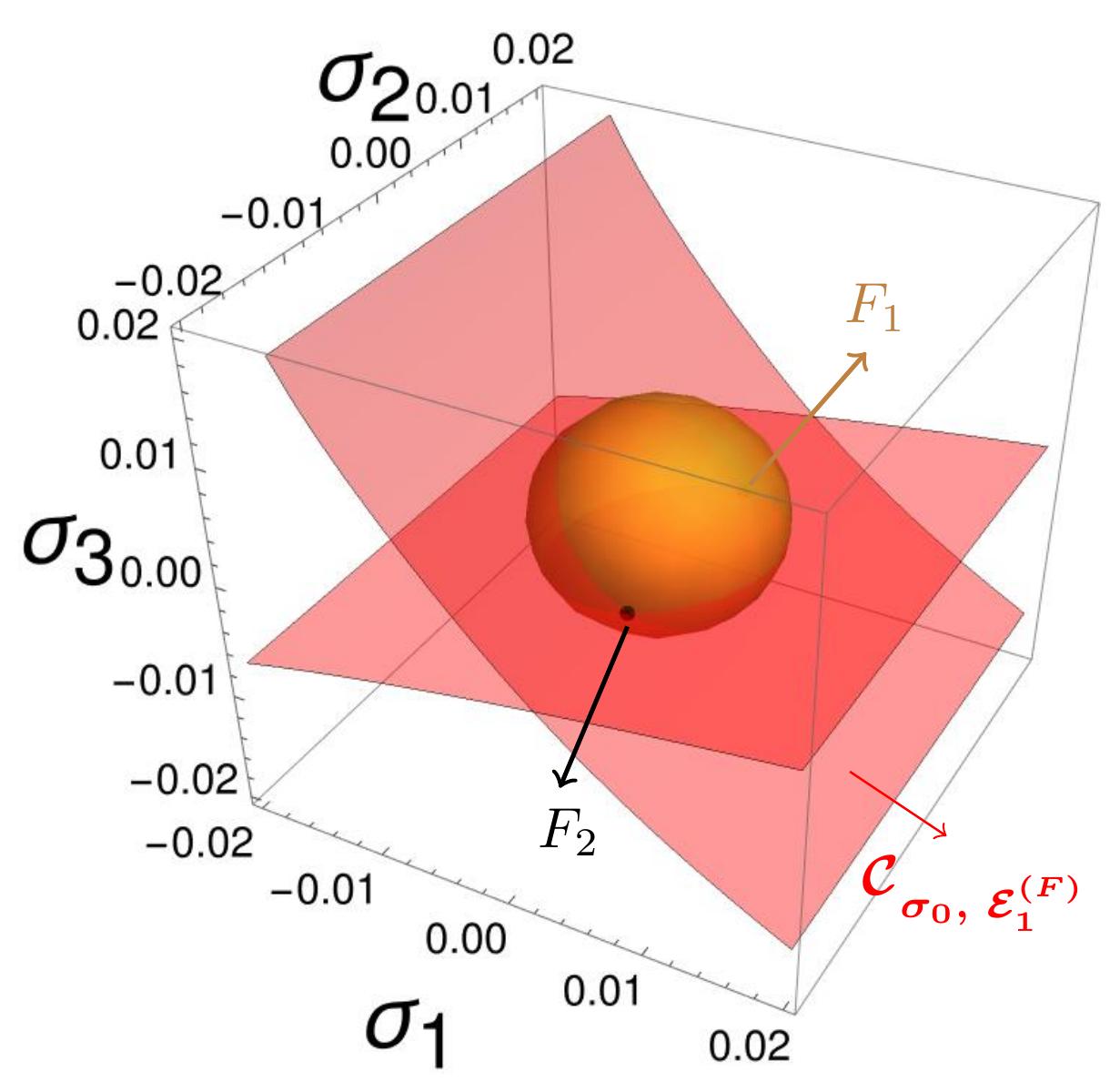}
\includegraphics[scale=1.]{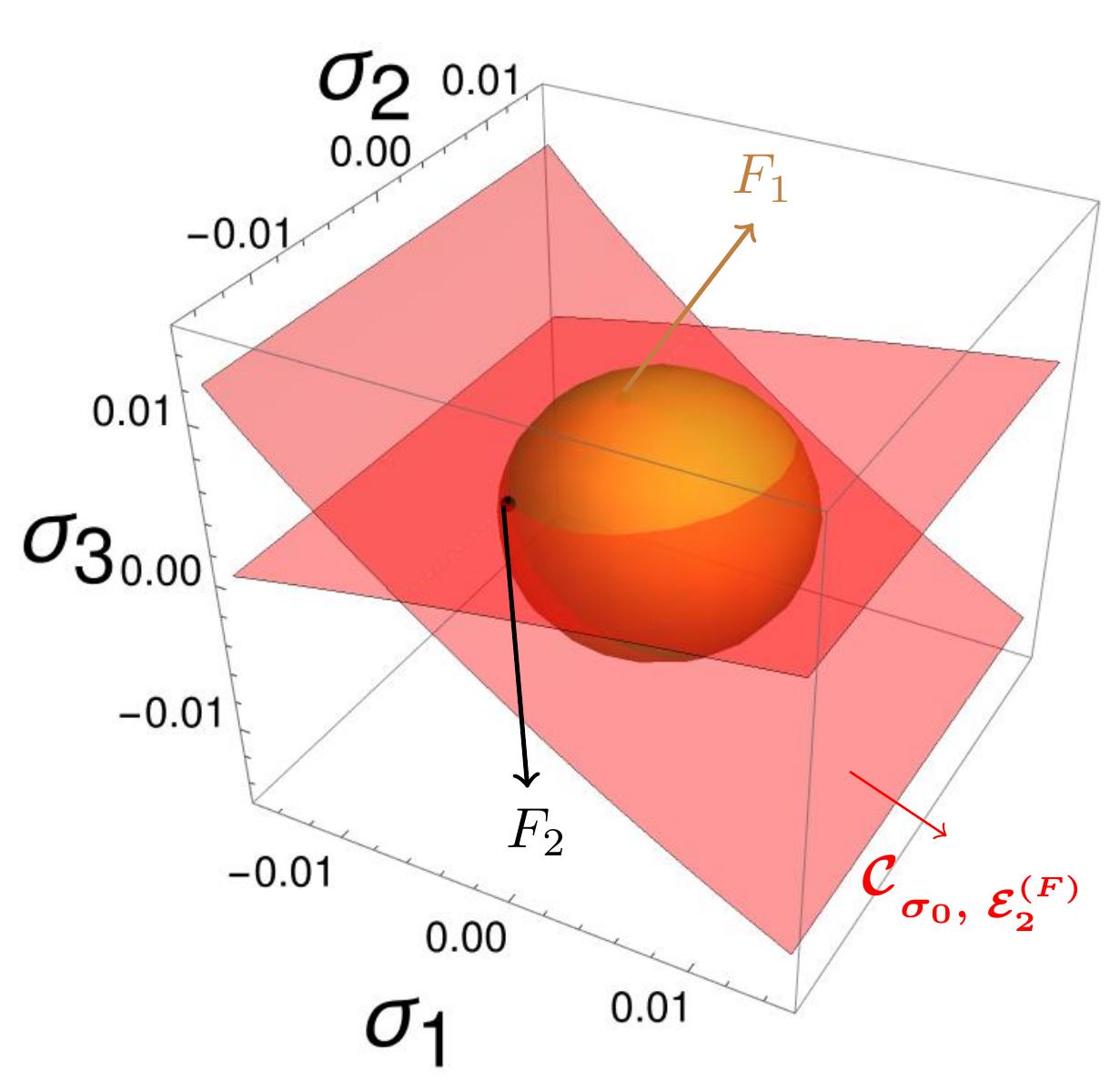}\\
\includegraphics[scale=1.]{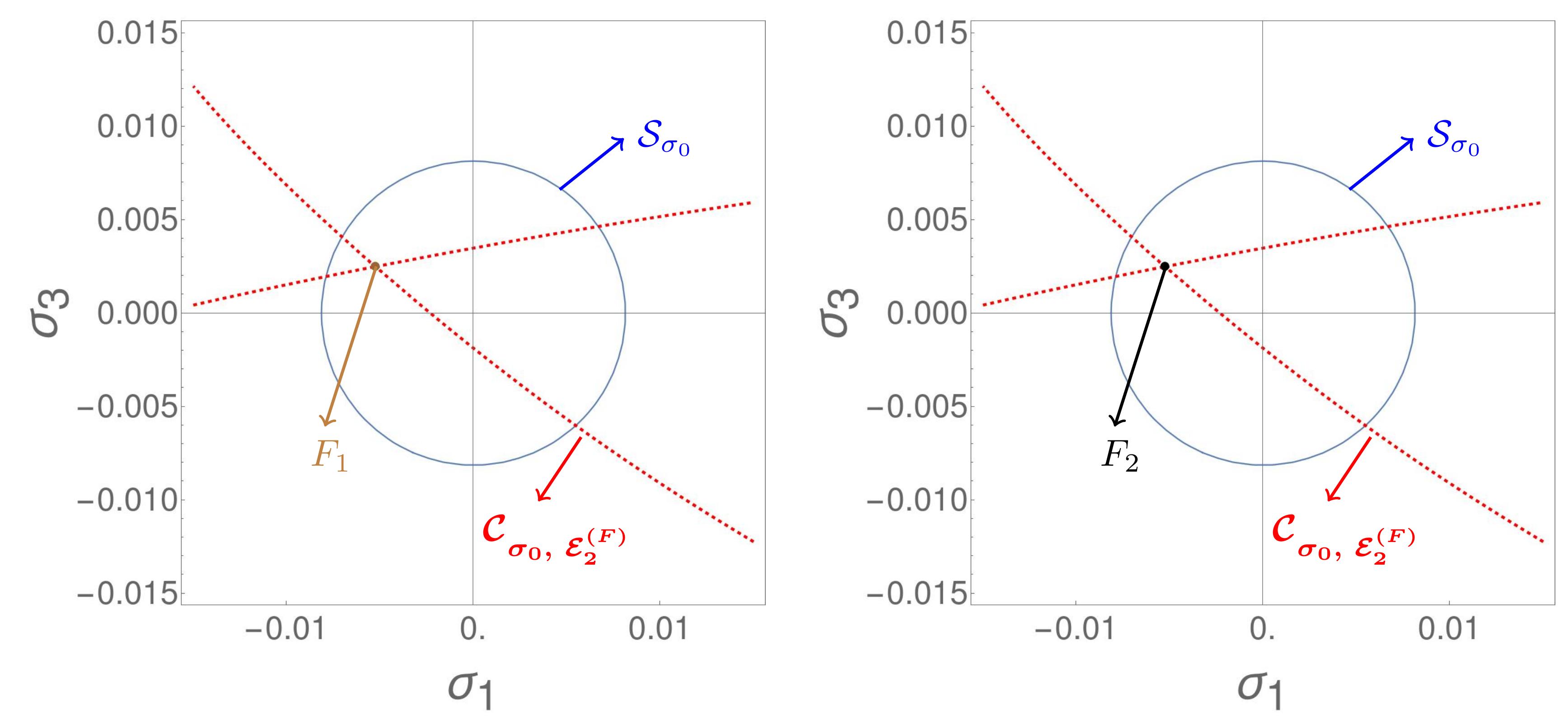}
\end{center}
\caption{Graphical representation of the critical points of second kind $F_1$ and $F_2$.  {Top} left: 3D representation of the sphere $\mathcal{S}_{\sigma_0}$ and the surface $\mathcal{C}_{\sigma_0,\, \Escr_1^{(F)}}$, in the case $\sigma_0=8.94\cdot 10^{-3}$, where $\Escr_1^{(F)}$ represents the energy of the fixed points $F_1$ and $F_2$. We can observe that the surface of the energy pierces the sphere in the $F$-modes.  {Top right}: As before, but in the case $\sigma_0=8.14\cdot 10^{-3}\,$. For a value of the energy, namely $\Escr_2^{(F)}$, $\Cscr_{\sigma_0,\Escr_2^{(F)}}$ pierces the sphere $\Sscr_{\sigma_0}$ in the points $F_1$ and $F_2$ of coordinates, respectively, $(\sigma_1^{(F)},\sigma_2^{(F)},\sigma_3^{(F)})$ and $(\sigma_1^{(F)},-\sigma_2^{(F)},\sigma_3^{(F)})$.  {Bottom}: The projection of the 3D  {top right} plot (case $\sigma_0=8.14\cdot 10^{-3}\,$) in the planes $\sigma_2=0$. }
\label{Fig.iperboli_INTEG_NORMAL_3D2D}
\end{figure}

The sequence of bifurcations found in Fig.~\ref{Fig.Poin_Hopf_INTEG_NORMAL} can be summarized as follows: The top-left frame shows the starting phase portrait of the (near-)planar regime with the usual tangency points A and B associated with anti-aligned and aligned ACR orbits. Decreasing the value of $\sigma_0$ (second frame of Fig.~\ref{Fig.Poin_Hopf_INTEG_NORMAL}, for $\sigma_0=1.023\cdot 10^{-2}$) gives rise to a similar saddle-node bifurcation as in Fig.~\ref{Fig.Poin_Hopf_INTEG}, with two new fixed points of the Poincaré map corresponding to periodic orbits $P_1$ and $P_2$. However, due to the fact that the curves $\mathcal{C}_{\sigma_0,\, \Escr}^{(\sigma_2=0)}$ are now hyperbola-like, we note that the stability type of each of the orbits $P_1$ or $P_2$ is inverted with respect to Fig.~\ref{Fig.Poin_Hopf_INTEG}. Thus, immediately after the bifurcation the tangent points belong to two nearby hyperbola-like curves, of which the curve containing the point $P_1$ yields an inner tangency (hence $P_1$ is unstable), while the curve containing the point $P_2$ yields an outer tangency (hence $P_2$ is unstable) (see a zoom in Fig.~\ref{Fig.zoom_Hopf_INTEG_NORMAL}). Decreasing further the value of $\sigma_0$, we arrive at a critical value ($\sigma_0\simeq 1.012\cdot 10^{-2}$, see fourth panel in Fig.~\ref{Fig.Poin_Hopf_INTEG_NORMAL}) where two independent hyperbola-like branches of the curve $\mathcal{C}_{\sigma_0,\, \Escr^{(P_2)}}^{(\sigma_2=0)}$ join each other into a saddle point $F$. This point is outside the disc limited by the curve $\mathcal{S}_{\sigma_0}^{(\sigma_2=0)}$, hence it represents no possible new fixed point of the flows. Decreasing, however, $\sigma_0$ still further, the saddle point $F$ moves towards the circle $\mathcal{S}_{\sigma_0}^{(\sigma_2=0)}$, and crosses it at a critical value of $\sigma_0$ which lies is in the interval $8.94\cdot 10^{-3}<\sigma_0<9.91\cdot 10^{-3}$ (see the transition between the fifth and sixth panel of Fig.~\ref{Fig.Poin_Hopf_INTEG_NORMAL}). At the crossing value we then have an inverse pitchfork bifurcation, whereby the point $P_1$ turns from unstable to stable, while for values of $\sigma_0$ below the critical one the saddle moves within the disc, giving rise to two unstable fixed points $F_1$ and $F_2$ of the  {CPII}-type. For still smaller values of $\sigma_0$, the saddle point $F$ moves further inside the disc towards the left. Figure \ref{Fig.iperboli_INTEG_NORMAL_3D2D} gives the 3D picture of the intersection of the surfaces $\mathcal{C}_{\sigma_0,\, \Escr^{(F)}}$ with the sphere 
$\mathcal{S}_{\sigma_0}$ for $\sigma_0=8.94\cdot 10^{-3}$ and for $\sigma_0=8.14\cdot 10^{-3}$, corresponding to the sixth and seven panels in Fig.~\ref{Fig.Poin_Hopf_INTEG_NORMAL}. Reducing $\sigma_0$ further, we then arrive at a second critical value lying in the interval $7.7\cdot 10^{-3}<\sigma_0<8.14\cdot 10^{-3}$ (see the transition between the seventh and eighth panel of Fig.~\ref{Fig.Poin_Hopf_INTEG_NORMAL}), where the saddle point $F$ crosses for a second time the circle $\mathcal{S}_{\sigma_0}^{(\sigma_2=0)}$, passing now from its interior to its exterior. At the second crossing we then have a second inverse pitchfork bifurcation turning the point B from stable to unstable. From this point on, decreasing $\sigma_0$ brings the saddle point $F$ outside the disc, hence no longer yielding feasible fixed points of the system. In the same time, the ACR points B and A are given by an inner and outer tangency, hence they are unstable and stable respectively. Finally, reducing further $\sigma_0$ brings the separatrices of the fixed point B very close to the tangency point A. Hence, the area occupied by the island of stability around the fixed point A is necessarily reduced, while the areas of the domains of stability around the fixed points $P_1$, and $P_2$ become large. As a consequence, the points $P_1$ and $P_2$ and their neighborhoods give now the dominant structures in the phase space (see 9th and 10th panel in Fig.~\ref{Fig.Poin_Hopf_INTEG_NORMAL}), while the influence of the ACR orbits A, B is reduced. Then, at a third critical value of $\sigma_0$ in the interval $7.33\cdot 10^{-3}<\sigma_0<7.49\cdot 10^{-3}$ the points A, B, as well as the hyperbolas which contain them, collide. Then, the points A, B disappear altogether through an inverse saddle-node bifurcation (see the transition from the 11th to the 12th (last) panel in Fig.~\ref{Fig.Poin_Hopf_INTEG_NORMAL}). From that point on, the only stable points surviving in the end of the sequence of bifurcations are $P_1$ and $P_2$.  \\
 {As an overall comment, the sequence of bifurcations produced by the integrable Hamiltonian $\Hscr_{int}$ differ considerably from those of the full secular Hamiltonian, the difference becoming more important as we approach towards the `Lidov-Kozai' regime of dynamics. In particular, the full secular Hamiltonian $\Hscr_{sec}$ contains a set of  harmonics $\cos(k_2w_2 +k_3w_3)$ with $k_2 + k_3 \neq 0$, which are not present in the integrable approximation $\Hscr_{int}$. Of the missing harmonics, the most important is $\cos(2w_2)$ ($k_2=2$, $k_3=0$), which is the prevalent harmonic at the Lidov-Kozai regime.}


\subsection{Secular normal form: octupole approximation $\tilde{\Hscr}_{int}^{(N_P=3,N_{bk}=4)}$}
\label{subsec:NormOct}
Of all multipolar truncations of the secular Hamiltonian, the first non-integrable one is the \textit{octupolar} approximation, corresponding to terms up to degree $N_{P}=3$ in the semi-major axis ratio $a_2/a_3$.  {More precisely, starting from the octupole approximation of the Hamiltonian, namely $\Hscr^{(N_P=3, N_{bk})}$ and decomposing it similarly as in Eq.~\eqref{hamdecompo}, we can apply the normal form described in Section~\ref{subsec:Norm_Hint1}, and obtain an integrable approximation $\widetilde{\Hscr}_{int}^{(N_P=3, N_{bk})}$ admitting $\sigma_0$ as a second integral. Moreover,} in this case, assuming an expansion up to order $N_{bk}=4$ in the eccentricities leads to the possibility of computing a secular normal form as in subsection \ref{subsec:Norm_Hint1}, whose expression is an exact quadratic form in the Hopf variables as in Eq.~\eqref{Ham.K_prel}.  {In turn, this allows to apply the analytical formul{\ae} reported in Section~\ref{subsec:formule_Anto_Gius}.} Explicit expressions for the normal form coefficients can be computed, leading to  {a Hamiltonian of the same form of Eq.~\eqref{Ham.K_prel}. More precisely}: 
\begin{equation}
\label{Ham_inter_norm_otto_C}
\widetilde{\Hscr}_{int}^{(N_P=3, N_{bk}=4)}
= {\t{A}}\sigma_1^2+ {\t{C}}\sigma_3^2+ {\t{B}}\sigma_1\sigma_3+ {\t{D}}(\sigma_0)\sigma_1+ {\t{E}}(\sigma_0)\sigma_3+ {\t{F}}(\sigma_0)\, ,
\end{equation}
where the coefficients (apart from  {additive} constants) are:
\begin{align*}
  {\t{A}} &=0\, ,\\
  {\t{C}} &= -\frac{2025 a_2^{9/2} \AMD \left(\sqrt{a_2} m_2+\sqrt{a_3} m_3\right)^2}{256 a_3^{19/2} {m_2} {m_3} {m_0}^2 (a+b)}+\frac{225 {a_2}^2 \AMD}{64 {a_3}^6 {m_0}^2 a} \left(\sqrt{\frac{a_2}{a_3}}\frac{m_3}{m_2}+\frac{{a_2}}{2 {a_3}}+\frac{{m_3}^2}{2 {m_2}^2}\right)
  \\
 &\phantom{=}
 +\frac{3a_2}{8 a_3^3 m_0 }\left(\frac{3}{2}\sqrt{\frac{a_2}{a_3}}- \frac{a_2 m_2 }{ a_3 m_3}+ \frac{m_3}{4 m_2}\right)\, , \\
  {\t{B}} & =  \frac{675 {a_2}^{13/4} \AMD \left(\sqrt{{a_2}} {m_2}+\sqrt{{a_3}} {m_3}\right)^2}{256  {a_3}^{33/4}{m_0}^2  {m_2}^{3/2} \sqrt{{m_3}}(a+b)}\left(\frac{{b}}{{a}}+3\right)+\frac{5 {a_2}^{9/4} }{128 {a_3}^{17/4} {m_0}}\left(57\sqrt{ \frac{{a_2 m_2}}{a_3 m_3} }-15\sqrt{ \frac{{m_3}}{m_2}}\right)\, ,\\
  {\t{D}}_1 & = \frac{ 2025 {a_2}^{13/4} \AMD \left(\sqrt{{a_2}} {m_2}+\sqrt{{a_3}} {m_3}\right)^2}{256 {a_3}^{33/4} {m_0}^2 {m_2}^{3/2} \sqrt{{m_3}} (a+b)}\left(\frac{b}{a}+3\right)-\frac{105 {a_2}^{9/4} }{128 {a_3}^{17/4} {m_0}}\left(9\sqrt{ \frac{{a_2 m_2}}{a_3 m_3} }+7\sqrt{ \frac{{m_3}}{m_2}}\right)\, ,\\
  {\t{\Delta}}_1 & = -\frac{ 675 {a_2}^{13/4} \AMD^2 \left(\sqrt{{a_2}} {m_2}+\sqrt{{a_3}} {m_3}\right)^2}{256 {a_3}^{33/4} {m_0}^2 {m_2}^{3/2} \sqrt{{m_3}} (a+b)}\left(\frac{b}{a}+3\right)+\frac{165 {a_2}^{9/4} \AMD \left(\sqrt{{a_2}} {m_2}+\sqrt{{a_3}} {m_3}\right)}{32 {a_3}^{19/4} {m_0} \sqrt{{m_2}} \sqrt{{m_3}}}\\
 &\phantom{=}-\frac{15 {a_2}^{11/4} \sqrt{\Gscr} \sqrt{{m_2}} \sqrt{{m_3}}}{16 {a_3}^{17/4} \sqrt{{m_0}}}\, , \\
  {\t{D}}_3 & = \frac{675 {a_2}^2 \AMD }{32 {a_3}^6 {m_0}^2 a}\left(\sqrt{\frac{a_2}{a_3}}\frac{ {m_3}}{ {m_2}}+\frac{{a_2}}{2 {a_3}}+\frac{{m_3}^2}{2 {m_2}^2}\right)+\frac{3 {a_2} }{4 {a_3}^3 {m_0}}\left(\frac{3 {a_2} {m_2}}{{a_3} {m_3}}-\frac{7 {m_3}}{4 {m_2}}\right)\, , \\
  {\t{\Delta}}_3 & = \frac{3 {a_2}^{3/2} \sqrt{\Gscr} }{8 {a_3}^{7/2} \sqrt{{m_0}}}\left(\sqrt{{a_2}} {m_2}-\sqrt{{a_3}} {m_3}\right)-\frac{225 {a_2}^2 \AMD^2 }{32 {a_3}^6 {m_0}^2 a }\left(\sqrt{\frac{a_2}{a_3}}\frac{ {m_3}}{ {m_2}}+\frac{{a_2}}{2 a_3}+\frac{{m_3}^2}{2 {m_2}^2 }\right)\\
 &\phantom{=}+\frac{3 {a_2} \AMD }{2 {a_3}^3 {m_0}}\left(\frac{m_3}{{m_2}}-\frac{{a_2} {m_2}}{{a_3} {m_3}}\right)\, ,
\end{align*}
 {where $\t{D}(\sigma_0)=\t{D}_1 \sigma_0 + \t{\Delta}_1$ and $\t{E}(\sigma_0)=\t{D}_3 \sigma_0 + \t{\Delta}_3$\, .}

The constants $a$ and $b$ correspond to secular frequencies (see Eq.~\eqref{Z0_1}) and they are given by
\begin{align*}
a &= -\frac{3 \sqrt{\Gscr} }{4 {a_3}^{7/2} \sqrt{{m_0}}}\left(2 {a_2}^{3/2} \sqrt{{a_3}} {m_3}+{a_2}^2 {m_2}\right)+\frac{3 {a_2} \AMD }{4 {a_3}^4 {m_0} {m_2} {m_3}}\left(6 \sqrt{{a_2 a_3}} {m_2} {m_3}+{a_2} {m_2}^2+5 {a_3} {m_3}^2\right)\, ,\\
b & = -\frac{3 \sqrt{\Gscr} }{4 {a_3}^{7/2} \sqrt{{m_0}}}\left( {a_2}^{3/2} \sqrt{{a_3}} {m_3}+2{a_2}^2 {m_2}\right)+\frac{3 {a_2} \AMD }{4 {a_3}^4 {m_0} {m_2} {m_3}}\left(6 \sqrt{{a_2 a_3}} {m_2} {m_3}+5{a_2} {m_2}^2+ {a_3} {m_3}^2\right) \, .
\end{align*}

In our numerical example,  {after dropping the additive constants}, as well as an irrelevant term depending only on $\sigma_0$, the secular normal form in the octupole approximation resumes the form of Eq.~\eqref{Ham_inter_norm_otto_C} with
\begin{align*}
& {\t{A}}\!=0, &\!\!\!\! &   {\t{C}}\!=-0.0283636, 
& {\t{B}}\!= -0.029019,  \\
& {\t{D}}(\sigma_0)\!=0.0011798 - 0.21118 \, \sigma_0, &\!\!\!\!
& {\t{E}}(\sigma_0)\!=0.00153399 - 0.270077\, \sigma_0\, .  &\!\!\!\!
\end{align*}
After performing the preliminary rotation described in Subsection~\ref{subsec:Norm_Hint1}, it takes the form of Eq.~\eqref{Ham.K}:
\begin{equation}
\label{Ham.rot.num.ex}
\widetilde{\Hscr}_{int}^{(N_P=3, N_{bk}=4)}=\Kscr_{I}(\t{\sigma}_1, \t{\sigma}_3; \sigma_0)= \AA\t{\sigma}_1^2+\CC\t{\sigma}_3^2+(\DD_1
{\sigma}_0 + \DELTA_1)\t{\sigma}_1+(\DD_3 {\sigma}_0 + \DELTA_3)\t{\sigma}_3 
\end{equation}
with
\begin{align*}
&\AA\!=0.00610734, & & \CC\!= -0.0344709,  & &
\DD_1\!=-0.089863, \\
&\DELTA_1\!= 0.000492281 &
& {\DD_3}\!=-0.330852,  & 
& {\DELTA_3}\!= 0.00187156,
\end{align*}
and Eqs.~\eqref{eq.CP1} and~\eqref{sol.CP2} can be solved explicitly yielding  the bifurcation values in the parameter $\sigma_0$ where critical points of the first and second kind appear. We find $\sigma_{0}^{( {CPI},1)}=0.00489265$ and $\sigma_{0}^{( {CPI},2)}=0.00655611$, while the values $\sigma_0^{( {CPII},1)}= 0.00623676$ and $\sigma_0^{( {CPII},2)}= 0.00497142$ correspond to the appearance/disappearance of  {CPII}, entering/leaving the circle $\sigma_1^2+\sigma_3^2=\sigma_0^2$ in the plane $\sigma_2=0$ (on the sphere, two CPs of the second kind bifurcate from/off a critical point of first kind  corresponding to a tangency on the plane $\sigma_2=0$).

\begin{figure}
\centering
\includegraphics[width=1\textwidth]{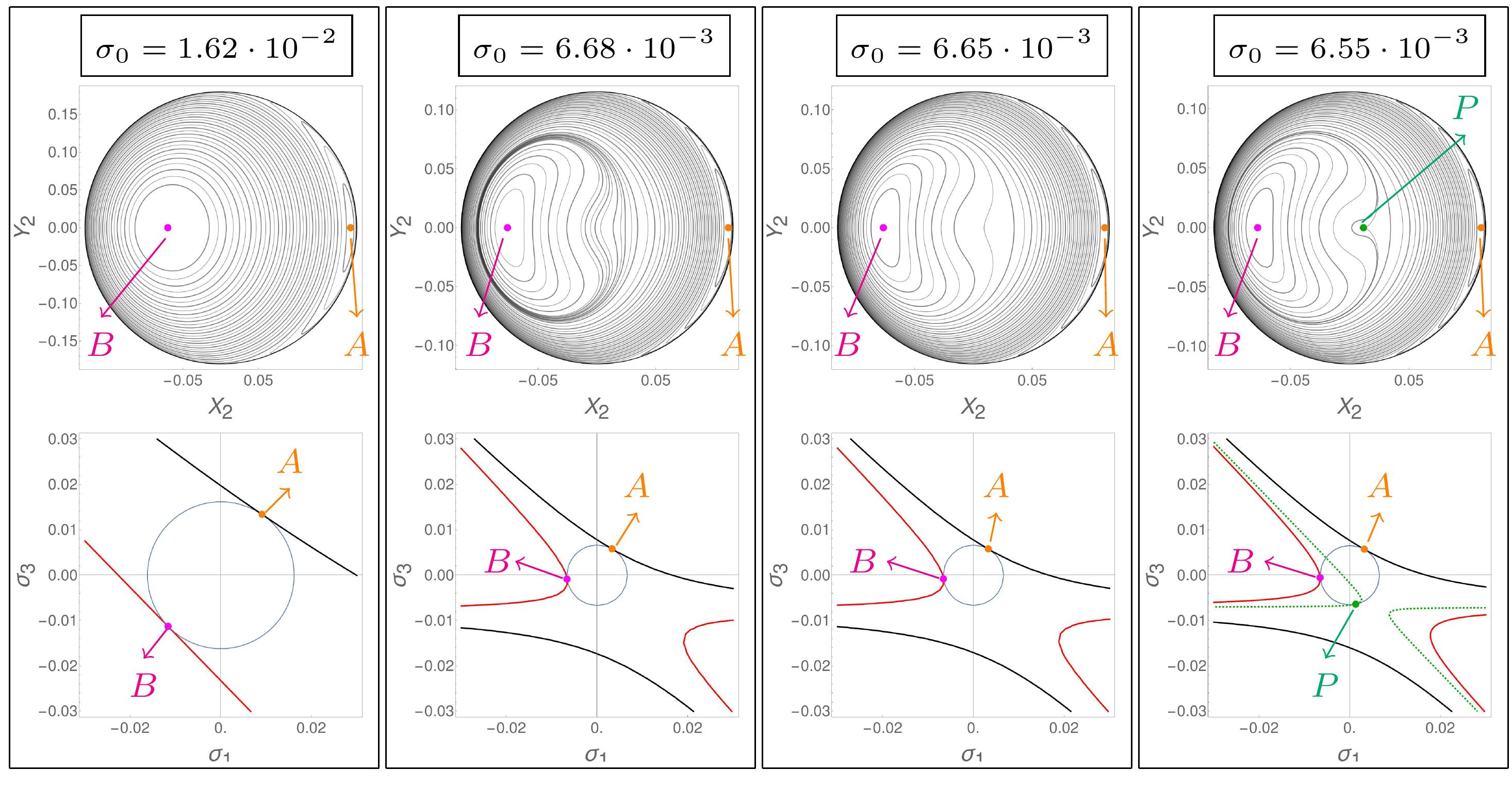}
\includegraphics[width=1\textwidth]{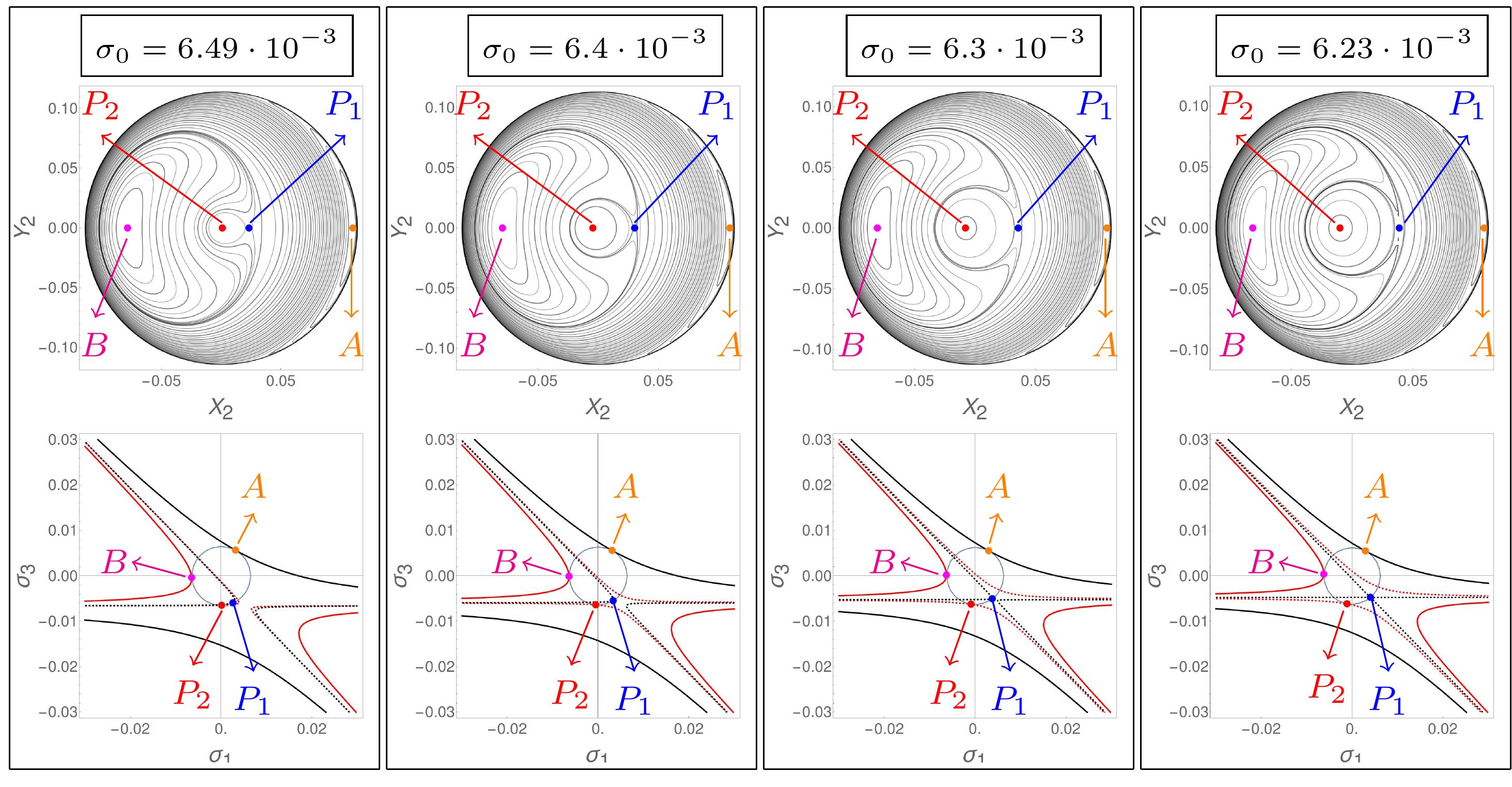}
\caption{
Phase portraits and analysis of the bifurcations (see text), same as in Figure~\ref{Fig.Poin_Hopf_INTEG_NORMAL}, but for the Hamiltonian $\widetilde{\Hscr}_{int}^{(N_P=3, N_{bk}=4)}$.}
\label{Fig.Poin_Hopf_INTEG_NORMAL_npol3eps4}
\end{figure}
\begin{figure}
\ContinuedFloat
\centering
\includegraphics[width=1\textwidth]{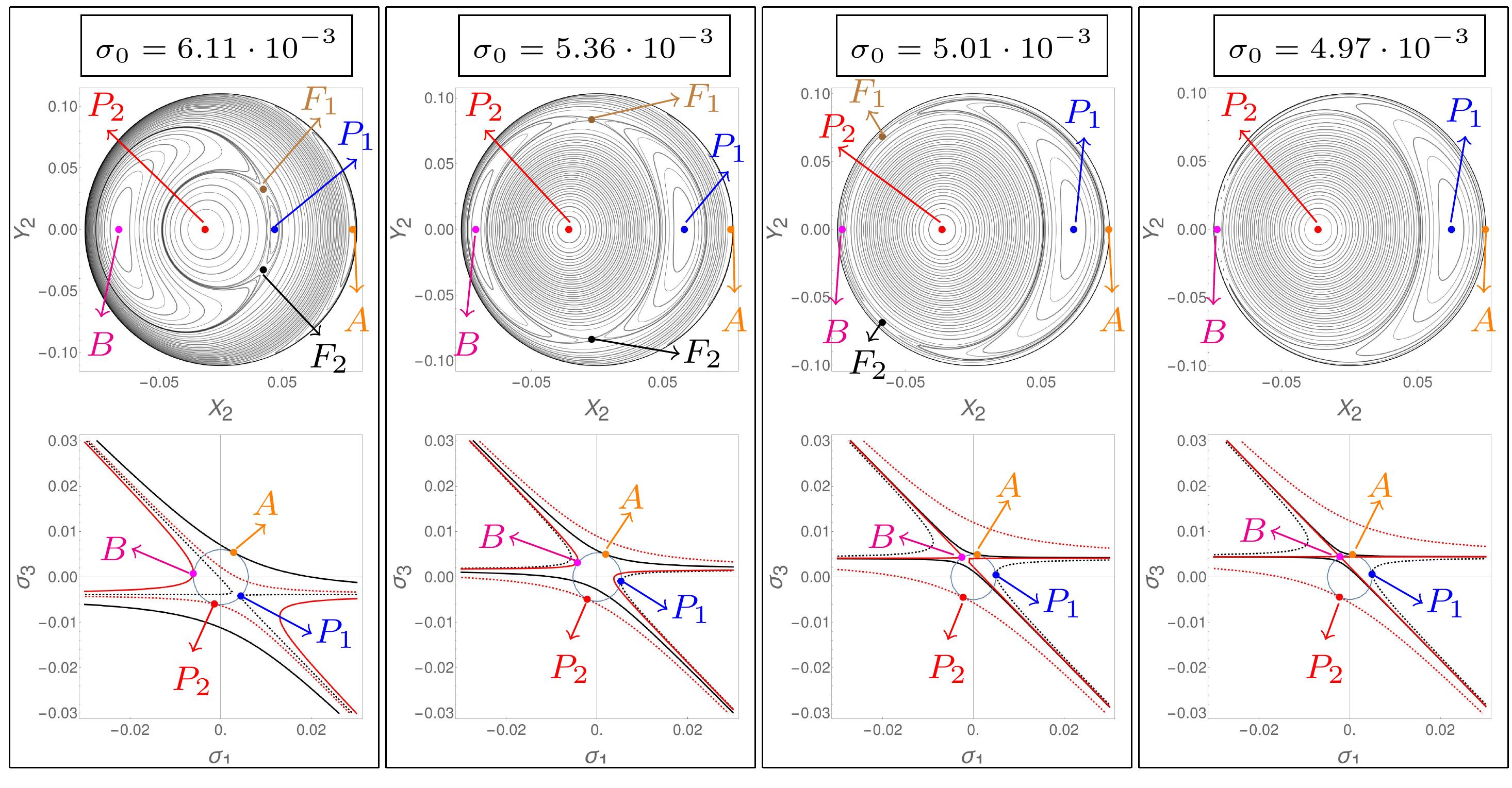}
\includegraphics[width=1\textwidth]{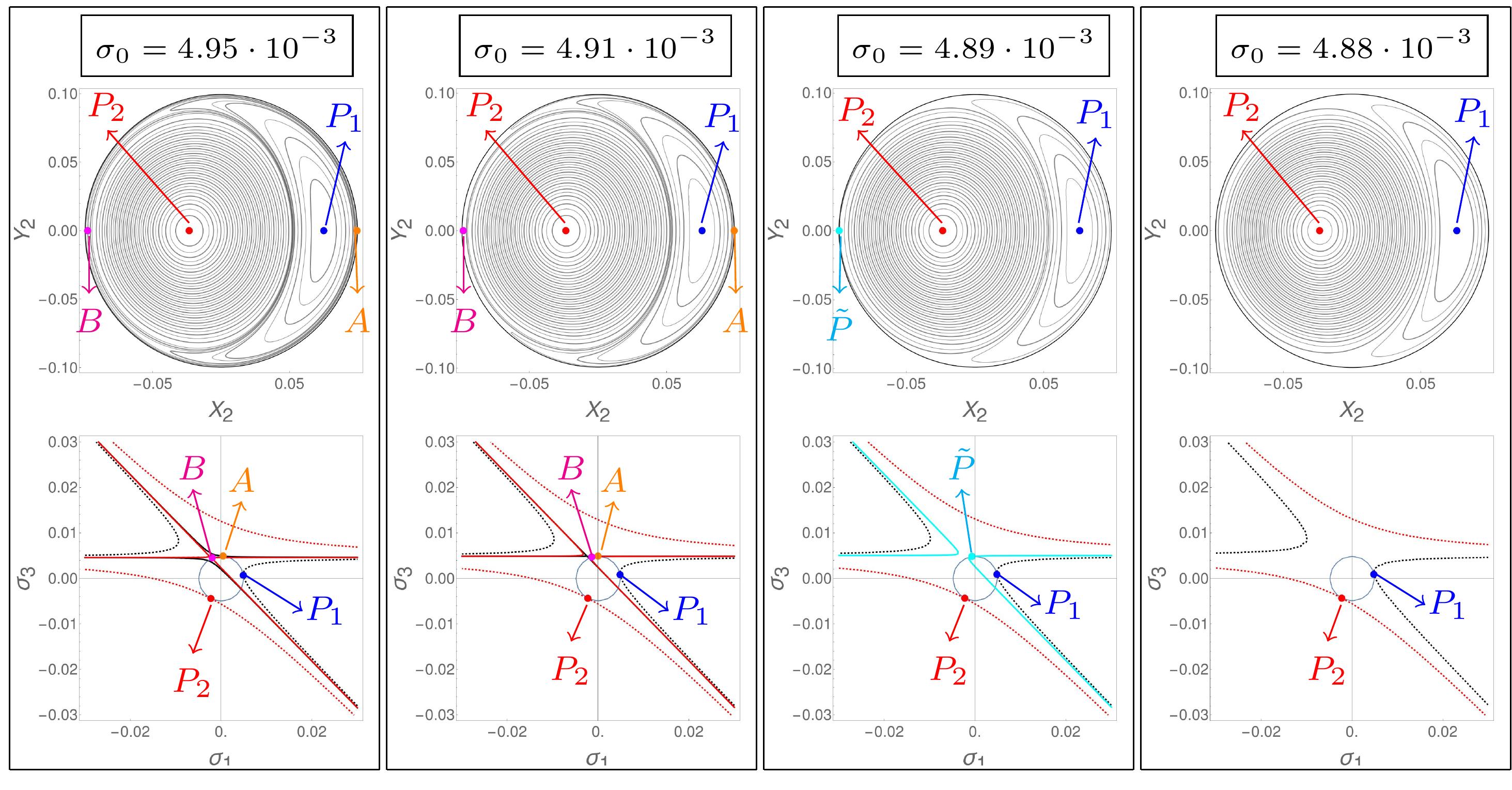}
\caption{Continued}
\end{figure}
The sequence of bifurcations observed for the Hamiltonian $\widetilde{\Hscr}_{int}^{(N_P=3, N_{bk}=4)}$ can now be analyzed using the same geometric method as for the Hamiltonians $\Hscr_{int}$ and $\Hscr_{int}^{(1)}$
(see Fig.~\ref{Fig.Poin_Hopf_INTEG_NORMAL_npol3eps4}). The usual ACR states A and B are given by external tangencies between $\mathcal{S}_{\sigma_0}$ and $\mathcal{C}_{\sigma_0,\, \Escr}$, the curves $\mathcal{C}_{\sigma_0,\, \Escr}^{(\sigma_2=0)}$ being hyperbola-like in this case as well. The fourth panel corresponds to the bifurcation value where one of the two analytical solutions of Eq.~\eqref{eq.CP1}  {is found}, $\sigma_0^{( {CPI}, 2)}=0.00655611$. Note that the analytical determination of all bifurcation values is due to the simplicity of the model, which leads to analytical expressions for all coefficients. Decreasing the value of $\sigma_0$ (fifth panel), the phase portrait exhibits two new periodic orbits $P_1$ and $P_2$. Looking in detail the passage from the fourth to the fifth panels, the bifurcation corresponds to the fact that the green hyperbola $\mathcal{C}_{\sigma_0^{( {CPI}, 2)},\, \Escr}^{(\sigma_2=0)}$, tangent to $\mathcal{S}_{\sigma_0^{( {CPI}, 2)}}^{(\sigma_2=0)}$ at the point $P$ in the fourth panel, splits into two hyperbolas in the fifth panel, one (red dashed) corresponding to an outer tangency yielding the stable fixed point $P_2$, and the other (black dashed) corresponding to an inner tangency yielding  the unstable fixed point $P_1$. Decreasing further the value of $\sigma_0$ (6th to 8th panel) the two branches of the hyperbola (black dashed) yielding the tangency at $P_1$  come closer and closer, until they merge at a bifurcation of a ` {CPII}' type ($\sigma_0^{( {CPII},1)}= 0.00623676$). Decreasing further $\sigma_0$ (9th panel) $P_1$ becomes stable by a pitchfork bifurcation, giving rise to two new unstable fixed points ( {CPII}) $F_1$ and $F_2$.
For still smaller $\sigma_0$ (10th and 11th panel) the fixed points $F_1$ and $F_2$ move away from the fixed point $P_1$, eventually colliding with the B mode (12th panel)  {at value $\sigma_0=\sigma_0^{( {CPII},2)}= 0.00497142$ which, again, can be computed analytically through Eq.~\eqref{sol.CP2}}. This terminates the $F$-family of periodic orbits by an inverse pitchfork bifurcation which renders the point B unstable.
 {In fact}, at still smaller $\sigma_0$ (13th panel), mode B  {is} unstable ( {having} internal tangency), while, as $\sigma_0$ decreases, the hyperbolas passing through the tangencies at B and A (red and black in the 13th and 14th panels) approach each other and eventually collide to a single hyperbola (15th panel, cyan) at $\sigma_0=\sigma_0^{( {CPI},1)}=4.89\cdot 10^{-3}$, computed also analytically. At the collision the points A and B collide to a single bifurcation point $\tilde{P}$. Then, for $\sigma_0<\sigma_0^{( {CPI},1)}$ the number of tangency points passes from four to two, i.e., the only stable points surviving at the end of the bifurcation sequence are $P_2$ and $P_1$.  

 {
We conclude observing than, notwithstanding its simplicity compared to the full secular Hamiltonian $\Hscr_{sec}$, the model
$\tilde{\Hscr}_{int}^{(N_P =3,N_{ bk} =4)}$ \textit{qualitatively} recovers the same sequence of bifurcations as in the full secular dynamics, in the entire range of energies up to $\Escr=\Escr_{C,2}$, i.e., before the birth of the Kozai-Lidov regime. Comparing Fig.~\ref{Fig.npol5eps12} and Fig.~\ref{Fig.Poin_Hopf_INTEG_NORMAL_npol3eps4}, despite the \textit{quantitative} difference in the form of the invariant curves and the precise values of the energy where bifurcations take place, the bifurcation sequences obtained by the two models are qualitatively the same. On the other hand, quantitatively comparable results with the two models are expected to hold in the case of hierarchical systems ($a_2\ll a_3$), in which the octupolar approximation of the Hamiltonian is sufficiently accurate. }

\section{Conclusions}
\label{sec:Conclusions}
In the present paper, we applied the geometric reduction method of \cite{kum1976, marpuc2016, Carrascoetal2018}, based on the representation of the phase space in Hopf variables \cite{vderMeer, cusbat1997, hansetal2000}, to address a central problem in planetary orbital theory, namely the analysis of all possible bifurcation sequences in the 3D secular planetary three-body problem leading to periodic orbital states other than the apsidal corotation resonances. Our purpose is twofold: i) to establish the efficiency of the method in producing analytical predictions for the emergence of periodic orbits of the system through bifurcations from the ACR states, and ii) to propose a suitable integrable secular normal form model able to qualitatively capture the sequence of bifurcations as observed in the complete (non-integrable) secular Hamiltonian. Our concluding remarks and results are summarized in the following:
\begin{enumerate}
\item Stemming from the definition of the basic invariant objects of the geometric reduction method for a generic Hamiltonian in 1:1 resonance, i.e., the sphere $\Sscr_{\sigma_0}$ and the energy surface $\Cscr_{\sigma_0, \Escr}$ (Section~\ref{sec:Bif}), the complete sequence of bifurcations of periodic orbits in the integrable approximation to the Hamiltonian can be recovered simply computing the tangencies or degenerate transverse intersections between the two surfaces $\Sscr_{\sigma_0}$ and $\Cscr_{\sigma_0, \Escr}$. In particular, depending on the three-body model's numerical parameters, the surface $\Cscr_{\sigma_0, \Escr}$ can be either an elliptic cylinder, or a set of two hyperboloid sheets. The degenerate limits of these surfaces are curves close to straight lines in the elliptic case, or two intersecting nearly planar surfaces in the hyperbolic case. Depending on the case considered, we are led to substantial differences as regards the resulting sequence of bifurcations, as well as the form and stability of periodic orbits stemming from the A or B-type ACR states. Both saddle-node and pitchfork bifurcations,  arising by changing a control parameter physically equivalent to the mutual inclination between the planetary orbits, can be interpreted by the above geometric representation. 

\item Section~\ref{sec:Apply} focuses on a comparison of the bifurcation sequences found in various levels of integrable Hamiltonian approximation to the full secular Hamiltonian $\Hscr_{sec}$ of the planetary three body problem. A main conclusion is that considering just $\Hscr_{int}$, i.e., the integrable part of $\Hscr_{sec}$, leads to a sequence of bifurcations  {at variance with} the one observed by the phase portraits of the complete secular Hamiltonian $\Hscr_{sec}$. However, using a  `book-keeping' method in conjunction with normalization by Lie series, we arrive at a secular normal form model $\Hscr^{(1)}_{int}$, which is able to qualitatively reproduce the same sequence of bifurcations as in the complete system. 

\item A special case of the previous analysis is the one of the octupolar approximation to the Hamiltonian truncated at order 4 in the orbital eccentricities (Subsection \ref{subsec:NormOct}). Computing a secular normal form as above for this last model leads to the Hamiltonian $\widetilde{\Hscr}_{int}^{(N_P=3, N_{bk}=4)}$, for which the bifurcation limits throughout the whole bifurcation sequence can be predicted analytically using the formulas of subsection \ref{subsec:formule_Anto_Gius}. This is particularly useful in cases of so-called hierarchical systems (see, for example,~\cite{miggoz2011},~\cite{foretal2000} and~\cite{naoetal2013a}) where the octupolar expansion is sufficiently precise. 
\end{enumerate}

As a final remark, we emphasize that the method here discussed is generic, i.e., applicable to any integrable Hamiltonian model which admits $\sigma_0$ (the total angular momentum) as a second integral and approximates the Jacobi-reduced 3D secular planetary dynamics. As regards the ability to provide explicit formulas for the critical bifurcation values of the periodic orbits of the system, this is only limited by the complexity of the formulas arising as higher order truncations are considered of the initial secular Hamiltonian. Thus, in cases of hierarchical systems, where a low order truncation of the Hamiltonian model is sufficiently precise, explicit formulas are easy to provide. In non-hierarchical systems, on the other hand, predictions by the method retain essentially their analytical character, the numerical operations to be performed being limited to computing the sign-definiteness of some suitably defined quadratic forms, as well as root-finding applied to the equations defining the tangencies (points  {CPI}) or transverse degenerate intersections (points  {CPII}) between the surfaces $\Sscr_{\sigma_0}$ and $\Cscr_{\sigma_0, \Escr}$.  {Note that, according to figures 17 and 18 of \cite{maseft2023}, the bifurcations dealt with in the present paper appear rather commonly for generic (non-hierarchical) values of the masses and/or semi-major axes of exoplanetary systems with mutually inclined planetary orbits. For example, apart from the case of the system $\upsilon-$Andromedae studied here, a simple comparison of the masses and semi-major axes shows that all the observed systems enlisted in Table 1 of \cite{libtzi2009} have a high probability to fall in the regime of an orbital configuration corresponding to one of the possible bifurcations  stemming from the basic ACR states A or B. The periodic orbits associated with such bifurcations are important also from the viewpoint that they represent the natural endstates of dissipative processes which are likely to occur in the period of formation of the exoplanetary systems. }\\
\\
\noindent {{\bf{Acknowledgments}}: G.P. acknowledges the support of INFN (Sezione di Roma2) and of GNFM/INdAM.} 

\appendix
\section{ {Numerical coefficients for the integrable Hamiltonian $\Hscr_{int}$}}
\label{appendix:num_val_H_int}

 {
Consider the integrable Hamiltonian $\Hscr_{int}$ (Eq.~\eqref{def.h0}). 
Up to terms of second order in the variables $\sigma_i$ (i.e. of fourth order in the eccentricities), apart from constants, we find the approximate formula 
$$
\Hscr_{int}\approx A\sigma_1^2+C\sigma_3^2+B\sigma_1\sigma_3+D(\sigma_0)\sigma_1+E(\sigma_0)\sigma_3+F(\sigma_0)\, ,
$$
that is equivalent to Eq.~\eqref{Ham.integ.} of Section~\ref{subsec:ApplyHint}.
For the parameters (masses, semi-major axes, and AMD value) as in the $\upsilon$-Andromed{\ae} system (see~\cite{maseft2023}) the coefficients read:
\begin{align*}
&A\!=0.00212824, &\!\!\!\! & C\!=0.00186469 ,  \\
&B\!= -0.00186482, &\!\!\!\!
&D(\sigma_0)\!=0.000165361 - 0.0159745 \, \sigma_0, \\ 
&E(\sigma_0)\!=0.0000214817 - 0.00532338\, \sigma_0,  &\!\!\!\!
&F(\sigma_0)\!=-0.00214065\,\sigma_0 - 0.108446 \,\sigma_0^2.
\end{align*}
The coefficients $A,B,C$ satisfy $B^2<4A C\,$, hence the quadratic form $A\sigma_1^2+C\sigma_3^2+B\sigma_1\sigma_3$ yields ellipses.
}

\section{ {Numerical coefficients for the secular integrable normal form $\Hscr_{int}^{(1)}$}}
\label{appendix:num_val_H_int_norm}
 {
Consider the integrable Hamiltonian $\Hscr_{int}^{(1)}$ (Eq.~\eqref{Ham_integ_norm}). 
Up to terms of second order in the variables $\sigma_i$ (i.e. of fourth order in the eccentricities), the secular normal form reads (apart from constants)
$$
\Hscr^{(1)}_{int}\approx A^{(1)}\sigma_1^2+C^{(1)}\sigma_3^2+B^{(1)}\sigma_1\sigma_3+D^{(1)}(\sigma_0)\sigma_1+E^{(1)}(\sigma_0)\sigma_3+F^{(1)}(\sigma_0)
$$ 
that is equivalent to Eq.~\eqref{Ham.integ.norm} of Section~\ref{subsec:Norm_Hint1}.
For the parameters (masses, semi-major axes, and AMD value) as in the $\upsilon$-Andromed{\ae} system (see~\cite{maseft2023}) the coefficients read:
\begin{align*}
&A^{(1)}\!=0.00912208, &\!\!\!\! & C^{(1)}\!=-0.0518887,  \\
&B^{(1)}\!= 0.00782589,  &\!\!\!\!
&D^{(1)}(\sigma_0)\!=0.000425947 + 0.00634477 \, \sigma_0, \\ 
&E^{(1)}(\sigma_0)\!=0.00101564 - 0.118017 \, \sigma_0,  &\!\!\!\!
&F^{(1)}(\sigma_0)\!=-0.0011129\,\sigma_0 -0.164326  \,\sigma_0^2.
\end{align*}
Being $(B^{(1)})^2>4A^{(1)} C^{(1)}\,$, the quadratic form $A^{(1)}\sigma_1^2+C^{(1)}\sigma_3^2+B^{(1)}\sigma_1\sigma_3$ yields hyperbolas.
}

\bibliographystyle{abbrvnat}
\bibliography{bibliography}

\end{document}